\documentclass[letterpaper,11pt]{article}
\pdfoutput=1
\pdfoutput=1
\pdfoutput=1

\usepackage{jheppub}
\usepackage{multirow}

\usepackage{subfig}
\usepackage{xspace}
\usepackage[countmax]{subfloat}

\usepackage{tikz}
\usetikzlibrary{math,calc,snakes, patterns,angles,quotes,arrows.meta,shapes.misc,decorations.pathmorphing,decorations.markings}

\usepackage{amssymb}
\usepackage{amsmath}
\usepackage{cancel}
\usepackage{tabu,booktabs}
\usepackage{color}
\usepackage{braket}
\usepackage{graphicx}
\graphicspath{ {./figures/} }
\usepackage{multirow}
\usepackage{verbatim}
\usepackage{amsthm}
\usepackage{slashed}
\usepackage{wasysym}
\usepackage{simplewick}
\usepackage{mathtools}
\usepackage{soul}
\usepackage{xspace}

\definecolor{verde}{RGB}{1, 113, 0}

\usepackage{enumitem}

\usepackage[only,llbracket,rrbracket]{stmaryrd}

\def\cC{\mathcal{C}}

\def\cJ{\mathcal{J}}

\def\cN{\mathcal{N}}

\def\cJ{\mathcal{J}}

\def\cX{\mathcal{X}}
\def\cY{\mathcal{Y}}

\def\nn{{\nonumber}}

\newcommand{\namedref}[2]{#1~\hyperref[#2]{\ref*{#2}}}
\newcommand{\secref}[1]{\namedref{Section}{#1}}
\newcommand{\appref}[1]{\namedref{Appendix}{#1}}
\newcommand{\tabref}[1]{\namedref{Table}{#1}}
\newcommand{\figref}[1]{\namedref{Figure}{#1}}

\def\be#1\ee{\begin{align}#1\end{align}}

\def\({\left(}
\def\){\right)}
\def\[{\left[}
\def\]{\right]}

\allowdisplaybreaks[3]

\newcommand{\avg}[1]{\left< #1 \right>}

\newcommand{\rhoavg}[1]{\llbracket #1 \rrbracket}

\newcommand{\eps}{\epsilon}

\newcommand{\Res}{\text{Res}}

\usepackage[font={small}]{caption}

\newcommand{\eeq}{\end{equation}}

\newcommand{\eeqq}{\end{equation*}}

\newcommand\eeqaa{\end{eqnarray*}}

\newcommand\eeqa{\end{array}}

\newcommand{\eea}{\end{eqnarray}}

\renewcommand{\Im}{\operatorname{Im}}

\newcommand{\Nmax}{N_{\text{max}}}

\newcommand{\G}{\Gamma}

\usepackage{graphicx} 

\usepackage[section]{placeins}

\title{The Stringy S-matrix Bootstrap:\\ Maximal Spin and Superpolynomial Softness}

\author[a,b]{Kelian H\"aring}
\author{and}
\author[a]{Alexander Zhiboedov}

\affiliation[a]{Theoretical Physics Department,
	CERN, 1211 Geneva 23, Switzerland}

\affiliation[b]{Fields and Strings Laboratory, Institute of Physics\\ École Polytechnique Fédéral de Lausanne (EPFL)
	\\ Route de la Sorge, CH-1015 Lausanne, Switzerland}

\abstract{We explore the space of meromorphic amplitudes with extra constraints coming from the shape of the leading Regge trajectory. 
This information comes in two guises: it bounds the maximal spin of exchanged particles of a given mass; it leads to sum rules obeyed by the discontinuity of the amplitude, which express the softness of scattering at high energies. We assume that the leading Regge trajectory is linear, and we derive bounds on the low-energy Wilson coefficients using the dual and primal approaches. For the graviton-graviton scattering in four dimensions, the maximal spin constraint leads to slightly more stringent bounds than those that follow from general constraints of analyticity, crossing, and unitarity. The exponential softness at high energies is manifest in our primal approach and is not used in our implementation of the dual approach. Nevertheless, we observe the agreement between the bounds obtained from both. We conclude that high-energy superpolynomial softness does not leave an obvious imprint on the low-energy observables. We exhibit a unitary three-parameter deformation of the Veneziano amplitude for the open string case. It has a novel, exponentially soft behavior at high energies and fixed angles. We generalize the previous analysis of this regime and present a stringy version of the lower bound on high-energy, fixed-angle scattering by Cerulus and Martin.}
\begin{document}

\begin{flushleft}
	\hfill \parbox[c]{40mm}{CERN-TH-2023-214}
\end{flushleft}

\maketitle

\section{Introduction}

One of the manifestations of locality in quantum field theory is a polynomial behavior of scattering amplitudes at high energies \cite{Eden:1971fm,Jaffe:1967nb,Polchinski:2001tt}. Stringy amplitudes famously violate this polynomiality and exhibit an exponential behavior at high energies and fixed angles \cite{Gross:1987kza,Gross:1989ge,Caron-Huot:2016icg}. Gravitational amplitudes are expected to violate the simple polynomial behavior as well due to black holes \cite{Amati:1987uf,Arkani-Hamed:2007ryv,Giddings:2009gj,Bah:2022uyz}.

This paper explores the possible effects of the amplitudes non-polynomiality at high energies on the low-energy observables, such as the Wilson coefficients. 
To make the problem tractable, we consider weakly coupled stringy scattering. In that context, let us introduce the leading Regge trajectory $j(t)$, which captures the basic high-energy properties of the amplitude. It is defined by taking the high-energy limit with momentum transfer $t$ kept fixed\footnote{For meromorphic amplitudes of interest, we take $s \to \infty$ with $\arg s >0$ kept fixed.}
\begin{equation}\label{eq:regge}
    T(s,t) \sim f(t) s^{j(t)} \,, \quad s\to \infty\,,~~~ t \text{ fixed}\, . 
\end{equation}
At positive $t$, it is related to the spectrum of the exchanged particles \cite{Caron-Huot:2016icg}. For negative $t$, it captures the behavior of the amplitude in the actual high-energy scattering experiment. This paper explores extra constraints on the low-energy observables drawn from some knowledge about $j(t)$.

The amplitudes of interest are described by meromorphic functions with extra constraints that we impose:
\begin{enumerate}[label=(\Alph*)]
    \item \label{ass:ACU} Standard bootstrap constraints: Analyticity (meromorphy), unitarity, and crossing symmetry (ACU). Because we effectively work at the tree level, unitarity is reduced to positivity, see \cite{Chen:2022nym,EliasMiro:2022xaa}. 
    \item \label{ass:ACU+J} Maximal spin constraint: $J(m^2)\leq j(m^2)$, where $J(m^2)$ is the maximal spin of the exchanged particle of mass $m$, and $j(m^2)$ is the leading Regge trajectory that we consider to be given. 
    \item \label{ass:ACU+J+RSR} Superpolynomial softness: we impose that the amplitude decays at high energies faster than any given power for negative enough $t<0$. In other words, for any $N \in \mathbb{Z_+}$ there exists $t<0$ such that $j(t)<-N$. These conditions are conveniently expressed in the \emph{Regge sum rules} (RSR), which will be introduced in the following.
\end{enumerate}
We, therefore, see that the additional assumptions \ref{ass:ACU+J} and \ref{ass:ACU+J+RSR} are related to the properties of the leading Regge trajectory $j(t)$ for positive and negative $t$ respectively. It is an interesting question to what extent the properties of $j(t)$ at negative and positive $t$ are related to each other, and we briefly comment on this question further in our conclusions. 

An example of the amplitude that satisfies \ref{ass:ACU}, but violates both \ref{ass:ACU+J} and \ref{ass:ACU+J+RSR} is given by $T(s,t)={1 \over (s-m^2)(t-m^2)}$, for which $j(t)=-1$ and $J(m^2)=\infty$. There are amplitudes that satisfy \ref{ass:ACU+J} and do not satisfy \ref{ass:ACU+J+RSR}, e.g., glueball scattering in large $N$ QCD, or recently constructed deformations of the Veneziano amplitude considered in \cite{Cheung:2023adk}. Finally, some amplitudes satisfy both \ref{ass:ACU+J} and \ref{ass:ACU+J+RSR}, such as for example the Veneziano amplitude \cite{Veneziano:1968yb} or the Coon amplitude \cite{Coon:1969yw,Baker:1970vxk,Coon:1972qz}.

In this paper, we focus on the case when the leading Regge trajectory is linear
\be
\label{eq:linearity}
j(m^2) = j_0 + \alpha' m^2 ,
\ee
where $\alpha'$ is the string tension and $j_0$ is the so-called Regge intercept. We derive bounds on the Wilson coefficients using both the so-called dual and primal approaches. 

In \emph{the dual approach}, reviewed in \appref{sec:dualMethod}, we derive bounds on the low-energy expansion of the amplitude without explicitly constructing the amplitude. A standard tool to do it is via dispersion relations \cite{Adams:2006sv}. In this case, \ref{ass:ACU+J} is implemented at the level of the discontinuity of the amplitude. On the other hand, \ref{ass:ACU+J+RSR} can be implemented using the \emph{Regge sum rules} that we introduce shortly below. In \emph{the primal approach}, we explicitly write down an ansatz for the amplitude that satisfies \ref{ass:ACU+J} and \ref{ass:ACU+J+RSR}, as well as analyticity and crossing, and we impose unitarity numerically.

For our dual results, the assumption \eqref{eq:linearity} about the linearity of the leading Regge trajectory can be easily relaxed, and any desired shape of the Regge trajectory (e.g., taken from the lattice data \cite{Lucini:2001ej}) could be put in. For the primal approach, our analysis could be generalized along the lines of \cite{Veneziano:2017cks} or \cite{Cheung:2023uwn}, which allow certain flexibility in the shape of the leading Regge trajectory.

We set the mass of the lightest massive state at $m_{\text{gap}}^2=1$. We will consider two types of amplitudes, which we call \emph{open} and \emph{closed}, following the example of fundamental strings. They are distinguished by the structure of poles, as well as by $\alpha'$ that appear in \eqref{eq:linearity}
\be
\label{eq:openclosed}
\alpha'_{\text{open}} = 1 , ~~~~~~\alpha'_{\text{closed}} = 2 \ .
\ee
For the closed string case, we consider the MHV scattering amplitude of gravitons in four spacetime dimensions. For the open string case, we consider the scattering of massless scalars in four spacetime dimensions.

\subsection{Review of the results}

Our paper is divided into two parts: the closed string case and the open string case. Apart from \eqref{eq:openclosed}, the difference between the two cases is that for the open string case, we assume that the amplitude has only poles in the $s$- and $t$- channels, whereas for the closed string case poles in all three channels are present.\\

\noindent {\bf Closed string case}\\[5pt]
We consider the MHV scattering amplitude of gravitons in four dimensions previously considered in \cite{Arkani-Hamed:2020blm,Bern:2021ppb,Caron-Huot:2022ugt,Chiang:2022jep}. We assume the linear Regge trajectory to be $j(t)=2+2t$, where $m_{\text{gap}}^2=1$. We derive bounds on the Wilson coefficients using the dual and primal approaches.

For the dual approach, this case was previously considered in \cite{Arkani-Hamed:2020blm,Bern:2021ppb,Caron-Huot:2022ugt,Chiang:2022jep}, where \ref{ass:ACU} was imposed. We find that imposing \ref{ass:ACU+J} leads to  slightly more stringent bounds excluding the small regions of parameter space around the amplitudes which involve particles of all spin at a given mass. 

We restate superpolynomial softness in terms of the \emph{Regge sum rules} (RSR) on the discontinuity of the amplitude, see \eqref{eq:RSRbasic} below.  However, due to the `oscillating' nature of the RSR, stemming from the fact that Legendre polynomials are not sign-definite for $t<0$, we find that these constraints \emph{are not used} in the numerics. Therefore, we do not get any difference between the dual bounds obtained from \ref{ass:ACU+J} and \ref{ass:ACU+J+RSR}. This issue is similar to the one described in \cite{Albert:2023jtd}, and we discuss it further in \appref{sec:RSRandDual}. 

At this point, however, we cannot be sure that this effect is not just a technical artifact of the current implementation of the dual bootstrap scheme, which, in particular, can impose only a finite number of RSR constraints. To make progress on this question, we develop a primal approach, where we explicitly construct amplitudes that satisfy \ref{ass:ACU+J+RSR}, see \eqref{eq:f_ClosedInf}. A remarkable fact about this ansatz is that it satisfies all the desired properties at finite $N_{\text{max}}$. We can, therefore, derive bounds on Wilson coefficients numerically first for finite $N_{\text{max}}$, and then extrapolate them to $N_{\text{max}} \to \infty$. We do not observe a clear gap between the primal and dual results within the available precision. We conclude that the extra constraints due to superpolynomial softness (not used in the dual approach and manifest in the primal approach) do not lead to stronger bounds.

Our results for various Wilson coefficients are summarized in Figures \ref{fig:beta3vsa00Nmax},\ref{fig:a2_SpectrumConstraints} and \ref{fig:a4at05_SpectrumAssumption}.\\

\noindent {\bf Open string case}\\[5pt]
The dual approach for the open string case leads to results very similar to the closed string case. Again we find that imposing the maximal spin condition leads to stronger bounds, whereas the superpolynomial softness, imposed through a finite number of RSR constraints, does not lead to visible effects. 

The situation with the primal approach, however, is very different. In this case, any truncation of the ansatz \eqref{eq:sumVenAnsatz} to a finite number of terms violates unitarity. Therefore we do not have a systematic way to derive the primal bounds in this case. Nevertheless, we identify an interesting
$N_{\text{max}}=\infty$ class of deformations which satisfy \ref{ass:ACU}, \ref{ass:ACU+J}, and \ref{ass:ACU+J+RSR}. They are conveniently given by the worldsheet integral \eqref{eq:MatsudaGen}.

A remarkable property of these amplitudes is a novel behavior at high energies and fixed angles, see \eqref{eq:fixedangleOS}. In particular, they go beyond the analysis of \cite{Caron-Huot:2016icg} in several respects, thus emphasizing the restricting nature of technical assumptions made in that paper. 

Based on these results, we propose a bound on the high-energy fixed (complex) angle behavior of the meromorphic stringy amplitudes and use it to derive a lower bound on high-energy fixed angle scattering \eqref{eq:CMstringy}, which is analogous to the old result by Cerulus and Martin in the context of gapped, relativistic QFTs.

Our results for various Wilson coefficients are summarized in \figref{fig:openG2DualJ}. In this case, our dual and primal bounds do not coincide. However, it is not very surprising given that our primal approach is not systematic, and further work is needed to clarify the interplay between \ref{ass:ACU+J} and \ref{ass:ACU+J+RSR} in that case.

\subsection{Connection to recent literature}

For the reader's convenience, let us comment on the relationship of this paper to the recent work on related topics. For the graviton scattering, bounds on low energy observables using the usual bootstrap axioms \ref{ass:ACU} were considered in \cite{Arkani-Hamed:2020blm,Bern:2021ppb,Caron-Huot:2022ugt,Chiang:2022jep}. Here, we consider the same observables and add extra constraints on the leading Regge trajectory \ref{ass:ACU+J} and \ref{ass:ACU+J+RSR}. In \cite{Arkani-Hamed:2020blm}, the authors considered a unitary deformation of the Virasoro-Shapiro amplitude with a single satellite term. In this work, we systematically constructed such deformations with an arbitrary number of satellite terms. 

For the open string case, bounds on low energy observables using the usual bootstrap axioms \ref{ass:ACU} were considered in \cite{Albert:2022oes,Fernandez:2022kzi,Li:2023qzs} in the context of large $N$ QCD. Here we imposed extra constraints on the leading Regge trajectory \ref{ass:ACU+J} and \ref{ass:ACU+J+RSR}. In \cite{Cheung:2023adk}, the authors derived a unitary deformation of the Veneziano amplitude. While this amplitude satisfies the maximal spin constraint \ref{ass:ACU+J}, it does not satisfy the Regge sum rules \ref{ass:ACU+J+RSR}. In this work, we find a different family of unitary deformations that satisfy both the maximal spin constraint \ref{ass:ACU+J} and superpolynomial softness \ref{ass:ACU+J+RSR}. These amplitudes have interesting high-energy, fixed-angle behavior and violate some of the technical assumptions made in \cite{Caron-Huot:2016icg}. We relax some of these assumptions and propose a new bound on the high-energy, fixed-angle scattering. 

Other approaches have been pursued in the literature to further restrict the space of stringy amplitudes. One interesting direction was followed in \cite{Huang:2020nqy,Chiang:2023quf,Berman:2023jys}, where the authors imposed that amplitudes satisfy certain monodromy relations stemming from the worldsheet representation of the amplitude. In this case, the space of allowed Wilson coefficients is drastically reduced. Note that the deformation of the Veneziano considered here \eqref{eq:MatsudaGen} does not satisfy the standard monodromy relation. 

Extensions of the open string amplitudes to different spectra and nonlinear leading Regge trajectories were also recently pursued. One notable deformation of the spectrum leads to the so-called Coon amplitude \cite{Coon:1969yw,Baker:1970vxk,Coon:1972qz} which has been explored recently \cite{Figueroa:2022onw,Geiser:2022icl,Chakravarty:2022vrp,Bhardwaj:2022lbz,Jepsen:2023sia,Geiser:2022exp,Cheung:2022mkw,Geiser:2023qqq}.
Keeping the spectrum of the Veneziano amplitude intact, the authors of \cite{Veneziano:2017cks} constructed an explicit amplitude that exhibits bending of the leading trajectory expected in large $N$ QCD. More recently, open string amplitudes with an arbitrary spectrum were constructed and explored in \cite{Cheung:2023uwn}.

Finally, this work explores constraints from the high-energy superpolynomial softness of the amplitude at $t<0$. A related exploration was done in \cite{McPeak:2023wmq}, where the authors studied the consequences of changing the Regge intercept $j_0$ instead. 
The authors have observed that lowering the Regge intercept below $j_0<1$ led to little or no improvement of the bootstrap bounds for the closed string case (in the presence of the $u$-channel poles).\footnote{For the open string case when no $u$-channel poles are present there is an improvement.} The nontrivial effect appeared when the Regge intercept was lowered further $j_0<0$, so no subtractions are needed in the dispersion relations. The same phenomenon was observed for nonperturbative amplitudes in \cite{EliasMiro:2022xaa}.

\subsection{Plan of the paper}

The plan of the paper is as follows. In \secref{sec:constraints}, we review the basic assumptions and constraints imposed. In \secref{sec:gravitons}, we explore the closed string case, namely the MHV scattering amplitude of gravitons. We derive both primal and dual bounds on the low-energy Wilson coefficients. In \secref{sec:OpenStrings}, we consider the open string case, where we take external particles to be massless scalars. We derive dual bounds on the low-energy observables and construct new explicit amplitudes with several remarkable properties. We conclude in \secref{sec:conclusion}, where we discuss the results of this work and mention some future directions. We provide various appendices which contain additional details and we refer to them throughout the text in places where they become relevant. Notably, the appendices contain a review of the dual method in \appref{sec:dualMethod}, examples of amplitudes in \appref{sec:amplitudeExample}, a bound on the asymptotic form of the amplitude in \appref{app:asybound} and a version of the Cerulus-Martin bound for stringy amplitudes in \appref{app:CMstringy}.

\section{Assumptions and constraints}\label{sec:constraints}

Let us start by reviewing the standard assumptions satisfied by the tree-level (or meromorphic) scattering amplitudes. We then explain the extra constraints imposed in this work in more detail.

Here we list properties of tree-level two-to-two scattering amplitudes of massless scalar particles in four spacetime dimensions. The Mandelstam variables satisfy $s+t+u=0$.
\begin{enumerate}[label=(\roman*)]
    \item \label{ac:Meromormpy}\textbf{Meromorphy: } The scattering amplitude is described by a meromorphic function of two variables $T(s,t)$ where all the singularities are simple poles. 
    \begin{equation}
        T(s,t)\underset{s\to m_n^2}{\sim} -\frac{R_n(t)}{s-m_n^2},
    \end{equation}
    and $m_n$ are masses of exchanged particles.
    \item \label{ac:crossing} \textbf{Crossing symmetry: } 
    For general external particles $A,B,C,D$, crossing symmetry is the requirement that
    \begin{equation}
        T_{AB\to CD}(s,t) = T_{A \bar C\to \bar B D}(t,s) = T_{A \bar D\to \bar B C}(u,t)\,.
    \end{equation}
    In this work, we consider different combinations of external particles, and the exact form of crossing symmetry will be specified for each case separately. This property was recently proven in the planar limit \cite{Mizera:2021fap}.
    
    \item \label{ac:unitarity}\textbf{Unitarity: } 
     The residues $R_n(t)$ can be decomposed in partial waves
    \begin{equation}\label{eq:residuesT}
     R_n(t)= -\underset{s=m_n^2}{\Res\,} T(s,t) = \sum_{J=0} c_{n,J} P_J\left(1+ \frac{2t}{m_n^2}\right)\ ,
    \end{equation} 
    where $P_J(x)$ are the usual Legendre polynomial. Unitarity is the statement that
    \begin{equation}
        c_{n,J}\geq 0 \ . 
    \end{equation}
    In the case of spinning particles, unitarity takes the form of a semi-definite matrix as reviewed in \cite{Hebbar:2020ukp,Bern:2021ppb} for the case of graviton scattering.
    \end{enumerate}
    This ends the list of the usual bootstrap assumptions for the tree-level scattering amplitudes. In this work, we want to impose extra constraints coming from the shape of the leading Regge trajectory. 

    \begin{enumerate}[resume,label=(\roman*)]
        \item \label{ac:maxSpin} \textbf{Maximal spin:} We require the residue \eqref{eq:residuesT} to be polynomial in $t$ whose maximal power is bounded by the leading Regge trajectory
    \begin{equation}
         R_n(t)= \sum_{J=0}^{j(m_n^2)} c_{n,J} P_J\left(1+ \frac{2t}{m_n^2}\right)\,.
    \end{equation}
    This condition is essentially imposing the finite energy sum rules (FESR) considered in the past in \cite{Igi:1962zz,Logunov:1967dy,Igi:1967zza,Gatto:1967zza,Dolen:1967zz,Dolen:1967jr,Ademollo:1967zz,Ademollo:1968cno}.\footnote{See also \cite{Mukhametzhanov:2018zja} for the rigorous formulation of the FESR using Tauberian theorems.}  This constraint effectively puts in the information about the shape of the leading Regge trajectory \eqref{eq:regge} for positive $t$. 
    For the linear trajectory \eqref{eq:linearity}, it prevents the appearance of an infinite tower of exchange particles of arbitrary high spin at a given mass.
    \item \label{ac:RSR} \textbf{Regge sum rules (RSR):} It is a statement about the softness of the amplitude at negative $t$ and is best derived starting from the contour integral 
    \begin{equation}
    \label{eq:contourbasic}
        \frac{1}{2\pi i}\oint_\mathcal{C} ds' (s')^n T(s',t) =0
    \end{equation}
    where $\cC$ is the contour described in \figref{fig:ContourRSR}. We can split the integral into two parts
    \begin{equation}
       0= \int_{\mathcal{C}_\infty} ds' (s')^n T(s',t) + \frac{1}{\pi} \int_{-\infty}^{\infty} ds' (s')^n T_s(s',t)\,.
    \end{equation}
    The integral over the large circle $\mathcal{C}_\infty$ can be computed using the known Regge behavior \eqref{eq:regge}
    \begin{equation}
        \int_{\mathcal{C}_\infty} ds' (s')^n T(s',t) \sim \int_{\mathcal{C}_\infty} \frac{ds'}{s'} (s')^{j(t) + 1+n} = 0 \,,\quad\text{if } j(t)<-1-n\,.
    \end{equation}
    We thus obtain the \emph{Regge sum rules}
    \begin{equation}
    \label{eq:RSRbasic}
         \textbf{RSR:}~~~ \int_{- \infty}^\infty ds'  (s')^n T_s(s',t) = 0, ~~~ j(t) < -1-n \, ,
    \end{equation}
    which conveniently express the superpolynomial softness of the amplitude in terms of the constraints on the discontinuity of the amplitude. In the case of meromorphic amplitudes, the integral reduces to a sum as $T_s \sim \delta(s-m_n^2)$ (plus the $u$-channel contribution). 
     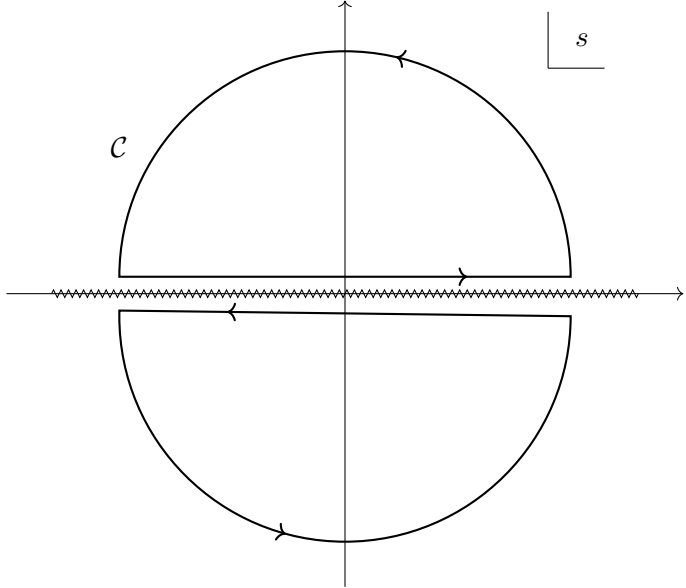
\begin{figure}
        \centering
    \begin{tikzpicture}[scale=1.5]
  \draw[->] (-3, 0) -- (3, 0);
  \draw[->] (0, -2.6) -- (0, 2.6);
  \draw[black, decoration = {zigzag, segment length = 1mm, amplitude = 0.5mm}, decorate] (-2.6,0) -- (2.6,0) ;

  \tikzmath{\ss = 2.8;}
  \draw (\ss-1,\ss-1+0.2)--(\ss-0.5,\ss-1 +0.2);
	\draw (\ss-1,\ss-1 +0.2)--(\ss-1 ,\ss-1 +0.7);
	\draw (\ss-1 +0.3,\ss-1 +0.45) node{$s$};
 
  \draw[thick, decoration={markings, mark=at position 0.3 with {\arrow{>}}, mark=at position 0.65 with {\arrow{>}}}, postaction={decorate}] (-2, 0.15) -- (2, 0.15) arc (0:180:2) -- cycle;
  
  \draw[thick, decoration={markings, mark=at position 0.1 with {\arrow{<}}, mark=at position 0.75 with {\arrow{<}}}, postaction={decorate}] (-2, -0.15) -- (2, -0.2) arc (0:-180:2) -- cycle;

  \draw (-2,1.3) node{$\mathcal{C}$};

\end{tikzpicture}
        \caption{The integration contour used to derive the Regge sum rules. Without loss of generality, we draw a cut along the full real axis.}
        \label{fig:ContourRSR}
    \end{figure}
\end{enumerate}

In the primal approach, we will explore the space of amplitudes by explicitly constructing them in a way that they obey all the above constraints. To write an explicit ansatz, we will have to choose a spectrum, and in this work, we consider amplitudes characterized by the equidistant spectrum.
\begin{enumerate}[resume,label=(\roman*)]
 \item \label{ac:linSpectrum} \textbf{Equidistant spectrum:} All particles in the spectrum have $m_n^2 = n\,$ with $n\in \mathbb{Z_+}$\,. 
\end{enumerate}
Let us emphasize that this is a \emph{technical} constraint, and we leave to future studies generalizations to more complicated spectra.
However, it is important to emphasize that we found that adding the equidistant spectrum assumption to the dual approach \emph{does not} affect the bounds on Wilson coefficients. In this sense, amplitudes with an equidistant spectrum are generic.

\section{Closed strings: MHV scattering of gravitons}\label{sec:gravitons}

In this section, we consider graviton scattering, which corresponds to the closed string case $\alpha'_{\text{closed}}=2$. In particular, we consider the two-to-two MHV amplitude
\begin{equation}\label{eq:MHVAmplitude}
        T_{++--}(s,t,u) = ([12]\langle 34\rangle )^4 f(s|t,u) \overset{\text{CMF}}{=} s^4 f(s|t,u) \ ,
\end{equation}
where crossing symmetry implies that $f(s|t,u) = f(s|u,t)$ and all other MHV amplitudes are described by the same function. By CMF we mean the center-of-mass reference frame, see \cite{Bern:2021ppb} for details. 

At low energy, the amplitude admits the following expansion
\begin{equation}\label{eq:f_lowEnergy}
        f(s|t,u)=  \frac{8 \pi G_N}{stu} + |\beta_{R^3}|^2 \frac{t u}{s} - |\beta_\phi|^2 \frac{1}{s} + \sum_{k\geq j\geq 0} a_{k,j} s^{k-j}t^j\,.
\end{equation}
The coefficients in this expansion define the so-called Wilson coefficients.  The first term is the well-known Einstein term, the second is the correction to the 3-pt coupling from the $R^3$ operator, and the third is due to the $\phi C^2$ coupling, which describes the massless scalar exchange. Finally, the $a_{k,j}$ are related to higher dimensional operators. We used the convention of \cite{Bern:2021ppb}, and we refer the reader to this reference for detailed computation of the low energy expansion \eqref{eq:f_lowEnergy} starting from the Lagrangian. Bounds on ratios of these Wilson coefficients were derived using the \emph{dual} method in \cite{Arkani-Hamed:2020blm,Bern:2021ppb,Caron-Huot:2022ugt,Chiang:2022jep} imposing only the standard bootstrap constraints (ACU) \ref{ass:ACU}. 

Next, we want to impose that the amplitude satisfies the well-known linear Regge behavior of string theory \eqref{eq:regge} with
\begin{equation}\label{eq:reggeMHVgrav}
    j(t) = 2+ 2t \,,\quad \forall t\,.
\end{equation}
As explained in \secref{sec:constraints}, we will impose this in two steps. First, it will be imposed for $t>0$ by bounding the maximal spin in the spectrum \ref{ass:ACU+J}. This can be easily done using the dual method reviewed in \appref{sec:dualMethod}. Second, we will also superpolynomial softness in the form of Regge sum rules at negative $t$. In this case, we observe that adding a finite number of Regge sum rules to our dual algorithm does not affect the bounds. To explore the space of Wilson coefficients when infinitely many Regge sum rules (or superpolynomial softness) are imposed, we will turn to the primal method, where these constraints are built-in. When constructing an ansatz, further assumptions have to be made about the spectrum, and we will assume an equidistant spectrum $m_n^2=n$. This assumption can also be made in the dual approach, and we did not observe any effects on the bounds by including it.

Note that in four spacetime dimensions, one-particle states are not good asymptotic states in gravity  \cite{Strominger:2017zoo}. This fact manifests itself through the IR divergences. There are two comments that we can make in this regard. First, all the basic ideas in the present paper are directly generalizable to $d>4$, and it would be interesting to do it explicitly. Second, in $d=4$, we expect that our conclusions should hold for the IR finite observables, e.g., for CFT correlators dual to gravitational theory in $AdS_4$, see \cite{Caron-Huot:2021enk,Chang:2023szz}. 

Next, we describe the ansatz in \secref{sec:ansatz_graviton} and the primal algorithm in \secref{sec:algorithmBound}, then we present the result for various Wilson coefficients in \secref{sec:result_gravitons}.

\subsection{Closed string ansatz}\label{sec:ansatz_graviton}

Our ansatz for closed string satisfying the constraint described in \secref{sec:constraints} is built out of the `Virasoro-Shapiro block'. It takes the following form\footnote{A similar ansatz was considered in the past in \cite{Altarelli:1969ck}, however, without imposing unitarity.}
\begin{align}\label{eq:f_ClosedInf}
        f(s|t,u) &= - 8 \pi G_N\frac{\G(-s)\G(-t)\G(-u)}{\G(1+s)\G(1+t)\G(1+u)} \nn\\
        &+ \sum_{c_s,c_t,c_u,d_s,d_t, d_u}^{'}\alpha_{c_s,(c_t,c_u),d_s,(d_t, d_u)} \frac{\G(c_s-s)\G(c_{t}-t)\G(c_{u}-u)}{\G(d_s+s)\G(d_{t}+t)\G(d_{u}+u)}\,,
\end{align}
where we used the symmetric notation $\alpha_{(i,j)} = \frac{1}{2}(\alpha_{i,j} + \alpha_{j,i})$ to emphasize that the function is $t-u$ symmetric. The first term is the well-known Virasoro-Shapiro amplitude for superstrings \cite{Virasoro:1969me,Shapiro:1970gy,Polchinski:1998rq}.  A unitary deformation with one satellite term $\alpha_{1,(1,1),2,(2, 2)} = - 8 \pi G_N \epsilon $, $0 \leq \epsilon \leq 1$ was recently considered in \cite{Arkani-Hamed:2020blm}. The $\sum'$ indicates that we only keep terms for which the residues are polynomial and which satisfy the Regge behavior \eqref{eq:reggeMHVgrav}
\begin{align}
    f(s|t,u)&\lesssim s^{-2+2t}\,\,, \quad s\to \infty\,,\, t \text{ fixed} \ , \\
     f(s|t,u)&\lesssim t^{2+2s}\,\,\,\,\,\,, \quad t\to \infty\,,\, s \text{ fixed}\,.
\end{align}
Not all the terms in the expansion above are independent and it is convenient to remove dependent terms. The dependence is nontrivial and we do not know of a general rule to select independent terms. We will discuss this point in more detail below.

Let us discuss the basic properties of this ansatz. It is obviously meromorphic with equidistant spectrum $m_n^2 = n \in \mathbb{Z}_+$ and crossing symmetry is built in. It also automatically satisfies the maximal spin constraint \ref{ass:ACU+J} because for a given $s$ only a finite number of terms in the ansatz contribute to the sum rule and each of them manifestly satisfies it. Regarding the superpolynomial softness, while each term satisfies it individually, the (infinite) sum might not. This point will be further discussed for the infinite sums of similar type in \secref{sec:RSR_OpenString}. Below, we consider the truncated sum for which superpolynomial softness will be manifest.  The only remaining constraint is thus unitarity, the latter is not automatic and imposes nontrivial constraints on the coefficients $\alpha$'s. 

Let us review here unitarity constraints for the MHV amplitude. As we are considering the scattering of spinning particles, we have a few different channels to consider. In the case at hand, the two independent channels are $(++\to ++)$ and $(+-\to +-)$ for which the residues \eqref{eq:residuesT} read\footnote{To label the $(++\to ++)$ amplitude, we use the all-in notation and call it $T_{++--}$ and similarly for the  $(+-\to +-)$ amplitude. }
\begin{align}
    -\underset{s=n}{\Res\,} T_{++--}(s,t) &= \sum_{J=0,2,\dots} c^{++}_{n,J}\, d_{0,0}^J\left(1+\frac{2t}{n}\right)\label{eq:resT++}\\
  -\underset{s=n}{\Res\,} T_{+--+}(s,t)  &= \sum_{J=0,2,\dots} c^{+-}_{n,J}\, d_{4,4}^J\left(1+\frac{2t}{n}\right)\,\label{eq:resT+-}
\end{align}
where the $d_{\mu,\nu}^J(x)$ are the usual Wigner small $d$-matrices (see for example \cite{Martin:1970hmp}) and $d_{0,0}^J(x)= P_J(x)$. The coefficients $c_{n,J}$ are square of coupling constants and are thus nonnegative
\begin{equation}
    c^{++}_{n,J}\geq 0\,, \quad c^{+-}_{n,J}\geq 0\, \ .
\end{equation}
These constraints restrict the allowed space of  $\alpha$'s. The space of amplitudes defined in this way is still infinite-dimensional and we will study its projection on the space of a few leading low-energy Wilson coefficients.

In order to explore the space of amplitudes numerically, we consider a truncated ansatz 
\begin{align}
        f_{\Nmax}(s|t,u) &= - 8 \pi G_N\frac{\G(-s)\G(-t)\G(u)}{\G(1+s)\G(1+t)\G(1+u)} \nn\\
        &+ \sum_{c_s,c_{tu}=0}^{\Nmax}\sum_{d_s,d_{tu}=1}^{2\Nmax} \theta_{\alpha_{\text{ind}}}(\alpha)\,\alpha_{c_s,c_{tu},d_s,d_{tu}} \frac{\G(c_s-s)\G(c_{tu}-t)\G(c_{tu}-u)}{\G(d_s+s)\G(d_{tu}+t)\G(d_{tu}+u)}\,, \label{eq:ansatz_f_Nmax}
\end{align}
where the limit of the sum is chosen such that the residues are polynomial and the Regge limit satisfies \eqref{eq:reggeMHVgrav}. This requires additional constraints on $\alpha$'s that we impose by inserting $\theta_{\alpha_{\text{ind}}}(\alpha)$ which is $1$ when $\alpha \in \alpha_{\text{ind}}$ and zero otherwise, it removes remaining dependent terms, see \appref{sec:ClosedAnsatzconstraintonAlpha}. This is performed in two steps. First, we realized that terms with $c_t\neq c_u$ and $d_t\neq d_u$ are redundant and thus can be removed from the sum as in \eqref{eq:ansatz_f_Nmax}. Second, we solve for the remaining dependence order by order in $\Nmax$. All in all, this ansatz contains $3N_{\text{max}}^2 +\Nmax-2$ free parameters. In what follows, we will explore \emph{primal} bound on Wilson coefficients numerically. This procedure will be explained in detail next in \secref{sec:algorithmBound}.
  
However, before going further, let us remind the reader that already in string theory, there exists a nontrivial $\alpha$ solution for $f(s|t,u)$ in \eqref{eq:f_ClosedInf}, namely the scattering of gravitons in heterotic string theory \cite{Kawai:1985xq}, where the MHV amplitude is
\begin{equation}
    f^{(hs)}(s|t,u) = - 8 \pi G_N\frac{\G(-s)\G(-t)\G(-u)}{\G(1+s)\G(1+t)\G(1+u)}\left(1-\frac{tu}{1+s}\right)
\end{equation}
which is simply generated from the ansatz with one nontrivial term $\alpha_{0,(1,1),2,(1,1)}= 8 \pi G_N$. We know that this amplitude satisfies all the constraints, the space of $\alpha$'s is thus nontrivial.

\subsection{Primal algorithm}\label{sec:algorithmBound} 

Here we describe the numerical primal algorithm used to find extremal stringy gravitational amplitudes and derive bounds on Wilson coefficients. 

Provided the ansatz \eqref{eq:ansatz_f_Nmax}, it is clear that the low-energy Wilson coefficients are linear combinations of the $\alpha$'s. Explicitly they take the form
\begin{equation}
    a_{k,j} = -x_0 8\pi G_N + \sum_{c_s,c_{tu}=0}^{\Nmax}\sum_{d_s,d_{tu}=1}^{2\Nmax}x_{c_s,c_{tu},d_s,d_{tu}}\alpha_{c_s,c_{tu},d_s,d_{tu}}\,,
\end{equation}
where $x_i\in \mathbb{R}$ and can be computed by expanding the ansatz at low energy. Similar relations hold for $ |\beta_{R^3}|^2$ and $|\beta_\phi|^2 $. The same is true for the partial wave coefficients  $c^{++}_{n,J},\,  c^{+-}_{n,J}$. The procedure of bounding Wilson coefficients can thus be efficiently implemented using Linear Programming, and in practice, we used SDPB \cite{Simmons-Duffin:2015qma,Landry:2019qug}.\footnote{We also tried linear solvers such as GLPK and Gurobi \cite{gurobi}. However, we observed that as we increase $\Nmax$, high precision was needed and we turned to SDPB, where arbitrary precision can be used.}  When imposing unitarity, this cannot be done numerically for all $n,J$, therefore we truncate the number of constraints by imposing
\begin{equation}
    c^{++}_{n,J}\geq 0\,, \quad c^{+-}_{n,J}\geq 0\,, \quad \forall\, n\leq n_{\text{max}}\,,\, \forall J\,.
\end{equation}
As we will see, in practice, the convergence in $n_{\text{max}}$ is fast. We impose the constraint for all spins, and due to the linearity of the Regge trajectory, the number of constraints scales as $\mathcal{O}(n_{\text{max}}^2)$. In practice, we computed the coefficients $c_{n,J}^{++},\, c_{n,J}^{+-}$ for each term in \eqref{eq:ansatz_f_Nmax} using \eqref{eq:resT++} and \eqref{eq:resT+-}. 

 We now give an example of the procedure of maximizing the quantity $A$ ($A$ is any ratio of Wilson coefficients and in the gravitational case we can normalize everything to $8 \pi G_N$)
\begin{enumerate}[label=(\alph*)]
    \item \label{step:nmax}At fixed $\Nmax$ for the ansatz \eqref{eq:ansatz_f_Nmax}, we maximize $A$ by increasing the number of unitarity constraints $n_{\text{max}}$. Experimentally, the extremal value converges to a plateau for $n_{\text{max}}\gtrsim 2\Nmax + 10$.
    \item \label{step:Nmax}We extremize $A$ for increasing size of the ansatz $\Nmax$. For each fixed $\Nmax$, the resulting amplitude satisfies all the constraints.
    \item \label{step:primalBound}We fit the extremal $A$ vs $\Nmax$ and when possible, we extrapolate to $\Nmax\to\infty$. As we will see, is it not always clear that $A$ converges to a finite value.
    The converged value is then a \emph{primal} bound on $A$.
\end{enumerate}
It is straightforward to extend this algorithm to explore a higher-dimensional space of parameters. This is done by fixing $(A_1,\dots, A_n)$ and maximizing $A_0$.

\newpage
\subsection{Bounds on Wilson coefficients}\label{sec:result_gravitons}
In this section, we present bounds on various ratios of Wilson coefficients. 

In section \secref{sec:BoundNormGn} we consider bounds on the Wilson coefficients normalized to $G_N$.
Some of the bounds of this type are known to suffer from IR divergencies in four dimensions. The simplest example concerns the correction to the graviton three-point coupling which is bounded as 
\cite{Camanho:2014apa}
\begin{equation}
    \frac{|\beta_{R^3}|^2}{8 \pi G_N}\lesssim \frac{\log(M_{\text{HS}} L_\text{IR})}{M_{\text{HS}}^4}\,,
\end{equation}
where $M_{\text{HS}}$ is the threshold for higher spin particles and $L_{\text{IR}}$ is an IR regulator. This bound was recently transformed to a sharp inequality in \cite{Caron-Huot:2022ugt} and derived by taking the flat space limit of AdS \cite{Caron-Huot:2021enk}. As the ansatz described above is tree-level, the amplitude is manifestly IR finite. The correction to the graviton three-point coupling $|\beta_{R^3}|^2$ is the first target for our \emph{primal} algorithm. We then proceed by deriving a bound on $a_{0,0}$, which corresponds to the contact term $R^4$ in the low-energy effective action. We also normalize it by $8 \pi G_N$, and the corresponding upper bound is again known to suffer the IR divergences \cite{Bern:2021ppb,Caron-Huot:2022ugt}. 

Then we consider various bounds on Wilson coefficients $a_{k,j}$ normalized by $a_{0,0}$ in \secref{sec:BoundNormBya00}. Such ratios are known to admit dual bounds when assuming ACU \ref{ass:ACU}. In this work, we explore the space of these coefficients where the extra information about the leading Regge trajectory $j(t)$ is put in. We use the primal method to impose this constraint for all $t$ (negative and positive), and the dual algorithm when it is effectively only imposed for positive $t>0$ \ref{ass:ACU+J}. 

\subsubsection{Bounds normalized by $G_N$}\label{sec:BoundNormGn}
\begin{figure}[t!]
    \centering
    \includegraphics[width=\linewidth]{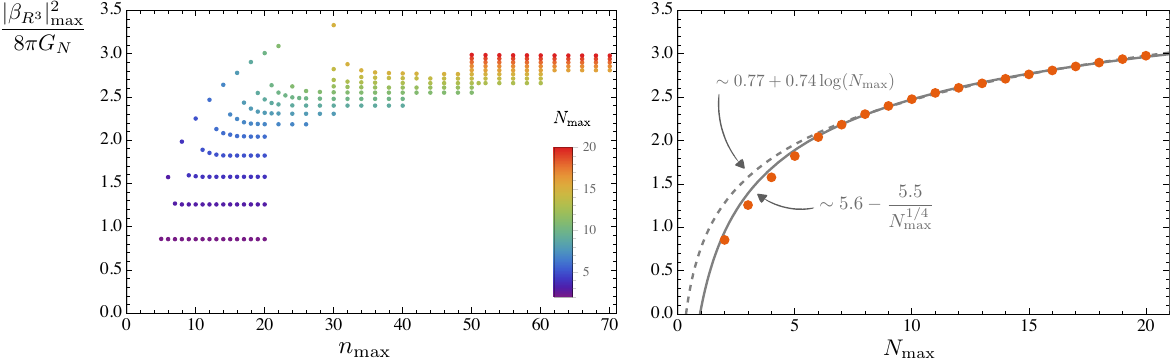}
    \caption{Maximum values of the correction to the 3-pt coupling $|\beta_{R^3}|^2$ normalized by $8 \pi G_N$. On the left panel, we present convergence in $n_\text{max}$, i.e., the number of massive states for which unitarity was imposed. We see that the convergence is fast and quickly stabilizes to a plateau. On the right, we show the converged value vs. $\Nmax$ along with two fits. The dashed line is a divergent $\log(\Nmax)$ fit, and the solid line is a convergent power-law fit.}
    \label{fig:beta3max}
    \vspace{5pt}
    \includegraphics[width=\linewidth]{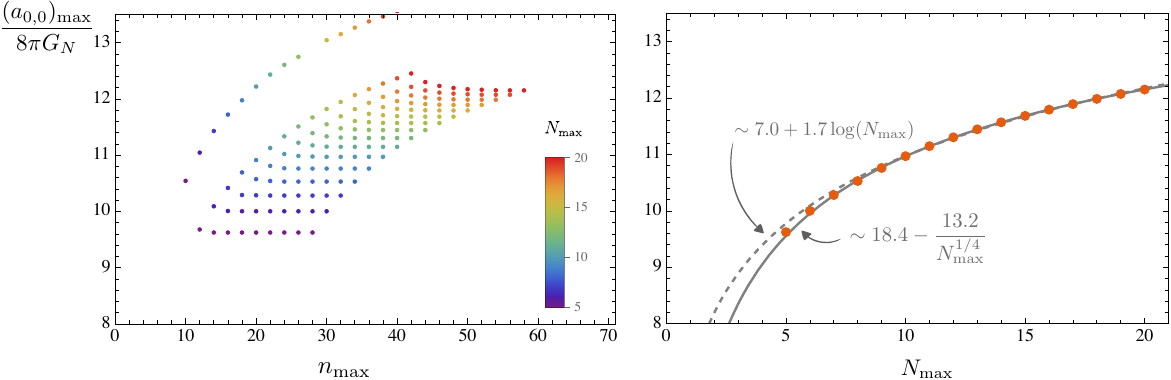}
    \caption{Maximum values of the first contact term correction $a_{0,0}$ normalized by $8\pi G_N$.  On the left panel, we present convergence in $n_\text{max}$. On the right panel, we show the converged value vs. $\Nmax$ along with two fits. The dashed line is a divergent $\log(\Nmax)$ fit, and the solid line is a convergent power-law fit. }
    \label{fig:a00max}
\end{figure}

\paragraph{Correction to the 3-pt coupling - $\mathbf{|\beta_{R^3}|^2}$}\mbox{}\\[2pt]
Let us start by considering the bound on correction to the graviton three-point coupling $|\beta_{R^3}|^2$, this example will also allow us to go through the numerical procedure described in \secref{sec:algorithmBound}. First, we extremize $|\beta_{R^3}|^2$ at fixed $\Nmax$ as we increase the number of constraints $n_{\text{max}}$, see step \ref{step:nmax}. We present the result in \figref{fig:beta3max} (left panel). This figure shows that at fixed $\Nmax$, $\frac{|\beta_{R^3}|^2}{8\pi G_N}$ converges to a plateau in $n_\text{max}$. The amplitude extracted at finite $\Nmax$ satisfies all the constraints listed in \secref{sec:constraints}, see step \ref{step:Nmax}. 

In this way, we can explicitly construct stringy tree-level amplitudes with  $|\beta_{R^3}|^2 \lessapprox 3\cdot 8\pi G_N$. To extract a primal bound, step \ref{step:primalBound}, we need to extrapolate in $\Nmax$, this is shown in \figref{fig:beta3max} (right panel). Clearly, the data does not allow us to determine if it converges to a finite value as $\Nmax\to \infty$. To highlight this point, we performed two fits, one using a power law (in gray) which converges to a finite value, whereas the second using a logarithm (in dashed) diverges. A large $\Nmax$ analysis is needed to distinguish between the two options. 

While deriving this bound, no assumption was made on $\beta_\phi$, and thus we allowed for a massless scalar exchange $|\beta_\phi|\geq 0$. Imposing the absence of massless scalar exchange $\beta_\phi=0$ does not change the qualitative behavior of the bound and the $\sim \log\Nmax$ behavior remains.

\paragraph{Dimension 8 operator -- $R^4$}\mbox{}\\[2pt]
Next, we consider the leading correction due to a contact term $R^4$ parameterized by $a_{0,0}$. It is well known from dispersion relations that this coefficient is positive (see \appref{sec:dualMethod} for a review). We follow the same procedure as for the correction to the three-point coupling and present the result in \figref{fig:a00max} for the upper bound. From the extrapolation, it is clear that it behaves similarly to the correction to the three-point coupling. As for the lower bound, the bound can be extrapolated and converge to $\frac{(a_{0,0})_{\text{min}}}{8\pi G_N}\approx 0$ and as such does not change compared to the dual bound based on assuming causality, unitarity, and crossing symmetry.

\begin{figure}
    \centering
    \includegraphics{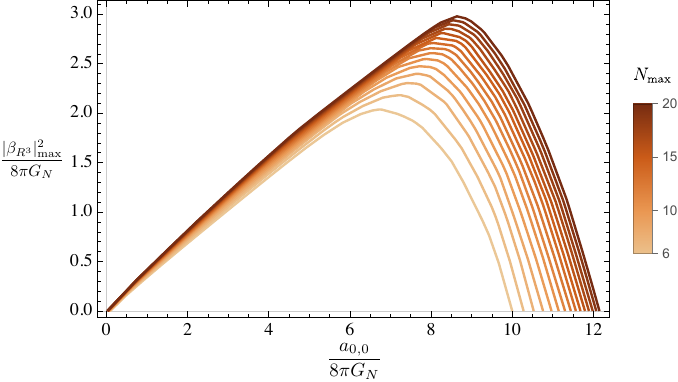}
    \caption{Allowed space of Wilson coefficients  $|\beta_{R^3}|^2$ vs. $a_{0,0}$ at finite $\Nmax$. Our results are consistent with the bounds diverging as $\log N_{\text{max}}$. Interestingly, this seems to be analogous to the presence of the IR regulator in the same bounds obtained in \cite{Caron-Huot:2022ugt}. }
    \label{fig:beta3vsa00Nmax}
\end{figure}

\newpage
\paragraph{$\mathbf{|\beta_{R^3}|^2}$ vs. $\mathbf{a_{0,0}}$}\mbox{}\\[2pt]
The result obtained for $a_{0,0}$ could have been anticipated. Indeed, it was shown in \cite{Bern:2021ppb} that $\frac{|\beta_{R^3}|^2}{a_{0,0}}\leq1$. And thus, if the upper bound for $|\beta_{R^3}|^2/(8\pi G_N)$ diverges as $\log(\Nmax)$, so must the upper bound on $a_{0,0}/(8\pi G_N)$. We can then bound the correction to the 3-point coupling $|\beta_{R^3}|^2$ at fixed $a_{0,0}$. This result is shown in \figref{fig:beta3vsa00Nmax} at various $\Nmax$. We do not perform an $\Nmax\to \infty$ extrapolation. The shape of the allowed region is similar to the one obtained in \cite{Caron-Huot:2022ugt} and is consistent with a $\sim \log \Nmax$ divergence in the large $\Nmax$ limit.

\subsubsection{Bounds normalized by the total cross-section moment $a_{0,0}$}\label{sec:BoundNormBya00}

In this subsection, we will consider bounds on $\frac{a_{k,j}}{a_{0,0}}$. Notice that the coefficients $a_{2k,0}$ measure moments of the total cross-section, see \appref{sec:dualMethod}, and are positive.

\paragraph{Dimension 12 operators -- $D^4R^4$}\mbox{}\\[2pt]
We start by looking at the coefficients of dimension 12 operators, namely $a_{2,0},a_{2,1}, a_{2,2}$, and normalize them by $a_{0,0}$. Out of the three coefficients, only two are independent because by crossing we have $a_{2,1}= a_{2,2}$. From previous works, we know that these ratios are bounded from causality and unitarity \ref{ass:ACU}, see \cite{Caron-Huot:2022ugt,Chiang:2022jep}.

\begin{figure}
    \centering
    \includegraphics{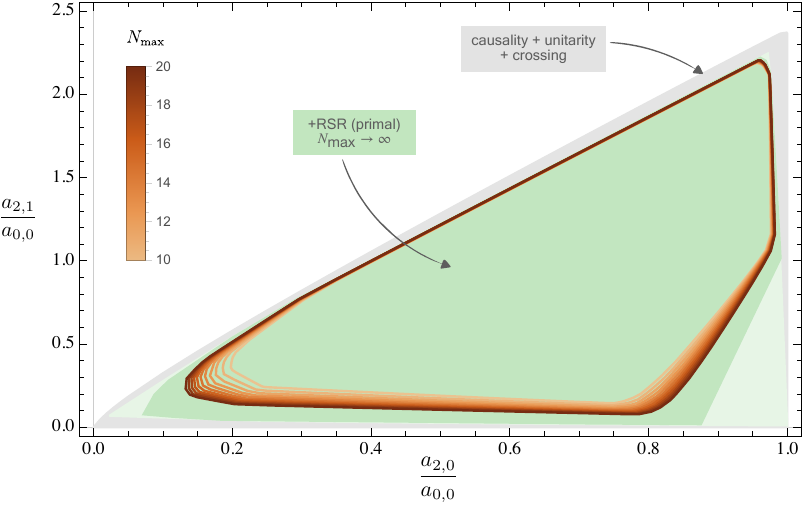}
    \caption{The allowed region for the space of Wilson coefficients $a_{2,0},a_{2,1}$. In light gray, we report the dual bound using only ACU \ref{ass:ACU} with $k_{\text{max}}=16$ null constraints. In red colors, we show primal results obtained using our ansatz for increasing $\Nmax$. The green region is the extrapolated $\Nmax\to\infty$ allowed region. In light green, we show a conservative estimate of the extrapolation error. }
    \label{fig:a21a20}
\end{figure}

\begin{figure}[t!]
    \centering
    \includegraphics[width=\linewidth]{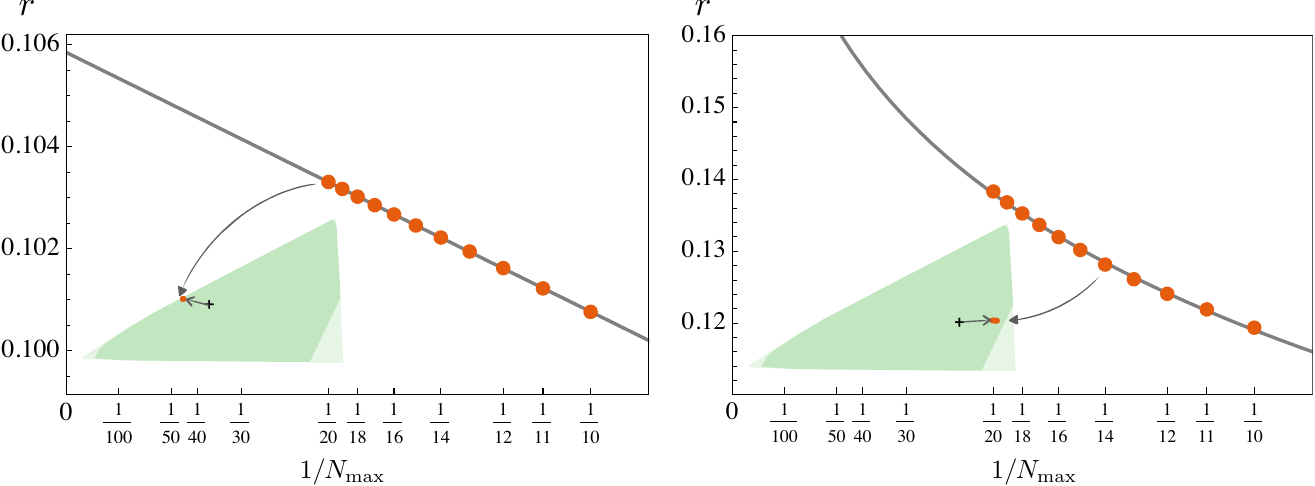}
    \caption{Examples of convergences in $1/\Nmax$ along the boundary of \figref{fig:a21a20}. $r$ is the distance between the `cross' and the points. On the left panel, we observe `good' linear convergence. In contrast, on the right panel, the convergence is extremely slow. }
    \label{fig:a2Extrapolation}
\end{figure}

Let us first derive the primal bound using the ansatz \eqref{eq:ansatz_f_Nmax}. We present the result in \figref{fig:a21a20} along with the bound assuming only ACU \ref{ass:ACU}. As before, the convergence in $n_{\text{max}}$ is fast and easy. Regarding the convergence in $\Nmax$ we observe different behaviors along the boundary. In some regions, we observe a `fast' convergence. This corresponds to the points on the boundary where the dark green goes up to the boundary in \figref{fig:a21a20}. The convergence is harder in other regions and seems to diverge even though we \emph{know} that a finite bound exists. We show examples of convergence in $\Nmax$ in \figref{fig:a2Extrapolation}. 

To overcome this issue, we use a fit of the form $r=a+ b \Nmax^c$. First, we included only points along the boundary that converge at least  linearly in $1/\Nmax$, then we used convexity to close the region. This leads to the boundary of the light green region in \figref{fig:a21a20}. To get an idea of the uncertainty in this procedure, we added points that `look' linear for $\Nmax \geq 15$ and used a linear fit. This leads to the darker green region in \figref{fig:a21a20} which can be thought of as an `optimistic' fit. To remain conservative, one should consider the full green region. 

Next, we bound the same Wilson coefficients with the dual method supplemented by the constraints on the leading Regge trajectory. For positive $t$, it leads to the maximal spin bound  on the spectrum \ref{ass:ACU+J}
\begin{equation}\label{eq:ClosedReggePositivet}
    J(m^2) \leq 2m^2 + 2\,.
\end{equation}
Imposing \ref{ass:ACU+J+RSR} in the form of a finite number of RSR does not lead to a stronger bound.

\begin{figure}[h!]
    \centering
    \includegraphics{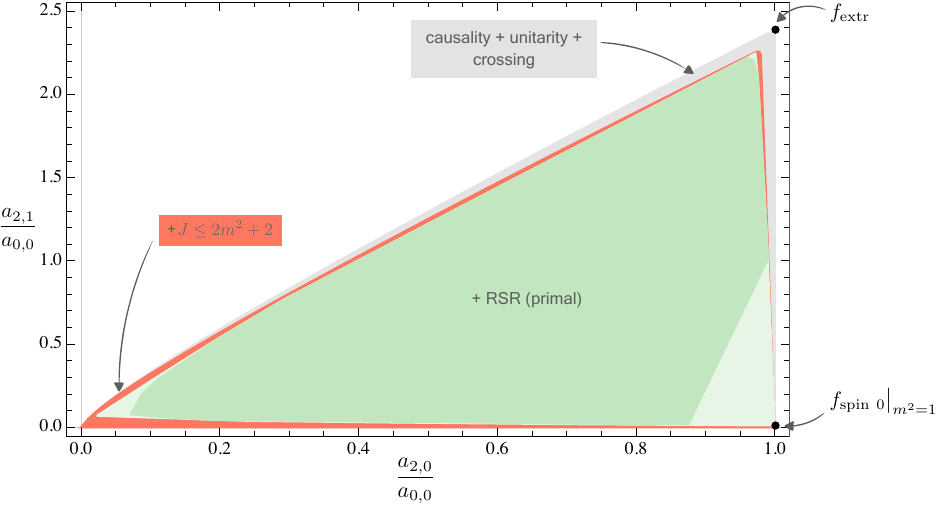}
    \caption{Comparison of various assumptions and the corresponding bounds for the Wilson coefficients $a_{2,0},a_{2,1}$. In gray, we imposed the standard bootstrap constraints \ref{ass:ACU} in the dual approach, in red, we further imposed the maximal spin constraint \ref{ass:ACU+J}. In green,  we show the bound obtained using the primal approach which manifestly satisfies \ref{ass:ACU+J+RSR}, see also \figref{fig:a21a20}. For the dual bounds, we used $k_{\text{max}}=6$. We also indicated special amplitudes: $f_{\text{extr}}$ given by \eqref{eq:fextr} with $\lambda, g$ given by \eqref{eq:lambdaAndg_Extr}, and the spin $0$ exchange at $m^2=1$ given by $f_{\text{spin}~0}$ in \eqref{eq:spin0f}.}
    \label{fig:a2_SpectrumConstraints}
\end{figure}

We compare the primal bound, the dual bound using only ACU, and the one with the maximal spin constraint in \figref{fig:a2_SpectrumConstraints}. From this plot, one clearly sees that RSR has little to no effect and the gap seems to close in most regions of the boundary. The dual bounds are also obtained with finitely many null constraints and the gap would close further as $k_{\text{max}}\to \infty$. Note that the primal ansatz has the extra assumption of equidistant spectrum $m_n^2=n$. We check that this constraint has no effect on the allowed region in the dual approach.

It is interesting to see where special amplitudes lie in this region. The simplest amplitude is perhaps the massive spin-$0$ exchange amplitude given by\footnote{In \appref{sec:amplitudeExample}, we show that this amplitude is unitary. Note that this amplitude has intercept $j_0=3$. However, it satisfies all the sum rules used in the dual approach. See \appref{sec:amplitudeExample} for further discussion.}
\begin{equation}\label{eq:spin0f}
    f_{\text{spin}~0}=\frac{8\pi G_N}{stu} + \frac{\lambda^2}{m^6}\frac{1}{m^2-s}\,,
\end{equation}
which leads to the ratios of Wilson coefficients
\begin{equation}
    \left( \frac{a_{2,0}}{a_{0,0}},\frac{a_{2,1}}{a_{0,0}}\right) = \left(\frac{1}{m^4},0\right) 
\end{equation}
and this populate the line $a_{2,1}=0$ as one vary the mass $m\in [1, \infty)$. Clearly, this amplitude satisfies \eqref{eq:ClosedReggePositivet} but not RSR.

We can also understand how the line $\frac{a_{2,0}}{a_{0,0}}=1$ is excluded by imposing \eqref{eq:ClosedReggePositivet}. From the sum rules for $a_{k,0}$, (see \eqref{eq:avg_ak0}), the only allowed spectrum allowed is at $m^2=1$. Then, we can explicitly construct the amplitude at the upper-right kink. It is given by an infinite tower of spins exchanged at $m^2=1$
\begin{equation}\label{eq:fextr}
 f_{\text{extr}}(s|t,u) = \frac{8\pi G_N}{stu}+ \frac{1}{(1-s)(1-t)(1-u)} + \frac{\lambda}{1-s} + \frac{g}{(1-t)(1-u)}\,,
 \end{equation}
 with 
 \begin{equation}\label{eq:lambdaAndg_Extr}
     \lambda =- \frac{2 \log (2)}{3}\,,\quad g = \frac{1628532-2349480 \log (2)}{7096320 \log (2)-4918777}\,.
 \end{equation}
Indeed, by computing the ratio of Wilson coefficients for this amplitude, we obtain
\begin{equation}
   \left( \frac{a_{2,0}}{a_{0,0}},\frac{a_{2,1}}{a_{0,0}}\right) = \left(1,\frac{1+g}{1+g+\lambda}\right) \overset{\eqref{eq:lambdaAndg_Extr}}{\approx} (1,2.367)\,,
\end{equation}
which is precisely the location of the upper-right corner.
In \appref{sec:amplitudeExample}, we show that this amplitude is unitary and has no spin $0$ and spin $5$ exchanges. Moreover, the line at $\frac{a_{2,0}}{a_{0,0}}=1$ is given by scanning over $\lambda$. As it is clear from the equation above, except at the spin $0$ point ($\lambda\to \infty$), the amplitudes on this line are given by an infinite tower of spin at $m^2=1$ and cannot satisfy the polynomial residue constraint $J\leq 2m^2+2$ \eqref{eq:ClosedReggePositivet}. This is exactly what we observe in \figref{fig:a2_SpectrumConstraints}. 

One advantage of the primal approach is that at any finite $\Nmax$, the amplitude is known explicitly. It is therefore interesting to study the physical properties of extremal solutions and how they evolve along the boundary. In particular, we can analyze the contribution of various spins and channels to the $a_{k,0}$ sum rule which reads
\begin{align}
    a_{k,0}&=\frac{1}{\pi}\int_{1}^{\infty} \frac{dm^2}{m^{2k+10}} \left(\sum_{J=0,2,\dots}^{\infty} \rho_J^{++}(m^2) +  \sum_{J=4,5\dots}^{\infty}(-1)^k\rho_J^{+-}(m^2)\right)\,,\\
    &=\sum_{J=0,2,\dots}^{\infty} \rhoavg{\rho_J^{++}}_k + (-1)^k\sum_{J=4,5,\dots}^{\infty} \rhoavg{\rho_J^{+-}}_k\,,
\end{align}
where in the second line we introduced the notation $\rhoavg{...}_k = \frac{1}{\pi}\int_{1}^{\infty} \frac{dm^2}{m^{2k+10}}(...)$ for the integral over $m^2$. 

\begin{figure}[h!]
    \centering
    \includegraphics[width=\linewidth]{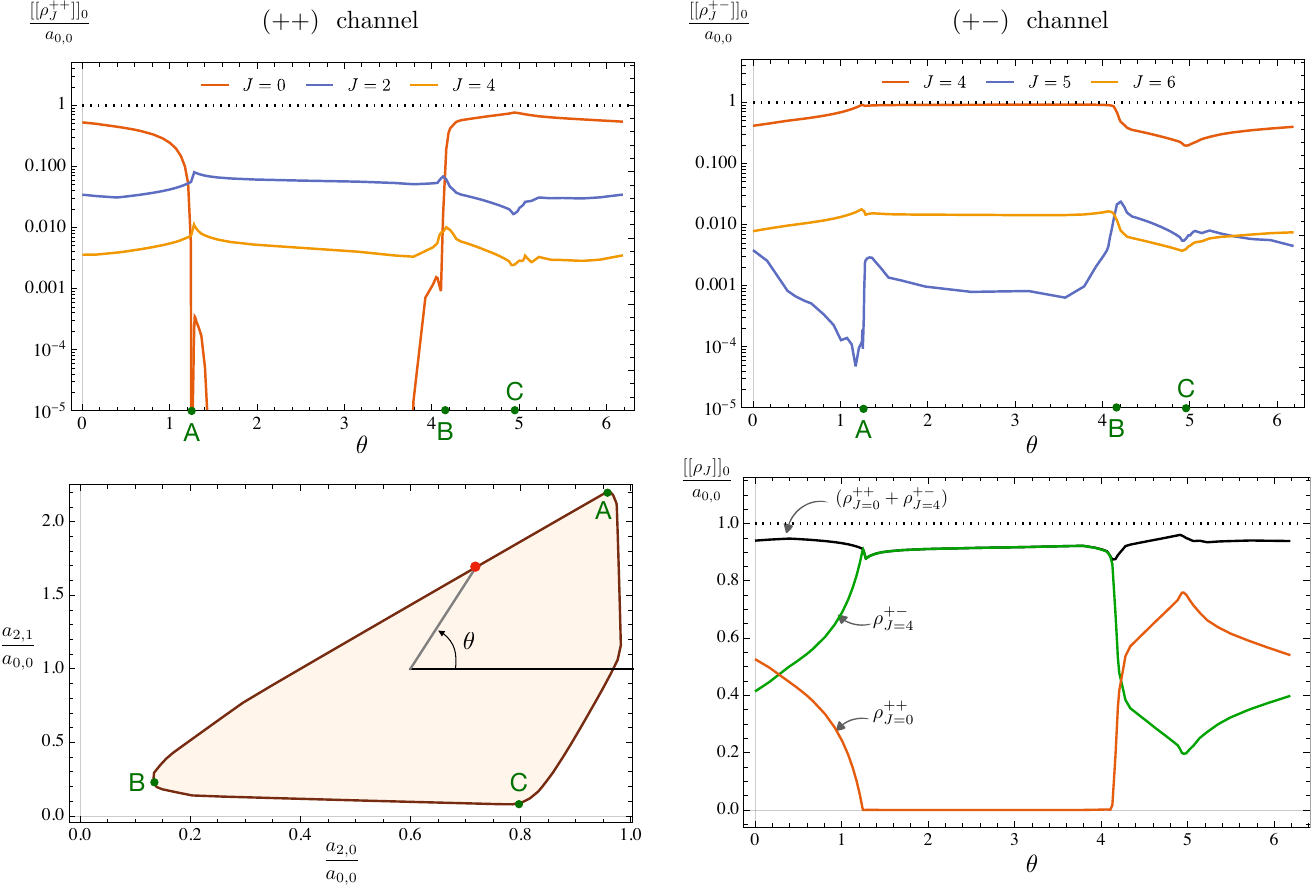}
    \caption{Spectral density moments along the boundary of the allowed region in the space of ratios of Wilson coefficients $\left(\frac{a_{2,1}}{a_{0,0}},\,\frac{a_{2,0}}{a_{0,0}}\right)$ for the amplitude with $\Nmax=20$. Points at the boundary are labeled by an angle $\theta$ defined in the lower-left panel. We indicated special points with $A,B,C$. From \eqref{eq:SRa00is1}, all contributions sum to $1$, which is indicated in the plots by a black dotted line. In the lower-right panel, we present the lowest spin contribution to the spectral density in each channel.}
    \label{fig:a2SpinPPPM}
\end{figure}

Setting $k=0$, we get
\begin{equation}\label{eq:SRa00is1}
    1 = a_{0,0}^{-1} \left(\sum_{J=0,2,\dots}^{\infty} \rhoavg{\rho_J^{++}}_0 + \sum_{J=4,5,\dots}^{\infty} \rhoavg{\rho_J^{+-}}_0\right)\,.
\end{equation}
In \figref{fig:a2SpinPPPM}, we show various contributions to the sum rule along the boundary.\footnote{Here we use the amplitude obtained with $\Nmax=20$.} To this end, we define an angle variable $\theta$ which spans the boundary, see \figref{fig:a2SpinPPPM}(bottom-left) for its definition. We observe that the lowest spin always dominates along the boundary in the $(+-)$ channel. This is not true in the $(++)$ channel where on the upper diagonal $J=2$ dominates. On the lower-right panel of \figref{fig:a2SpinPPPM}, we highlight that along the boundary $\sim 90\%$ of the sum rule comes from the lowest spin contribution in each channel.

\begin{figure}[h!]
    \centering
    \includegraphics[width=0.95\linewidth]{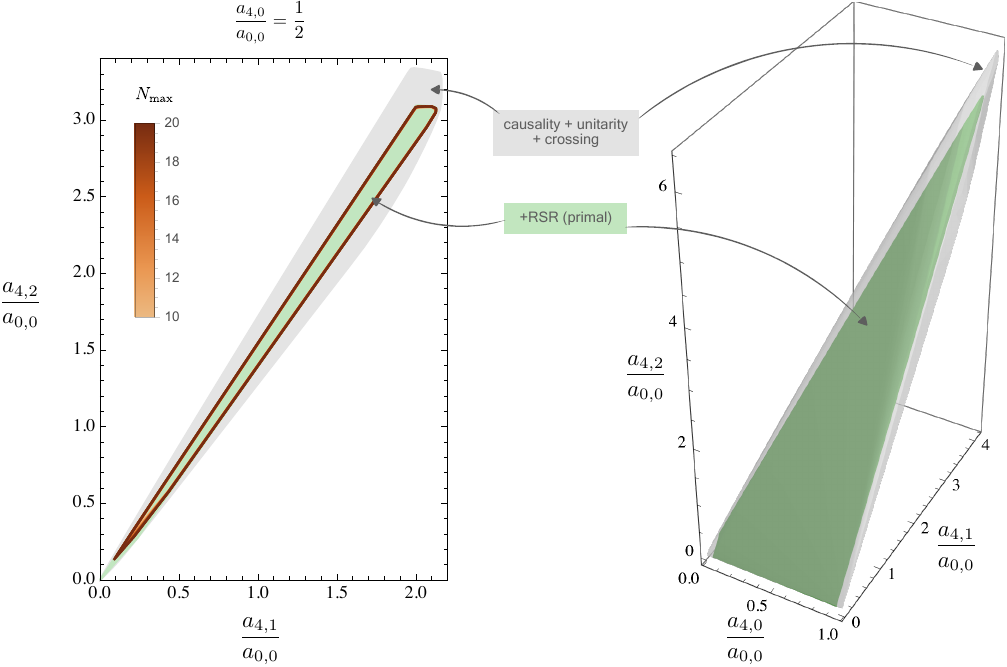}
    \caption{The allowed region for the Wilson coefficients $a_{4,0},a_{4,1},a_{4,2}$.  In light gray, we report the dual bound using only causality, unitarity, and crossing (ACU) \ref{ass:ACU}. In color, we present the primal bounds that also satisfy \ref{ass:ACU+J} and \ref{ass:ACU+J+RSR} using increasing $\Nmax$. The green region is the extrapolated bound $\Nmax\to\infty$. On the right, we show the result for the 3d region and, on the left, we present an example of the section at $a_{4,0}/a_{0,0}=1/2$.  At finite $\Nmax$, the convergence is fast except close to the origin. The 3d region is built for $0.05\leq \frac{a_{4,0}}{a_{0,0}}\leq 0.95$ at $\Nmax=20$ $+$ extrapolation was used when necessary.}
    \label{fig:a4} 
\end{figure}

\paragraph{Dimension 16 operators -- $D^8R^4$}\mbox{}\\[2pt]
We next consider bounds on the $a_{4,j}$ coefficients normalized by $a_{0,0}$. At this level, there are 3 independent coefficients $a_{4,0},a_{4,1}$ and $a_{4,2}$ which carve a finite region in 3d space. We proceed similarly as in the case of dimension 12 operators. In \figref{fig:a4}, we present a section of the allowed space at $a_{4,0}/a_{0,0}=1/2$ and the entire 3d region using the primal ansatz. The convergence in $\Nmax$ is fast except at the origin $a_{4,2} = a_{4,1} = 0$.

In the existing literature, the dual bounds using only causality, unitarity, and crossing symmetry were never presented. Instead, various authors \cite{Arkani-Hamed:2020blm,Bern:2021ppb,Chiang:2022jep}, considered bounds on the homogeneous ratios $\frac{a_{4,1}}{a_{4,0}}$ vs. $\frac{a_{4,2}}{a_{4,0}}$. In this space, \cite{Bern:2021ppb} highlighted that all theories populate a smaller region dubbed the \emph{low-spin dominance} region obtained assuming that higher spin contributions to the spectral densities $\rho_J$ are suppressed. The same effect appears for the $a_{2,j}$ coefficients, and in a subsequent work \cite{Caron-Huot:2022ugt}, the authors emphasized that bounds on the homogeneous ratios are dominated by a small region close to the free theory point when considering the inhomogeneous ratios. Here, we observe the same effect. A similar observation was made in \cite{Chiang:2022jep}, where the authors realized that by fixing the value $\frac{a_{4,0}}{a_{0,0}}$, the bound on ratios of homogeneous coefficients shrinks significantly (this corresponds to a section in the 3d region in \figref{fig:a4}).

\begin{figure}[h!]
    \centering
\includegraphics{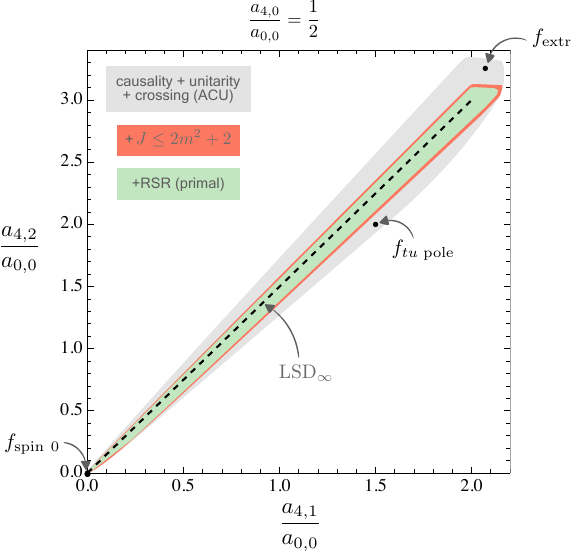}
    \caption{Bounds for $a_{4,1},a_{4,2}$ at $\frac{a_{4,0}}{a_{0,0}}= 1/2$ from various assumptions. In gray, the usual bootstrap assumptions are considered (ACU) \ref{ass:ACU}. In red, we further imposed the maximal spin constraint \ref{ass:ACU+J}, and, in green, superpolynomial softness \ref{ass:ACU+J+RSR} is imposed using the primal approach \figref{fig:a4} (left). For the dual bounds, we used $k_{\text{max}}=6$. We also indicated special amplitudes: $f_{\text{extr}}$ given by \eqref{eq:fextr} (but with the mass of the exchanged tower of particles given by $m^8=2$), the $tu$-pole amplitude, and the spin $0$ exchange amplitude with mass $m^8=2$, see \tabref{tab:fgravExample}.}
    \label{fig:a4at05_SpectrumAssumption}
\end{figure}

Second, we bound the section $\frac{a_{4,0}}{a_{0,0}}=\frac{1}{2}$ using the dual method by imposing the maximal spin constraint \ref{ass:ACU+J} with \eqref{eq:ClosedReggePositivet}. We present the result in \figref{fig:a4at05_SpectrumAssumption}. They present a clear overlap between the assumptions \ref{ass:ACU+J} and \ref{ass:ACU+J+RSR}, i.e., RSR has little to no effect. We also checked that imposing equidistant spectrum $m^2_n=n$ in addition to $J\leq 2m^2+2$ does not change the shape of the allowed region and cannot create a gap between the primal and dual regions. In this \figref{fig:a4at05_SpectrumAssumption}, we also draw the line of low spin dominance LSD${}_\infty$ defined by
\begin{equation}
    \text{LSD}_\alpha: \qquad \frac{ \rhoavg{\rho_0^{++}}_k}{ \rhoavg{\rho_{J>0}^{++}}_k}\geq \alpha \quad \text{and} \quad \frac{ \rhoavg{\rho_4^{+-}}_k}{ \rhoavg{\rho_{J>4}^{+-}}_k}\geq \alpha\,.
\end{equation}

\begin{figure}[h!]
    \centering
    \includegraphics[width=\linewidth]{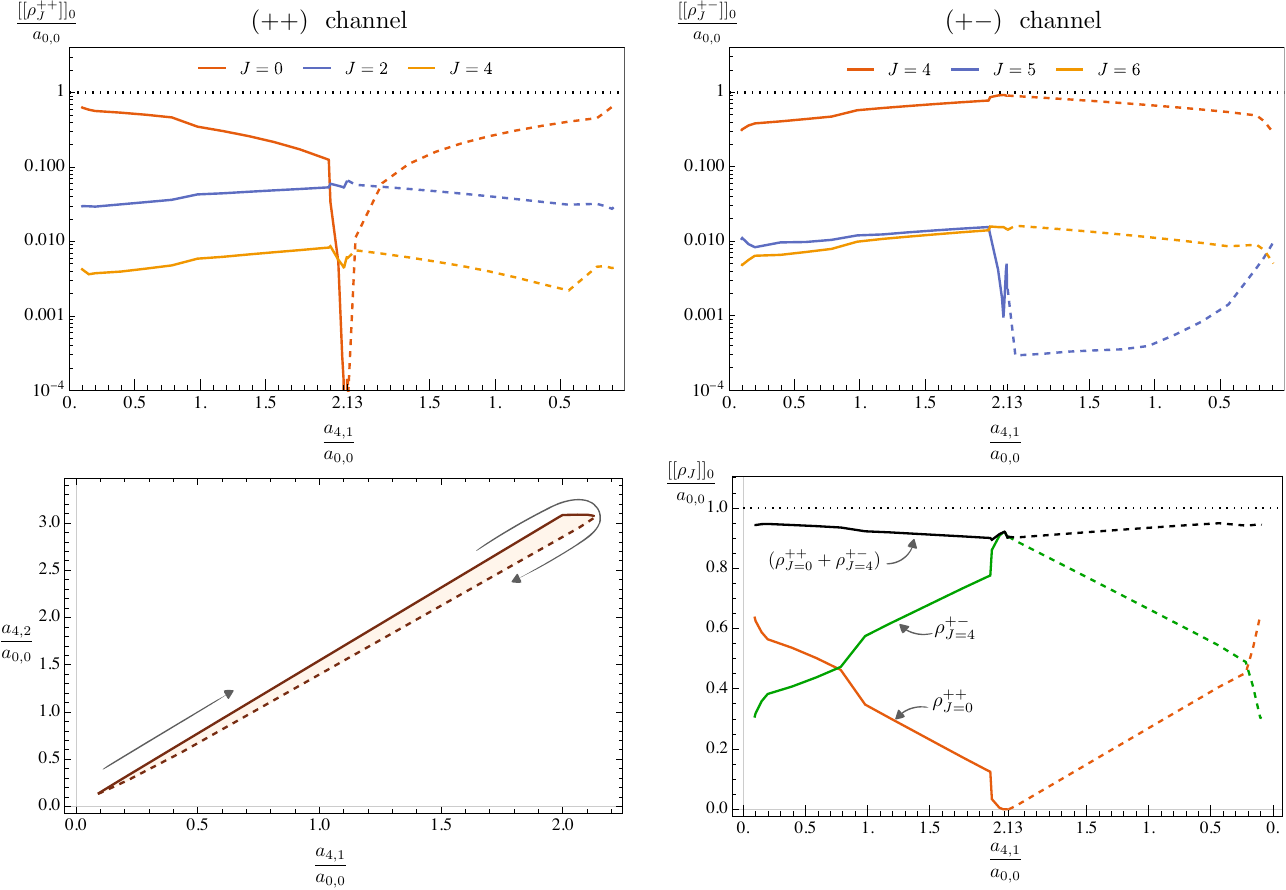}
    \caption{The spectral density moment along the boundary of the allowed region in the space $\left(\frac{a_{4,2}}{a_{0,0}},\,\frac{a_{4,1}}{a_{0,0}}\right)$ at fixed $\frac{a_{4,0}}{a_{0,0}}=\frac{1}{2}$ for the amplitude with $\Nmax=20$. Points at the boundary are labeled by $\frac{a_{4,1}}{a_{0,0}}$. The upper branch is depicted in solid and the lower in dashed, as shown in the lower-left panel. From \eqref{eq:SRa00is1}, all contributions sum to $1$, indicated in the plots by a black dotted line. In the lower-right panel, we present the lowest spin contribution in each channel.}
    \label{fig:a4SpinPPPM}
\end{figure}

It is also interesting to study the content of the extremal primal amplitudes and we proceed similarly to the case of $a_{2,k}$. We present the result in \figref{fig:a4SpinPPPM}, and except for the region close to the upper-right corner, the lowest spin always dominates in each channel.
Furthermore, as for the case of $a_{2,k}$, the sum of the lowest spin spectral density in each channel constitutes  $\sim 90\%$ of the sum rules all along the boundary.


\section{Open strings: scattering of massless scalars}\label{sec:OpenStrings}

In this section, we consider the same problem but for open strings: we assume that $T(s,t)$ has poles \emph{only} in the $s$- and $t$-channels; it obeys the crossing relation $T(s,t) = T(t,s)$; in the Regge limit, it takes the form \eqref{eq:regge} with 
\begin{equation}\label{eq:jtOpenCase}
    j(t)=t,
\end{equation}
where as before we set $m_{\text{gap}}^2=1$.

The main difference compared to the case of closed strings is that we find that unitarity excludes any ansatz with a finite number of satellite terms, see \appref{sec:noFiniteSumAnsatz}. Due to this fact, we were not able to set up a systematic primal bootstrap scheme to derive bounds on Wilson coefficients for the open string case.

There are however nontrivial solutions with infinitely many satellite terms which satisfy unitarity. 
We find a three-parameter family of such unitarity amplitudes which can be represented through a simple worldsheet integral. They exhibit novel high-energy, fixed-angle behavior.

\subsection{Ansatz}

As before we consider the amplitudes with exactly equidistant spectrum $m^2=n$, polynomial residues, that satisfy crossing. For the open string case where the amplitude only has poles in the $s$- and $t$-channels we get the following ansatz
\begin{align}
        T(s,t) &= \sum_{c_s,c_t=0}^{\infty}\sum_{c_u = \max(c_s,c_t)}^{c_s+c_t} \alpha_{c_s,c_t,c_u}\frac{\Gamma(c_s-s)\Gamma(c_t-t)}{\Gamma(c_u-s-t)}\,, \quad \alpha_{c_s,c_t,c_u}=\alpha_{c_t,c_s,c_u}
\end{align}
where the lower bound on $c_u$ comes from imposing the Regge behavior $s^{t+(c_s-c_u)}$ and the upper bound from imposing that residues are polynomials. 

Not all the terms in the ansatz above are independent. Eliminating the redundancies, we can write a simpler ansatz 
\be
        T(s,t) &= \sum_{i=0}^{\infty}\sum_{k=0}^{i}b_{ik}\frac{\Gamma(i-s)\Gamma(i-t)}{\Gamma(i+k-s-t)} \label{eq:sumVenAnsatz} ,
\ee
which was considered by Khuri in \cite{khuri1969derivation}.

The term $k=i=0$ corresponds to the Veneziano amplitude. Notice that the amplitude \eqref{eq:sumVenAnsatz} automatically satisfies the maximal spin constraint. The reason is that taking the discontinuity in $s$ automatically truncates the sum over $i$ and one can  trivially check that $T_s \sim s^t$. In \appref{app:completeness}, we argue that this ansatz is complete for amplitudes with an equidistant spectrum and linear trajectories.

\subsection{Regge sum rules}\label{sec:RSR_OpenString}

Checking RSR for the amplitude \eqref{eq:sumVenAnsatz} is more subtle. Here we can distinguish two cases: when the sum over $i$ truncates to $\leq N_{\text{max}}$; when the sum over $i$ goes all the way to infinity.

Let us first discuss the case $i \leq N_{\text{max}}$. In this case, superpolynomial softness is trivial because for given $i$ and $k$, the amplitude behaves in the Regge limit as $T(s,t) \sim s^{t-k}$. 
However, we show in \appref{sec:noFiniteSumAnsatz} that all such amplitudes violate unitarity. Thus, we conclude that  \emph{no unitary deformations of the Veneziano amplitude with $i \leq N_{\text{max}}$ exist.} This makes the method used to derive bounds on Wilson coefficients in the previous section inapplicable, see \secref{sec:algorithmBound}.

Next, we consider the case when the sum over $i$ goes all the way to infinity. In this case, a class of unitary deformation was recently found by Cheung and Remmen in \cite{Cheung:2023adk}.
In particular, they found an amplitude depending on the parameter $r$
\begin{equation}\label{eq:CRAmplitude}
        T_{\text{CR}}(s,t) = \sum_{i=0}^\infty {1 \over i!} {r \over r+i} \frac{\Gamma(i-s)\Gamma(i-t)}{\Gamma(i-s-t)} = \frac{\Gamma(-s)\Gamma(-t)}{\Gamma(-s-t)}{}_3F_2(-s,-t,r;-s-t,1+r;1)
\end{equation}
which for $r=0$ reduces to the Veneziano amplitude. Unitarity imposes an additional constraint on $r$. For example, in $d=4$, it requires that $r\geq -1/2$. In the Regge limit, this deformation takes the form
\be
T_{\text{CR}}(s,t) = s^t + {r \over (1+t) s} + ... \ ,
\ee
where the second term explicitly violates RSR. The mechanism by which this term emerges is interesting: the Regge limit and sum over $i$ above do not commute. Therefore even though each term in the sum \eqref{eq:CRAmplitude} satisfies RSR, the full amplitude given by an infinite sum does not.

To the best of our knowledge, the existence of unitary deformations of the Veneziano amplitudes that satisfy RSR has not been explicitly demonstrated so far, and it is what we will show next. We will not try to be exhaustive and it would be very interesting to classify all such deformations. We leave this problem for future work.

\subsection{Unitary amplitudes}

We do not know what is a complete set of unitary amplitudes \eqref{eq:sumVenAnsatz} that satisfy RSR. Here we consider a three-parameter family of amplitudes and explore it in detail. The easiest way to define them is via the worldsheet-like integral 
\begin{equation}
        T_{c_0,c_1,\lambda}(s,t)=\int_{0}^{1}dz\, z^{-s-1} (1-z)^{-t-1} (1-4 \lambda  (1-z) z)^{c_0 + c_1(s+t)} , ~~~ 0 \leq \lambda \leq {1 \over 2} ,
        \label{eq:MatsudaGen}
\end{equation}
where the restriction on $\lambda$ comes from imposing RSR.\footnote{We do not have a rigorous derivation of this fact and we cannot with full confidence exclude the possibility that there are interesting amplitudes that satisfy RSR beyond that range. Notice that analyticity of the amplitude constraints $\lambda \leq 1$ for real $\lambda$. Similarly, we do not analyze here the case of complex $(c_0, c_1, \lambda)$, or the case where we insert multiple deformation factors into the worldsheet integral.} Moreover, the Regge behavior \eqref{eq:jtOpenCase} further requires that $c_1 \lambda < {1 \over 4}$ with the leading Regge behavior given by $T_{c_0,c_1,\lambda}(s,t) \simeq \Gamma(-t) (1-4 c_1 \lambda)^t (-s)^t$.

The integral can be evaluated explicitly and the result takes the following form
\begin{equation}
        T_{c_0,c_1,\lambda}(s,t)=\frac{\Gamma (-s) \Gamma (-t)}{\Gamma (-s-t)} \, _3F_2\left(-s,-t,-c_0- c_1(s+t) ;-\frac{s+t}{2},\frac{1-s-t}{2}; \lambda \right)\, .
\end{equation}
For $\lambda=0$ it becomes the Veneziano amplitude $T_{0,0,0}=\frac{\Gamma (-s) \Gamma (-t)}{\Gamma (-s-t)}$. Similar amplitudes have been considered in the past: Matsuda \cite{Matsuda:1969zz} considered the case $c_1=1/2\,,\, c_0=0$; Mandelstam \cite{Mandelstam:1968czc} considered the case $c_1=0$. For these particular cases, expansion coefficients in \eqref{eq:sumVenAnsatz} can be found explicitly 
\begin{itemize}
        \item For $c_1=1/2\,,\, c_0=0$ (the Matsuda case), the coefficients read
        \begin{equation}
            b_{ik}= \frac{\lambda^i}{i!}\frac{(-1)^k 2^{i-k} \Gamma (i+k+1)}{\Gamma (k+1) \Gamma (i-k+1)}\,.
        \end{equation}
        \item For $c_1=0$ (the Mandelstam case), the coefficients read
        \begin{equation}
                b_{ik} = \left\{\begin{array}{ll}
                        b_{ii}&=\frac{\lambda^i}{i!}\frac{4^i \Gamma \left(i-c_0\right)}{\Gamma \left(-c_0\right)}\\
                        b_{ik}&= 0\,\,, k\neq i
                \end{array}\right. \,.
        \end{equation}
\end{itemize}
Let us also mention a couple of special cases which further simplify dramatically
\be
T_{0,1/2,1/2}(s,t) &= {1 \over 2} {\Gamma(-{s \over 2}) \Gamma(-{t \over 2}) \over \Gamma(-{s+t \over 2})} \ , \\
T_{-1/2,1/2,1/2}(s,t) &= {1 \over 2} {\Gamma(-{s \over 2}) \Gamma(-{t-1 \over 2}) + \Gamma(-{s-1 \over 2}) \Gamma(-{t \over 2}) \over \Gamma(-{s+t-1 \over 2})} \ .
\ee
These amplitudes satisfy unitarity, however they have $c_1 \lambda = {1 \over 4}$ and their Regge limit $T(s,t) \sim s^{t/2}$ differs from \eqref{eq:jtOpenCase}.

\begin{figure}
    \centering
    \includegraphics[width=\linewidth]{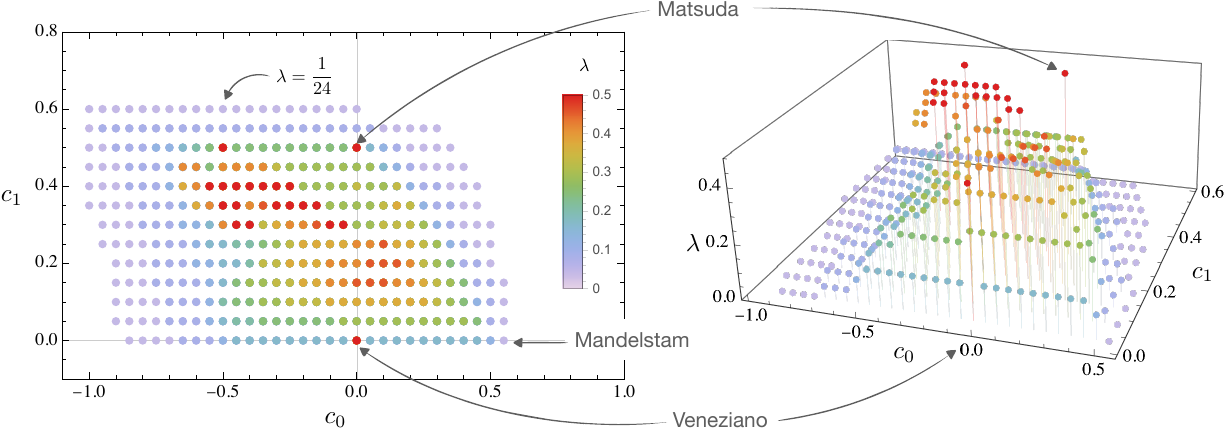}
    \caption{Region in the $(c_0, c_1 , \lambda)$ space, where unitarity is satisfied. $\lambda=0$ corresponds to the Veneziano amplitude. The smallest $\lambda$ for which unitarity was checked is $\lambda=\frac{1}{24}$. Notice that points $(0,1/2,\lambda)$ and $(-1/2,1/2,\lambda)$ are unitary for any $0 \leq \lambda \leq {1 \over 2}$ (the same is trivially true for $(0,0,\lambda)$). On the right panel, the lines indicate that unitarity is expected to hold for $\lambda \leq \lambda_*$, where $\lambda_*$ is marked by a dot.}
    \label{fig:unitRegionc0c1lambda}
\end{figure}

Our next step is to impose unitarity. It imposes further nontrivial constraints on the allowed values of $(c_0, c_1, \lambda)$. We analyzed unitarity numerically by choosing a grid in the space of parameters $(c_0, c_1 , \lambda)$ and explicitly checking unitarity up to level $100$. We then further checked unitarity at level $200,300,400$. The results are shown in \figref{fig:unitRegionc0c1lambda}. In particular, we find that unitarity implies that $c_1 \geq 0$. As $\lambda\to 0$, the number of levels needed to check unitarity increased and we do not exclude that the lowest level (in blue-violet in \figref{fig:unitRegionc0c1lambda}) might be reduced further as the number of levels goes to infinity. We provide the list of points satisfying unitarity in an ancillary file linked to this publication.
We conclude that there is a finite region of unitary amplitudes that satisfy RSR in the three-dimensional space $(c_0, c_1 , \lambda)$.

Verifying unitarity for stringy amplitudes is a famously difficult problem because they have infinitely many poles. Even for the Veneziano amplitude, the original proof is via the no-ghost theorem \cite{Goddard:1972iy}. This was recently revisited in \cite{Arkani-Hamed:2022gsa} and proven for all superstring amplitudes in $d\leq 6$ directly for the residues. Here, we checked unitarity numerically up to a certain maximal mass by explicitly computing the residues. 

\subsection{High-energy, fixed-angle scattering}\label{sec:highEnergyFixedAngle}

Here we consider high-energy $s,t \to \infty$, fixed-angle ($s/t - \text{fixed}$) behavior of the amplitude. Let us consider first the universal limit \cite{Caron-Huot:2016icg} when both $s,t>0$.\footnote{To avoid the poles we as usual go slightly in the complex direction $(s,t) \to (s(1+i \epsilon),t (1+i \epsilon)) $.} In this limit, the amplitude is large and we find that its leading asymptotic takes the following form
\be
\lim_{s,t \to \infty} \log T_{c_0,c_1,\lambda}(s,t) &= (s+t) \log (s+t) - s \log s - t \log t \nn \\
&+ c_1 \Big( t \log {1 \over 2} \Big( 1 - \tilde \lambda {s-t \over s+t} +\sqrt{1- \tilde \lambda} \sqrt{1 -  \tilde \lambda {(s-t)^2 \over (s+t)^2} } \Big) + \{ s \leftrightarrow t \} \Big) ,
\label{eq:fixedangleOS}
\ee
where it was convenient to introduce the following effective coupling
\be
0 \leq \tilde \lambda = 4 \lambda (1 - \lambda) \leq 1 .
\ee
This expression can be derived for example by evaluating \eqref{eq:MatsudaGen} using a saddle point approximation as in \cite{Gross:1987kza}. Let us comment on several features of this result. First, the leading term does not depend on $c_0$. Second, for $\lambda = 1/2$ or $\tilde \lambda = 1$ the result simplifies dramatically and we get
\be
\lim_{s,t \to \infty} \log T_{c_0,c_1,1/2}(s,t) &=(1-c_1) \Big[ (s+t) \log (s+t) - s \log s - t \log t \Big] .
\ee
Third, for unitary amplitudes, namely $c_1 \geq 0$, we find that 
\be
\label{eq:bounasym}
\lim_{s,t \to \infty} \log T_{c_0,c_1,\lambda}(s,t) &\leq (s+t) \log (s+t) - s \log s - t \log t ,
\ee
and we further comment on this below. Finally, let us define the asymptotic Regge trajectory $j_{\text{asy}}(t)$ 
\be
\lim_{s \to \infty} \lim_{s,t \to \infty} \log T_{c_0,c_1,\lambda}(s,t) \simeq j_{\text{asy}}(t) \log s + ...  \ .
\ee
For $0 \leq \lambda < 1/2$ we find that
\be
j(t) = j_{\text{asy}}(t) = t ,
\ee
however for $\lambda = {1 \over 2}$ we find that 
\be
\label{eq:asymptoticreggevsnormalregge}
j(t) = t, ~~~ j_{\text{asy}}(t) =(1-c_1) t. 
\ee
where recall that $j(t)$ is the Leading Regge trajectory defined by $\lim_{s \to \infty} \log T_{c_0,c_1,\lambda}(s,t) \simeq j(t) \log s$. Therefore we see that for $\lambda=1/2$, the two limits are not continuously related.

In the language of \cite{Caron-Huot:2016icg},  $j(t)$ counts the total number of the \emph{excess zeros} $z_i(t)$, and $j_{\text{asy}}(t)$ counts those excess zeros that do not escape to infinity, namely $\lim_{t \to \infty} {z_i(t) \over t} < \infty$, as we take the limit $t \to \infty$. Therefore $j_{\text{esc}}(t) \equiv j(t) - j_{\text{asy}}(t) \geq 0$ measures the fraction of the \emph{escape zeros}. For this picture to be consistent with \eqref{eq:asymptoticreggevsnormalregge} we need $c_1 \geq 0$. This is precisely the condition that we found when imposing unitarity!

The result \eqref{eq:fixedangleOS} sheds interesting light on the bootstrap analysis of \cite{Caron-Huot:2016icg}, where the behavior of stringy amplitudes at high energies $\lim_{s,t \to \infty} \log T(s,t)$ was constrained on general grounds. In particular, we see that two assumptions made in that paper are too restrictive:
\begin{itemize}
    \item The asymptotic Regge limit assumption $j(t) = j_{\text{asy}}(t)$ made in \cite{Caron-Huot:2016icg} is explicitly violated by the amplitudes with $\lambda = 1/2$. In other words, there are amplitudes for which the number of escape zeros is large so that $j_{\text{esc}}(t) \sim t$.
    \item The assumption about the support of zeros of Legendre polynomials made in \cite{Caron-Huot:2016icg} related to the support of ${\rm Disc}_\beta \partial_\beta \log T(s, s \beta)$ being restricted to an ellipse extended between $-1 \leq \beta \leq 0$ is explicitly violated by the amplitudes with $0<\lambda<1/2$. In this case the support of zeros is given by $ -1 - {2 \sqrt{\tilde \lambda} \over 1 - \sqrt{\tilde \lambda}}\leq \beta \leq 0$.
\end{itemize}

We see therefore that already at the level of amplitudes with equidistant spectrum and exactly linear Regge trajectories, the result of \cite{Caron-Huot:2016icg} was based on too restrictive assumptions. It is a very interesting question: which extra properties of the amplitude lead to the asymptotic uniqueness of the Veneziano amplitude? For example, the emergent $s-u$ asymptotic crossing property discussed in \cite{Sever:2017ylk} is not satisfied by the amplitudes $0<\lambda< 1/2$ and $c_1 \neq 0$. Similarly, it would be very interesting to understand upon which extra assumptions the property $j(t) = j_{\text{asy}}(t)$ holds.

\subsection{Results for Wilson coefficients}
As the finite sum ansatz is never unitary, we could not use the same primal approach that we adopted for the closed string amplitudes in  \secref{sec:result_gravitons}. However, it is still interesting to see the region in the space of Wilson coefficients that is covered by the unitary amplitude \eqref{eq:MatsudaGen}. The status of this exercise is very different compared to what we have done in the previous section because it could be that by generalizing our model further, a larger region of the parameter space could be covered.

To define the Wilson coefficients, we expand the general open string amplitude \eqref{eq:sumVenAnsatz} at low energy\footnote{Here, we follow the convention of \cite{Albert:2022oes}.}
\begin{equation}
    T(s,t) = b_{0,0} \frac{u}{s t} + g_{1,0}(s+t) + g_{2,0}(s^2 + t^2) + 2 g_{2,1} st + \dots
\end{equation}
As before, we consider bounds on ratios of Wilson coefficients. Here, we normalize everything by $g_{1,0}$ and define\footnote{Since the Regge intercept is $j_0 = 0$, the coefficient $g_{1,0}$ is dispersive (in other words, it can be expressed in terms of the discontinuity of the amplitude). } 
\begin{equation}
    \tilde{g}_2 =\frac{g_{2,0}}{g_{1,0}}\,,\qquad \tilde{g}'_2 =\frac{2 g_{2,1}}{g_{1,0}}\,.
\end{equation}
Furthermore, note that only the $i=k=0$ term contributes to the massless pole. 

\begin{figure}
    \centering
    \includegraphics[scale=1]{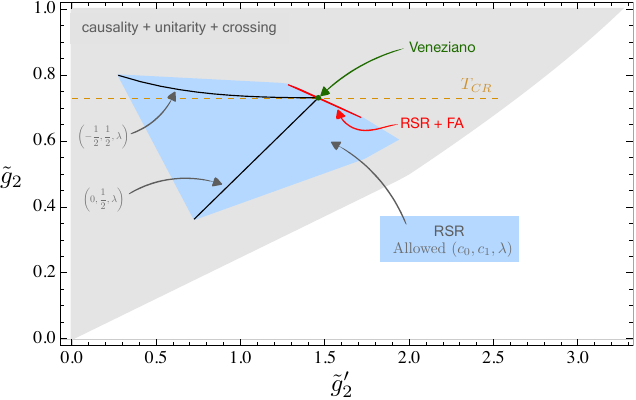}
    \caption{The space of Wilson coefficients $\{ \tilde{g}_2,  \tilde{g}'_2\}$ covered by the unitary amplitudes \eqref{eq:MatsudaGen} (in blue). We highlighted a few special lines. For example, $\left(0, \frac{1}{2}, \lambda\right)$ corresponds to the Matsuda amplitude \cite{Matsuda:1969zz}.  In red, we highlighted the only region that has the same high-energy, fixed-angle behavior as the Veneziano amplitude -- see \secref{sec:highEnergyFixedAngle}. Finally, in orange, we show the line covered by the Cheung-Remmen amplitude \eqref{eq:CRAmplitude}. The gray region corresponds to the usual bootstrap constraints (ACU) \ref{ass:ACU} using $k_{\text{max}}=15$.}
    \label{fig:open_g2}
\end{figure}

In \figref{fig:open_g2}, we present the region covered by the amplitude \eqref{eq:MatsudaGen} and compare it with the dual bound found in \cite{Albert:2022oes} and obtained using only causality, unitarity, and crossing symmetry \ref{ass:ACU}. The amplitudes \eqref{eq:MatsudaGen} cover a portion of the allowed space. 

In a recent work \cite{Fernandez:2022kzi}, the authors pointed out that the spin $0$ contribution can be removed from string amplitude to generate new unitary amplitudes. The scalar contribution can be removed from the amplitude by considering
\begin{equation}
    T_{>0}(s,t) = T(s,t) - \sum_{n=1}^{\infty} \left(\frac{c_n}{s-n} +\frac{c_n}{t-n}  \right)\,,
\end{equation}
where $c_n$ are fixed to remove the spin $0$ contribution for all $n$. The second term, however, clearly violates RSR and so will the resulting amplitude. It is thus not possible to remove such contributions without changing the shape of the leading Regge trajectory for negative $t$.

\begin{figure}
    \centering
    \includegraphics{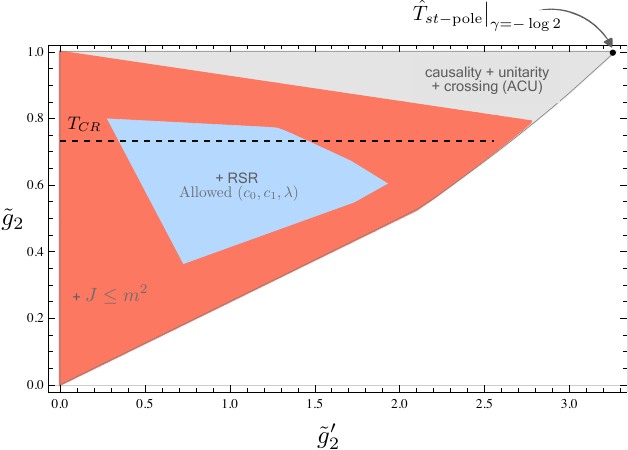}
    \caption{Comparison of the spaces of allowed Wilson coefficients $\{ \tilde{g}_2,  \tilde{g}'_2\}$ coming from different assumptions (in gray and red), and the space covered by the unitary amplitudes \eqref{eq:MatsudaGen} (in blue). The gray region corresponds to the usual bootstrap constraints \ref{ass:ACU}. In red, we further imposed maximum spin $J\leq m^2$ in the dual approach \ref{ass:ACU+J}.  The number of null constraints used to produce this plot is $k_{\text{max}}=9$.}
    \label{fig:openG2DualJ}
\end{figure}

We can now impose linearity of the leading Regge trajectory for positive $t$ \ref{ass:ACU+J} using the dual formalism. We present the results in \figref{fig:openG2DualJ}. Adding a finite number of RSR constraints in the dual approach does not lead to stronger bounds. Similar to what we observed in the closed string case, the maximum spin constraint removes part of the region in the vicinity of $\tilde{g}_2=1$. This can be expected since the line $\tilde{g}_2=1$ only allows for exchange particles of mass $m=1$ and is populated by the following amplitudes 
\begin{equation}
    \hat T_{st-\text{pole}} = \frac{1}{(1-s)(1-t)} +  \gamma \left(\frac{1}{1-s} + \frac{1}{1-t}\right)
\end{equation}
with $\gamma\geq -\log 2$ to satisfy unitarity.  The line $\tilde{g}_2=1$ is described by varying $\gamma\in[-\log 2, \infty)$. The upper-right kink saturates this inequality, at which point the spin $0$ contribution to the residue of the amplitude vanishes. Clearly, all these amplitudes violate the maximal spin constraint for the linear leading Regge trajectory.

\section{Conclusions}\label{sec:conclusion}

Charting out the space of stringy tree-level amplitudes is largely an open problem. Among other things, this space is important because it contains large $N$ QCD and weakly coupled UV completions of Einstein gravity. 

In this paper, we developed the S-matrix approach to this problem. We utilized extra knowledge about the leading Regge trajectory which we considered to be linear. On one hand, it puts an upper bound on the maximal spin of exchanged particles at a given mass.
On the other hand, scattering amplitudes in this class exhibit superpolynomial softness: they decay faster than a polynomial at high energies and fixed angles. This condition can be conveniently restated as an infinite set of Regge sum rules \eqref{eq:RSRbasic} that the discontinuity of the amplitude has to obey. 

Our basic conclusion is that superpolynomial softness does not lead to any obvious low-energy imprint as exhibited by the low-energy Wilson coefficients. In contrast to that, the maximal spin constraint leads to slightly more stringent bounds compared to the standard bootstrap scheme based on causality and unitarity.\footnote{Strictly speaking, to strengthen this conclusion it would be desirable to extend our primal ansatzes both for the closed and open string cases. For the closed string case, it would be interesting to construct amplitudes that do not satisfy RSR more systematically. For example, including terms considered in \cite{Veneziano:2017cks} would be an obvious way to do it. For the open string case, on the opposite, we would like to have a better understanding of the landscape of the amplitudes that do satisfy RSR, beyond the example considered in the present paper.}

Let us provide a simple, intuitive explanation of these results. The low-energy Wilson coefficients are dominated by the contribution from the lightest degrees of freedom that were integrated out. The maximal spin constraint puts a bound on the maximal spin of these lightest massive degrees of freedom and thus affects the low-energy Wilson coefficients. The UV softness, however, can kick in at energies $s_{\text{UV}} \gg m_{\text{gap}}^2$ and therefore leave very little imprint on the low-energy observables. It is not obvious that it is possible to construct amplitudes with the separation of these two scales (the mass gap, and the energy scale at which the UV soft behavior becomes visible). For example, in the standard string amplitudes, the UV softness can be already seen at energies $s_{\text{UV}} \sim m_{\text{gap}}^2$ and not just asymptotically. Our primal ansatz achieves precisely that: it delays the kick-in of the UV soft behavior to higher energies and thus effectively hides it from a low-energy experimentalist.

In the statements above, we effectively assumed that properties of the leading Regge trajectory $j(t)$ for positive and negative $t$ are unrelated. It is quite probable that this is not the case. For example, in all known examples, $j(t)$ is a convex function of $t$.\footnote{Convexity of the leading Regge trajectory can be proven for nonperturbative CFTs \cite{Costa:2017twz}, it is not known if it holds for planar CFTs which are dual to weakly coupled stringy scattering in AdS.} On a related note, for the amplitudes of the type considered in \cite{Veneziano:2017cks} (assuming they could be made unitarity for the closed string case as well) we can make the leading Regge trajectory at negative $t$ arbitrarily flat while keeping it intact for positive $t$. 

For the closed string case, we focused on the MHV scattering of gravitons in four dimensions with the leading Regge trajectory $j(t)=2+2t$, where we work in the units $m_{\text{gap}}^2=1$.\footnote{Here $m_{\text{gap}}$ is the mass of the lightest massive state that appears in the amplitude.} We put forward a primal bootstrap scheme, see \secref{sec:algorithmBound}, which is analogous to the one for the nonperturbative case put forward in \cite{Paulos:2017fhb}. In this scheme, analyticity and crossing are manifest, but unitarity is a nontrivial constraint that is imposed numerically. We then derived bounds on Wilson coefficients using both the primal and dual methods. The results are presented in Section \ref{sec:result_gravitons}. We found bounds that are slightly more stringent than the ones that follow simply from analyticity, unitarity, and crossing. We also observed an approximate agreement between the primal and the dual bounds. This fact is interesting because in our dual implementation, only a finite number of RSR could be added and they are not used in the numerics. Our primal ansatz, on the other hand, satisfies infinitely many RSR. 

For the open string case, we considered the scattering of massless scalars and took the leading Regge trajectory to be $j(t)=t$. As opposed to the closed string case, we showed in Appendix \ref{sec:noFiniteSumAnsatz} that there are no unitary ansatzes with a finite number of satellite terms in this case. There are, however, unitarity amplitudes that satisfy Regge sum rules and have infinitely many satellite terms. We constructed a three-parameter deformation of the Veneziano amplitude \eqref{eq:MatsudaGen} and we showed that it obeys unitarity in a finite region of the parameter space $(c_0, c_1, \lambda)$. We then explored the space of Wilson coefficients covered by this family of amplitudes and compared them to the bounds derived using the dual method. 

We also found that \eqref{eq:MatsudaGen} exhibits nontrivial behavior in the high-energy, fixed-angle region which goes beyond the analysis of \cite{Caron-Huot:2016icg} in several interesting ways. In particular, we observed that a technical assumption about the support of excess zeros made in \cite{Caron-Huot:2016icg} needed to prove the uniqueness of the high-energy limit of the amplitude does not follow from basic principles, and is thus genuinely an extra assumption. It would be very interesting to remove it completely.

In Appendix \ref{app:asybound},\footnote{We thank Miguel Correia for collaboration on this topic.} we set up a dual bootstrap version of the analysis \cite{Caron-Huot:2016icg} \emph{assuming} the asymptotic distribution of zeros of Legendre polynomials is supported along the negative axis. It leads to the following bound 
\be
\label{eq:bounasymG}
\lim_{s,t \to \infty} \log T(s,t) &\lesssim \alpha' \left( (s+t) \log (s+t) - s \log s - t \log t \right) ,
\ee
where $\alpha'$ is the slope of the Regge trajectory $j(t) \simeq \alpha' t$ at large positive $t$. This bound should be understood as either the statement about the residues of the amplitude or as the statement about the high-energy limit taken slightly away from the real axis.

Using \eqref{eq:bounasymG},\footnote{In the original work of Cerulus and Martin \cite{Cerulus:1964cjb}, the analog of \eqref{eq:bounasymG} is played by the assumption of polynomial boundedness needed for the Mandelstam representation to hold.} in Appendix \ref{app:CMstringy}, we derived a lower bound on the behavior of stringy amplitudes at high energies and fixed physical angles, namely $s,t \to \infty$, $s/t$ fixed and $t<0$,
\be
\label{eq:CMstringy}
\underset{|z| \leq z_0}{\rm max} \Big| T \Big(s, t=-{s \over 2}(1-z) \Big) \Big| \gtrsim e^{- \alpha' s \log {1+\sqrt{1-z_0^2} \over z_0}} \ . 
\ee
This generalizes the Cerulus-Martin bound \cite{Cerulus:1964cjb} derived in the context of gapped QFTs to the case of tree-level stringy amplitudes.

There are many open directions that we think are worth exploring further. These are naturally related to relaxing various assumptions made in the paper. Stringy amplitudes that exhibit an accumulation point in the spectrum were analyzed in \cite{Coon:1969yw,Baker:1970vxk,Coon:1972qz,Figueroa:2022onw,Geiser:2022icl,Chakravarty:2022vrp,Bhardwaj:2022lbz,Jepsen:2023sia,Geiser:2022exp,Cheung:2022mkw,Geiser:2023qqq}. Stringy amplitudes with the spectrum different from linear were constructed recently in \cite{Cheung:2023uwn}. Stringy amplitudes, with linear spectrum and no accumulation point that exhibit power-like behavior at high energies and fixed angles, were explored in \cite{Veneziano:2017cks,Cheung:2023adk}. Stringy amplitudes that satisfy monodromy relations were studied in  \cite{Huang:2020nqy,Chiang:2023quf,Berman:2023jys}. It would also be very interesting to generalize our analysis to general number of spacetime dimension $d$,\footnote{Bound on Wilson coefficient assuming (ACU) \ref{ass:ACU} of the graviton amplitude in higher dimension were recently derived in \cite{Caron-Huot:2022jli}.} as well as to the scattering of gauge bosons as in, for example, \cite{Bachu:2022gof}. Finally, the soft behavior of the amplitudes at high energies and fixed angles is essential for celestial holography \cite{Arkani-Hamed:2020gyp}, and it would be interesting to explore the models studied in the present paper in that context.

An important problem in the stringy S-matrix bootstrap program is the construction and consistency of multi-point amplitudes. Once these are constructed a consistent factorization must be checked. In the case of string theory, the factorization of multi-particle amplitudes reveals the degeneracy of states not visible at the level of the $2 \to 2$ amplitudes and the Hagedorn growth of their density with energy \cite{Fubini:1969qb}. Similarly, multi-particle amplitudes with satellite terms studied in this paper can be constructed, and their factorization can be analyzed \cite{Gross:1969db}.\footnote{In \cite{Geiser:2023qqq}, the factorization of the multi-point Baker-Coon-Romans amplitude was explored. In \cite{Bianchi:2020cfc}, multi-point amplitudes, which generalize the Lovelace-Shapiro model of pion scattering, were analyzed.} It would be very interesting to revisit this question and explore it in conjunction with unitarity and, in particular, for the concrete unitary models studied in the present paper. 

\section*{Acknowledgments}

We thank Jan Albert, Simon Caron-Huot, Clifford Cheung, Miguel Correia, Felipe Figueroa, David Gross, Leonardo Rastelli, Grant Remmen, Amit Sever, Piotr Tourkine, and Alessandro Vichi for useful discussions. We also thank Alessandro Vichi for comments on the draft. We would also like to thank the organizers and participants of the KITP program ``Bootstrapping Quantum Gravity" for the numerous discussions that motivated this work. This project has received funding from the European Research Council (ERC) under the European Union’s Horizon 2020 research and innovation program (grant agreement number 949077). The work of KH is supported by the Simons Foundation grant 488649 (Simons Collaboration on the Nonperturbative Bootstrap) and by the Swiss National Science Foundation through the project 200020 197160 and through the National Centre of Competence in Research SwissMAP.

\appendix
\section{Review of the dual method}\label{sec:dualMethod}
In this appendix, we review the dual method of \cite{Caron-Huot:2020cmc} used to bound Wilson coefficients using causality, unitarity, and crossing symmetry. Here, we will focus on the case of the MHV scattering amplitude of gravitons. For a review of this method in open string scattering, see for example \cite{Albert:2022oes}. See also \cite{EliasMiro:2022xaa,Li:2023qzs} for a  detailed explanation of the dual bootstrap as an SDP problem. 

\subsection{Dispersion relation and Wilson coefficients}
We start by writing a dispersion relation for $f(s|t,u)$
\begin{equation}\label{eq:dispRel_f}
\begin{split}
    f(s|t,u)&= \oint \frac{ds'}{2\pi i} \frac{f(s'|t,-s'-t)}{s-s'} = \frac{8\pi G_N}{stu} + |\beta_{R^3}|^2 \frac{t u}{s} - |\beta_{\phi}|^2 \frac{1}{s}  \\
    &- \frac{1}{\pi}\int_{m_{\text{gap}}^2}^\infty dm^2\left(\sum_{J=0}^{\infty} \frac{1+ (-1)^J}{2} \frac{\rho_{J}^{++} (m^2) d_{0,0}^{J}\left(1+\frac{2t}{m^2}\right)}{m^8 (s-m^2)}\right.\\
    &\qquad\qquad\qquad\qquad\left.+\sum_{J=4}^{\infty} \frac{\rho_{J}^{+-} (m^2) d_{4,4}^{J}\left(1+\frac{2t}{m^2}\right)}{(t+m^2)^4 (-s-t-m^2)}\right)
\end{split}
\end{equation}
where we recall that the spectral densities $\rho^{++}_J(m^2),\rho^{+-}_J(m^2)$ are the imaginary part of the partial amplitude
\begin{align}
        \Im T_{++--}(s,t,u) &= \sum_{J=0}^{\infty} \frac{1+ (-1)^J}{2}\rho_{J}^{++} (m^2) d_{0,0}^J\left(1+ \frac{2t}{s}\right) \label{eq:imTppmmPW}\\ 
        \Im T_{+--+}(s,t,u) &= \sum_{J=4}^{\infty}\rho_{J}^{+-} (m^2) d_{4,4}^J\left(1+ \frac{2t}{s}\right)\,,\label{eq:imTpmmpPW}
\end{align}
and from unitarity $\rho^{++}_J(m^2),\rho^{+-}_J(m^2) \geq 0$. In writing the dispersion relation for $f(s|t,u)$, we used that the intercept of the amplitude is $j_0 =2$ and thus  $f(s|t,-s-t)\lesssim 1/|s|^2$ and the arc at infinity can be dropped.

By expanding the dispersion relation \eqref{eq:dispRel_f} at low energy and comparing it with the low-energy expansion \eqref{eq:f_lowEnergy}, it is straightforward to obtain dispersive representation for the $a_{k,j}$ and for example
\begin{align}
    a_{k,0} &= \avg{\frac{1}{(m^2)^{4+k}}}_{++} + \avg{\frac{(-1)^k}{(m^2)^{4+k}}}_{+-} \label{eq:avg_ak0}\\
    a_{2,1}&= \avg{\frac{\cJ}{m^{14}}}_{++} + \avg{\frac{22-\cJ}{m^{14}}}_{+-}\\
    a_{4,1}&= \avg{\frac{\cJ}{m^{18}}}_{++}+\avg{\frac{24-\cJ}{m^{18}}}_{+-}\\
    a_{4,2}&= \avg{\frac{\cJ (\cJ-2)}{4 m^{18}}}_{++} +\avg{\frac{\cJ (62-\cJ) + 864}{4 m^{18}}}_{+-}
\end{align}
where $\cJ = J(J+1)$ and we used the notation
\begin{align}
    \avg{(\dots)}_{++} &= \frac{1}{\pi}\int_{m_{\text{gap}}^2}^\infty \frac{dm^2}{m^2}\sum_{J=0}^{\infty} \frac{1+ (-1)^J}{2} \rho_{J}^{++} (m^2) (\dots) \label{eq:avg++}\\
      \avg{(\dots)}_{+-} &= \frac{1}{\pi}\int_{m_{\text{gap}}^2}^\infty \frac{dm^2}{m^2}\sum_{J=4}^{\infty}  \rho_{J}^{+-} (m^2) (\dots)\,,\label{eq:avg+-}
\end{align}
to represent the moment with positive measures. It is also convenient to name the function inside the brackets
\begin{equation}
    a_{k,j} = \avg{a^{++}_{k,j}(m^2, \cJ)}_{++}+ \avg{a^{+-}_{k,j}(m^2, \cJ)}_{+-}\,.
\end{equation}
Clearly, for even $k$ \eqref{eq:avg_ak0} imposed positivity of $a_{k,0}$ and the ordering $ a_{0,0}\geq a_{2,0}\geq a_{4,0}\geq \dots $. No simple statement can be made for the other coefficients and we will use a numerical method as explained in the next subsections.

\subsection{Crossing symmetry and null constraints}
The function $f(s|t,u)$ is symmetric in $t-u$. However, the dispersion relation \eqref{eq:dispRel_f} is performed at fixed $t$ and makes this symmetry not manifest. By imposing the RHS of \eqref{eq:dispRel_f} to be symmetric in $t-u$, we obtain extra constraints. Explicitly we obtain the `master null constraint'
\begin{equation}\label{eq:master_nullConstraint}
    \begin{split}
       & \avg{\frac{d_{0,0}^{J}\left(1+\frac{2t}{m^2}\right)}{m^6 (s-m^2)}}_{++}  + \avg{\frac{m^2 \,d_{4,4}^{J}\left(1+\frac{2t}{m^2}\right)}{(t+m^2)^4 (u-m^2)}}_{+-}\\
        =&  \avg{\frac{d_{0,0}^{J}\left(1+\frac{2u}{m^2}\right)}{m^6 (s-m^2)}}_{++}  + \avg{\frac{m^2 \,d_{4,4}^{J}\left(1+\frac{2u}{m^2}\right)}{(u+m^2)^4 (t-m^2)}}_{+-} \,.
    \end{split}
\end{equation}
By expanding \eqref{eq:master_nullConstraint}, at low $s,t$, we obtain a sequence of null constraints $\cX_{k,j}=0$. They are labeled similarly as the coefficients $a_{k,j}$ in \eqref{eq:f_lowEnergy}
\begin{equation}\label{eq:NC_Expansion}
    0 = \sum_{k\geq j\geq 0} \cX_{k,j} s^{k-j}t^j\,.
\end{equation}

They are yet another set of null constraints. Indeed, the same function also appears in a third amplitude 
\begin{align}
    T_{+-+-}(s,t,u)= ([13]\langle 24\rangle )^4  f(t|s,u) = t^4 f(t|s,u)  \label{eq:Tpmpm}\\
    \Im T_{+-+-}(s,t,u)=\sum_{J=4}^{\infty} \rho_J^{+-}(s) (-1)^J d^J_{4,-4}(1+2t/s)\,,
\end{align}
and thus the function has another dispersion representation. As the intercept for the gravitational amplitude is $j_0=2$, we write a 3SDR for $f(t|s,u)$ using \eqref{eq:Tpmpm}
\begin{equation}
\begin{split}
    &f(t|s,u)= \oint \frac{ds'}{2\pi i} \frac{s^3 f(t|s',-s'-t)}{(s')^3(s-s')} = \frac{8\pi G_N}{stu} + |\beta_{R^3}|^2 \frac{s u}{t}-  |\beta_{\phi}|^2 \frac{1}{t} +c_0 (s^2 + u^2) + c_1 su + f_0(t)\\
    & ~~~- \frac{1}{\pi} \int_{m^2_{\text{gap}}}^\infty dm^2 \left(\sum_J^\infty (-1)^J \rho_J^{+-}(m^2) \frac{d_{4,-4}^J\left(1+\frac{2t}{m^2}\right) s^3}{t^4} \left(\frac{1}{m^6 (s-m^2)} + \frac{1}{(m^2 +t)^3(-s-t-m^2)}\right)\right)\,.
\end{split}
\end{equation}
The coefficients $c_0,\, c_1$ and the function  $f_0(t)$ are unknown subtraction terms.
The Mandelstam dependence is only in the kernel and we can thus write
\begin{equation}
\begin{split}
     f(s|t,u) &= \frac{8\pi G_N}{stu} + |\beta_{R^3}|^2 \frac{t u}{s}-  |\beta_{\phi}|^2 \frac{1}{s} +  c_0 (t^2 + u^2) + c_1 tu + f_0(s)\\
     &- \avg{(-1)^J \frac{d_{4,-4}^J\left(1+\frac{2s}{m^2}\right) t^3}{s^4} \left(\frac{1}{m^4 (t-m^2)} + \frac{m^2}{(m^2 +s)^3(-s-t-m^2)}\right)}_{+-}
\end{split}
\end{equation}
Equating with \eqref{eq:dispRel_f}, we obtain a second `master null constraint'. By expanding at low energy, we get a second sequence of null constraints $\cY_{k,j}$. As the subtraction terms are unknown, it implies that these null constraints are valid for $k-j\geq 3$ and $j\geq 1$. We emphasize here that we need 3 subtractions as the intercept is exactly $j_0=2$. If one considers constraints on an EFT where the UV is nonperturbative, two subtractions are enough \cite{Haring:2022cyf}, see \cite{Chiang:2022jep} where such constraints were imposed. 

\subsection{Dual bootstrap algorithm}\label{sec:DualBootstrapEqu}
To derive dual bounds, we start by writing the \emph{bootstrap equation}. Without loss of generality, we will explain the case where the Wilson coefficients are normalized by $a_{0,0}$ and we set $m_{\text{gap}}=1$ for clarity. We will consider here carving out a 2d region $\{g, \lambda\}$ where $g, \lambda$ can by any of the $a_{k,j}$. Let us define the vectors 
\begin{align}
    \vec v^{++} (m^2,J) &= \left(a_{0,0}^{++} (m^2, J) , g^{++}(m^2, J), \lambda^{++}(m^2, J) , \vec{n}^{++}(m^2, J)\right)\\
    \vec v^{+-} (m^2,J) &= \left(a_{0,0}^{+-} (m^2, J) , g^{+-}(m^2, J), \lambda^{+-}(m^2, J) , \vec{n}^{+-}(m^2, J)\right)\\
    \vec v_{o} &= (-1,0,0,\vec{0})\\
    \vec v_{g} &= (0,-1,0,\vec{0})\\
    \vec v_{\lambda} &= (0,0,-1,\vec{0})
\end{align}
where $\vec n(m^2,J)$ is a vector of null constraint $\vec n = (\vec \cX, \vec \cY)$ and thus 
\begin{equation}
    \vec 0 = \avg{\vec{n}^{++}(m^2, J)}_{++} + \avg{\vec{n}^{+-}(m^2, J)}_{+-}\,.
\end{equation}
We denote the number of null constraints used by $k_{\text{max}}$, the highest value of $k$ in \eqref{eq:NC_Expansion}.
We can then write the \emph{bootstrap equation}
\begin{equation}
0=    a_{0,0}\,\vec v_{0} + g\, \vec v_g + \lambda\, \vec v_{\lambda} + \avg{\vec v^{++} (m^2,J)}_{++} +\avg{\vec v^{+-} (m^2,J)}_{+-} \,.
\end{equation}
The corresponding bootstrap problem is to find a functional $\vec \alpha$ such that 
\begin{itemize}
    \item $\vec\alpha$ is normalized by $\vec \alpha \cdot \vec v_g = \left\{\begin{array}{ll}
        +1&,\,\text{ for upper bound}\\
        -1&,\,\text{ for lower bound}\\
    \end{array}
    \right.$
    \item $\vec\alpha$ maximize $\vec \alpha \cdot \left(\vec v_o + \frac{\lambda}{a_{0,0}} \vec v_\lambda\right)$. We call the result of this optimization $A(\lambda)$.
    \item $\vec\alpha$ is positive on the spectrum: 
    \begin{equation}\label{eq:PosOnalpha}
        \begin{array}{ll}
            \vec \alpha\cdot \vec v^{++}(m^2, J)\geq 0 \, & \text{for all } (m,J)\in \text{  spectrum}\\
            \vec \alpha\cdot \vec v^{+-}(m^2, J)\geq 0\, &\text{for all }(m,J)\in \text{  spectrum}\,.
        \end{array}
    \end{equation}
\end{itemize}
In this case, without specific spectrum assumption \ref{ass:ACU}
\begin{equation}\label{eq:spectrumACU}
    \text{spectrum}: \left\{\begin{array}{ll}
    (++)\text{ channel}:&~~~m\geq m_{\text{gap}}, J=0,2,\dots\\
    (+-)\text{ channel}:&~~~m\geq m_{\text{gap}}, J=4,5,\dots
    \end{array}\right.
\end{equation}
For $\vec \alpha$ solution to the bootstrap problem, applying the functional $\vec \alpha$ to the bootstrap equation and using linearity of the average $\avg{\dots}$ we obtain 
\begin{equation}
 \vec \alpha \cdot \left(\vec v_o + \frac{\lambda}{a_{0,0}} \vec v_\lambda\right)  \pm \frac{g}{a_{0,0}}\leq 0
\end{equation}
which leads to the two-sided bound
\begin{equation}
    A^{-}(\lambda) \leq \frac{g}{a_{0,0}} \leq -A^{+}(\lambda)\,.
\end{equation}
This procedure can be efficiently implemented in SDPB \cite{Simmons-Duffin:2015qma,Landry:2019qug}. It led to all the dual bounds using only causality, unitarity, and crossing symmetry presented in this work (gray regions in our plots). This procedure can be extended to carve a 3d region as shown in \figref{fig:a4}.

In practice, we need to truncate the number of constraints in spin. However, as already observed in \cite{Caron-Huot:2020cmc}, the convergence in spin is fast and we truncated at $J_{\text{max}} = 100$.

\subsection{Maximal spin constraint in the dual approach}
Let us explain next how the maximal spin constraint \ref{ass:ACU+J} is imposed in the dual approach. It changes the sum over spins in \eqref{eq:dispRel_f} into 
\begin{equation}
    \sum_{J}^{\infty}\to \sum_{J}^{j(m^2)}\,.
\end{equation}
This change propagates all the way to the definition of the averages \eqref{eq:avg++} and \eqref{eq:avg+-}.

It has the effect of changing the spectrum in the constraint \eqref{eq:PosOnalpha} on the functional $\alpha$. Instead of \eqref{eq:spectrumACU}, we now have
\begin{equation}\label{eq:spectrumACU+J}
    \begin{gathered}
           \text{spectrum with}\\[-6pt]
         \text{the maximal spin constraint}
        \end{gathered}: \left\{\begin{array}{ll}
    (++)\text{ channel}:&~~~m\geq m_{\text{gap}}, J=0,2,\dots,j(m^2)\\
    (+-)\text{ channel}:&~~~m\geq m_{\text{gap}}, J=4,5,\dots,j(m^2)
    \end{array}\right.\,,
\end{equation}
and can be efficiently implemented in SDPB. To do so, we invert the relation $j(m^2) \to m^2(j)$.  Then the constraint has to be applied on all $J$, and $m^2\geq m^2(J)$. Finally, by a change of variable $m^2 = m^2(J)+x$, the constraint can be written as a polynomial in $x$ and imposed for all $x\geq 0$.

Similarly, we can impose a discrete spectrum. Following the same steps, we have (for example with an equidistant spectrum)
\begin{equation}\label{eq:spectrumACU+J+n}
    \begin{gathered}
           \text{equidistant spectrum with}\\[-6pt]
         \text{the maximal spin constraint}
        \end{gathered}: \left\{\begin{array}{ll}
    (++)\text{ channel}:&~~~m^2= n \in \mathbb{Z}_+, \,J=0,2,\dots,j(n)\\
    (+-)\text{ channel}:&~~~m^2= n \in \mathbb{Z}_+, \, J=4,5,\dots,j(n)
    \end{array}\right.\,.
\end{equation}
This can also be implemented in SDPB with the difference that the spectrum in $m^2$ cannot be implemented as a polynomial in $x$ and we have to choose a grid for $n$. In practice we chose a grid of the form $n=1,2,\dots, n_{\text{max}}$ and added some points at large $n\sim10^5, 10^6,\dots$. The convergence in the size of the grid was fast. 

With these extra constraints on the spectrum, the convergence in the number of spin constraints is slower than by considering \eqref{eq:spectrumACU}. However, in practice, it is sufficient to add several constraints at large spin $J\sim 10^5,10^6$ to obtain the final result.\footnote{Similar observation was made in \cite{Albert:2022oes} in a different context.}

\subsection{RSR as null constraints}\label{sec:RSRandDual}
Here we will discuss how RSR can be added to the dual method described above and why it does not change the bound in the present formalism. For simplicity, we will describe here the case of the open string but the same argument applies for closed string.

Following the same procedure as above, the Wilson coefficients can be written using dispersion relations 
\begin{equation}\label{eq:gnlOpenSR}
    g_{n,\ell} = \avg{\frac{2^\ell}{\ell!}\frac{P_J^{(\ell)}(1)}{m^{2n}}}
\end{equation}
where $P_J^{(\ell)}(x)$ is the $\ell$-derivative of the Legendre polynomial and the average is defined via
\begin{align}
    T_s(s,t) &= \sum_{J=0}^\infty \rho_J(s)P_J\left(1+ \frac{2t}{s}\right)\,,\\
    \avg{\dots} &= \frac{1}{\pi}\sum_{J=0}^\infty\int_{m_{\text{gap}}^2}^{\infty}\frac{dm^2}{m^2} \rho_J(m^2) (\dots)\,.
\end{align}
Null constraints that follow crossing symmetry are easily obtained using $g_{n,\ell} = g_{n,n-\ell}$ and we  denote them $\cX_{n,\ell}$.\footnote{See for example \cite{Albert:2022oes} for detailed expressions. A second set of null constraints can also be obtained but does not influence the argument in this section.} 

Let us now write the RSR in a similar form. Starting from \eqref{eq:RSRbasic}, we obtain
\begin{equation}\label{eq:RSRSumRuleDual}
    \text{RSR:} \quad \mathcal{R}_n(t)\equiv \avg{m^{2n} P_J\left(1+ \frac{2t}{m^2}\right)}=0\,,~~ j(t) < -n,\, n\geq 2\,.
\end{equation}
For $n=0,1$, one would pick the constant and pole of the amplitude at $s=0$. This is a new family of null constraints.  Let us focus on the linear trajectory $j(t)=t$, which implies that $t<-n$. Looking at the argument of the Legendre polynomials, we see that 
\begin{equation}
    1+\frac{2t}{m^2} <-1\,,\quad \text{for}\quad -t< m^2<m_{\text{gap}}^2\,,
\end{equation}
this region always exists for $t<-n$ and $m_{\text{gap}}=1$. Using properties of the Legendre polynomials, it implies that at large $J$ and any fixed $m_*$ in this interval
\begin{equation}\label{eq:RSRLargeSpin}
    \left[\mathcal{R}_n(t)\right](m_*^2,J)\sim m_*^{2n} \,c^J (-1)^J \,,~~~ c>1\,,
\end{equation}
and thus grows exponentially with an oscillating sign. This contrasts with the sum rules for the Wilson coefficients (and thus also the usual null constraints $\cX$), which grows  as a polynomial in $J$. Let us see what it implies for the bootstrap algorithm. As in \appref{sec:DualBootstrapEqu}, we build the vector $\vec v$ and add one RSR constraint
\begin{equation}\label{eq:vecWithRSR}
    \vec{v} (m^2,J) = \left(g_{1,0}(m^2,J), g_{n,\ell}(m^2.J), \vec \chi(m^2,J), [\mathcal{R}_{n}(t)](m^2,J)\right)
\end{equation}
at large $J$, it behaves as 
\begin{equation}\label{eq:vecLargeJRSR}
    \vec{v}(m^2,J)\sim (0,0, \vec 0, c^J (-1)^J) + O(J^k)\,.
\end{equation}
Thus, the constraints on the functional $\vec{\alpha}\cdot\vec{v}(m^2,J)\geq 0$ at large $J$ imposes that the last coefficient of $\vec \alpha$ is set to zero. Thus, the constraint $\mathcal{R}_{n}(t)=0$ is not used by the dual algorithm. The conclusion does not change for any finite number of RSR constraints included as the coefficient $c$ in \eqref{eq:RSRLargeSpin} depends on $t, m_*$. 

When supplemented by the maximal spin constraint \ref{ass:ACU+J}, the argument presented above fails as one forbids arbitrary large spin at fixed $m_*$. However, it is easy to see how a single RSR cannot be used in this case as well. Let us consider the large mass behavior of \eqref{eq:vecWithRSR}. Clearly, from \eqref{eq:gnlOpenSR}, all Wilson coefficients and null constraints decay at large $m^2$. In contrast, the RSR \eqref{eq:RSRSumRuleDual} grows and we get
\begin{equation}
    \vec v(m^2,J)\underset{m^2\gg1}{\sim} \left(0,0,\vec{0}, m^{2n}P_J\left(1+\frac{2t}{m^2}\right)\right) + O(m^{-2})\,.
\end{equation}
For large but finite mass, the argument of the Legendre polynomial is $x=1-\delta<1$ and oscillates in $J$. Provided that $j(m^2)$ is larger than this oscillation period, the constraint on the functional $\vec{\alpha}\cdot\vec{v}(m^2,J)\geq 0$ will also set the last coefficient of $\vec \alpha$ to zero. Note that this mechanism is `softer' as the growth is polynomial and not exponential as in \eqref{eq:vecLargeJRSR}. However, a single constraint still cannot be used. It is less clear that a large or infinite number of constraints \eqref{eq:RSRSumRuleDual}  could not be used, for example, by using a single $n$ and various $t$.

With infinitely many constraints included, the argument above fails, and it might be that the sum rules could be used. It is also possible that by applying a `smart' functional to this constraint, they could be included (for example, by `smearing' in $t$?). We leave this investigation to future work. Instead, in the present paper, we use a primal approach and build an ansatz that satisfies all constraints $R_n(t)$ by construction. 

\section{Constraints on the closed string sum}\label{sec:ClosedAnsatzconstraintonAlpha}

In this section, we present constraints on the $c_i,\, d_i$ range that appear in \eqref{eq:ansatz_f_Nmax}. First, we impose that all the residues are polynomials. This implies that
\be
c_s+c_{tu} \geq d_{tu}, ~~~ 2 c_{tu} \geq d_s \ .  
\ee
Second, we impose that the leading trajectory is $j(t)=2+2 t$. This leads to the Regge boundedness conditions for the satellite terms 
\be
c_{tu} \leq d_{tu} + 1 , ~~~2+c_s - d_s \leq d_{tu} - c_{tu} \ .
\ee
Finally, we only consider terms with $c_s, c_{tu} \geq 0$, which have poles at nonnegative integer mass square. The inequalities above then imply that and $c_{tu}, d_{tu}\geq 1$ and $d_s \geq 2$.

The minimal solution to these inequalities, namely $d_s =2$, $c_{tu}=d_{tu}=1$, and $c_s=0$ is precisely the deformation corresponding to the heterotic string amplitude.
\section{Examples of amplitudes}\label{sec:amplitudeExample}
This appendix lists various meromorphic amplitudes of massless particles satisfying unitarity and crossing symmetry. They are listed in \tabref{tab:openAmp} for the scalar amplitudes and \tabref{tab:fgravExample} for the gravitational amplitudes. We also show that an amplitude built out of a sum of single, double, and triple poles is consistent with unitarity for the gravitational amplitude.
\begin{table}[h!]
    \centering
    \centerline{
    \begin{tabular}{c|c|c}
        \toprule
        Name&Amplitude&Regge behavior\\ \midrule
        Veneziano & $T_V = \frac{\G(-s)\G(-t)}{\G(-s-t)}$&$j(t) = t$ \\
        Cheung-Remmen&$T_{CR}=\frac{\Gamma(-s)\Gamma(-t)}{\Gamma(-s-t)}{}_3F_2(-s,-t,r;-s-t,1+r;1)$&$j(t) =\left\{\begin{array}{cc}
            t & t\geq -1 \\
            -1 & t< -1\\
        \end{array}\right.$ \\[8pt]
        Matsuda& $T_{0,1/2,\lambda} =\frac{\Gamma(-s)\Gamma(-t)}{\Gamma(-s-t)} {}_2F_1(-s,-t;\frac{1-s-t}{2};\lambda) $& $j(t)=t$\\[8pt]
        Mandelstam & $T_{c_0,0,\lambda} =\frac{\Gamma(-s)\Gamma(-t)}{\Gamma(-s-t)} {}_3F_2(-s,-t, -c_0; -\frac{s+t}{2},\frac{1-s-t}{2};\lambda) $&$j(t)=t$\\[6pt]
        Spin 0 exchange & $T_{\text{spin 0}} = \frac{m^2}{m^2-s} + \frac{m^2}{m^2-t}$&$j(t) =0$\\[6pt]
        $st$-pole & $T_{st-\text{pole}} = \frac{M^4}{(m^2-s)(m^2-t)}$&$j(t) =-1$\\[6pt]
        \bottomrule
    \end{tabular}
    }
    \caption{Here we list examples of unitary amplitudes with $T(s,t)= T(t,s)$ and no $u$-channel poles. More amplitudes with spin $1$ and spin $2$ exchanges can be found in \cite{Albert:2022oes,Fernandez:2022kzi}.}
    \label{tab:openAmp}
\end{table}

\begin{table}[h!]
    \centering
    \begin{tabular}{c|c|c|c|c}
        \toprule
        Name&Amplitude&Regge behavior\\ \midrule
        Virasoro-Shapiro & $f_{VS} = -\frac{\G(-s)\G(-t)\G(-u)}{\G(1+s)\G(1+t)\G(1+t)}$&$j(t) = 2+2t$ \\[6pt]
        Spin 0 exchange & $f_{\text{spin 0}} = \frac{\lambda^2}{m^6}\frac{1}{m^2-s}$&$j(t) =3$\\[6pt]
        $tu$-pole& $f_{tu-\text{pole}} = \frac{g}{m^4}\frac{1}{(m^2-t)(m^2-u)}$&$j(t) =3$\\[6pt]
        $stu$-pole& $f_{stu-\text{pole}} = \frac{\kappa}{m^2}\frac{1}{(m^2-s)(m^2-t)(m^2-u)}$&$j(t) =2$\\
        \bottomrule
    \end{tabular}
    \caption{Here we write examples of the MHV graviton amplitude using the functions $f(s|t,u)$. If not explicit, one needs to add the graviton pole to have a gravitational amplitude. We removed it here for brevity. }
    \label{tab:fgravExample}
\end{table}

\paragraph{A pole amplitude for the MHV gravitational amplitude}\mbox{}\\
Let us show here that an amplitude built as a sum of $1,2,3$ poles is consistent with unitarity (positivity).
We start with the following combination of terms
\begin{equation}
    f_{\text{poles}}(s|t,u) =  \frac{\lambda}{m^6}\frac{1}{m^2-s} + \frac{g}{m^4}\frac{1}{(m^2-t)(m^2-u)}+  \frac{\kappa}{m^2}\frac{1}{(m^2-s)(m^2-t)(m^2-u)}
\end{equation}
We will show below that this amplitude is unitary for
\begin{equation}\label{eq:fpoles_Coefsconstraint}
   \lambda \geq -\kappa\frac{2 \log (2)}{3}\,,\quad g \geq -\kappa \frac{12 (195790 \log (2)-135711)}{7096320 \log (2)-4918777}\,,\quad  \kappa\geq 0\,.
\end{equation}
At fixed $\kappa$, saturation of $\lambda$ removes spin $0$ in the $(++)$ channel and saturation of $g$ removes spin $5$ in the $(+-)$ channel. We can also immediately see that any single term is independently unitary for positive coefficients.

To show it, recall that we can invert \eqref{eq:imTppmmPW} and \eqref{eq:imTpmmpPW} 
\begin{align}
    \rho_J^{++}(s) &= a_J\,s^4 \int_{-1}^1 dx\, f_s(s|t(x),u(x)) P_J(x)\\
    \rho_J^{+-}(s) &= a_J\, s^4 \int_{-1}^1 dx\, (1+x)^4  f_s(u(x)|s,t(x)) d_{4,4}^{J}(x)\,,
\end{align}
where $a_J = \frac{1}{2}(2J+1)$.\footnote{Compared to \cite{Correia:2020xtr}, we absorbed the prefactor $n_J$ in $\rho_J$ in \eqref{eq:imTppmmPW}. Thus $a_J = n_J \cN_4/2$ in their convention. }
Plugging  $f_{\text{poles}}(s|t,u)$ into the formula above, we obtain for the $(++)$ channel 
\begin{align}
    \rho_J^{++}(s) &=  a_J \pi \delta(s-m^2) \int_{-1}^1 dx \left(\frac{4 \kappa}{9-x^2} + \lambda\right) P_J(x)\,.
\end{align}
Performing the integral for the spin $0$, we obtain the constraint on $\lambda \geq -\kappa\frac{2 \log (2)}{3}$. For higher spins, we use the Froissart-Gribov formula (see for example \cite{Gribov:2003nw,Correia:2020xtr}) to see that all partial waves are proportional to Legendre $Q$-function $\rho_J^{++}(s)\sim \kappa \delta(s-m^2) Q_J(3)$ and are nonnegative for $\kappa\geq 0$. This analysis was performed analogously in \cite{Caron-Huot:2020cmc} in the context of scalar amplitudes.

Let us now turn our attention to the other channel. We obtain
\begin{align}
    \rho_J^{+-}(s) &=  a_J \pi \delta(s-m^2) \int_{-1}^1 dx (1+z)^4 d_{4,4}^{J}(x) \left(\frac{4}{9-x^2}+\frac{2g}{3-x}\right) \\
    &=  a_J \pi \delta(s-m^2) \int_{[-1,1]} \frac{dz}{\pi i} (1+z)^4 e^J_{4,4}(z)\left(\frac{4}{9-z^2}+\frac{2g}{3-z}\right)\,.
\end{align}
In the second line, we wrote the integral as a  counterclockwise contour in the complex plane along the axis $z=[-1,1]$ using the Wigner $e$-function \cite{wignerE,Martin:1970hmp}.\footnote{See \cite{Haring:2022sdp} for a recent use of the Wigner $e$-functions in a different context.}  It is defined such that its discontinuity in $x\in [-1,1]$ is given by the Wigner $d$ matrices. They are analogous to the Legendre $Q$-functions but for spinning particles in $d=4$. Explicitly, we have
\begin{equation}
	\begin{aligned}
		e^{\, J}_{\lambda \mu} (z) &= \frac{(-1)^{\lambda - \mu}}{2} \left[\Gamma(J+\lambda +1)\Gamma(J-\lambda +1)\Gamma(J+\mu +1)\Gamma(J-\mu +1)\right]^{\frac{1}{2}} \left(\frac{1+z}{2}\right)^{\frac{\lambda+\mu}{2}}\\
		&\times\left(\frac{1-z}{2}\right)^{-\frac{\lambda-\mu}{2}}
		 \left(\frac{z-1}{2}\right)^{-J -\mu - 1} \frac{1}{\Gamma(2J +2)} {}_2 F_1\left(J+\lambda+1, J +\mu +1, 2 J +2, \frac{2}{1-z}\right)~, 
	\end{aligned}
\end{equation} 
for $\lambda + \mu \geq 0$ and $\lambda - \mu \geq 0$.\footnote{The other ranges are defined through the identities
\begin{equation}
	e^{\, J}_{\lambda \mu} (z) = (-1)^{\lambda-\mu}e^{\, J}_{\mu \lambda} (z) = (-1)^{\lambda-\mu} e^{\, J}_{-\lambda, -\mu} (z) 
\end{equation}}

We can now deform the contour and pick the pole at $z=\pm 3$ to obtain\footnote{Note that the Wigner $e$-function has extra singularities at $z=\pm 1$ but they are precisely canceled by the prefactor $(1+x)^4$.} 
\begin{equation}
\rho_J^{+-}(s) =  \delta(s-m^2) \frac{4 \pi a_J}{3}\left[ (1+3g)4^4e^J_{4,4}(3) - 2^4e^J_{4,4}(-3)\right]
\end{equation}
which is positive if\footnote{Here we used that $e_{\mu,\lambda}^J(-z) = (-1)^{1+J+\mu-2\lambda}e_{\mu,-\lambda}^J(z)$, see for example \cite{Martin:1970hmp}. }
\begin{equation}
    16 (1 + 3 g) e^J_{4,4}(3) + (-1)^Je^J_{4,-4}(3)\geq 0\,.
\end{equation}
Using positivity properties of the Wigner $e$-function $e^{\, J}_{4,\pm 4} (z)\geq 0$ for $z>1\,,J\geq 4$, the strongest constraint comes from odd spins. Finally, using that $\frac{e^J_{4,4}(3)}{e^J_{4,-4}(3)}$ is a growing function of $J$, the strongest constraint comes from $J=5$ which leads to the second constraint in \eqref{eq:fpoles_Coefsconstraint}
\begin{equation}
 g \geq -\kappa \frac{12 (195790 \log (2)-135711)}{7096320 \log (2)-4918777}\,.
\end{equation}
Saturation of this constraint removes the spin $5$ exchange in the $(+-)$ channel.

\paragraph{Comment on the Regge behavior}\mbox{}\\
Finally, let us comment on the Regge behavior of amplitudes presented in \tabref{tab:fgravExample}. Two of the functions for the MHV amplitude presented above grow too fast in the Regge limit, namely they have the Regge intercept $j_0=3$. For the $tu$-pole amplitude, this can be cured by considering the triple product with different mass $\frac{1}{(M^2-s)(m^2-t)(m^2-u)}$. Indeed, performing the same analysis as above shows that for $M\geq m$, this amplitude is unitary. Thus we can define the improved $tu$-pole amplitude by
\begin{equation}
    f_{tu-\text{pole}}^{\text{improved}} = \frac{g}{m^2}\frac{1}{(M^2-s)(m^2-t)(m^2-u)} \,,\quad M\gg m\,.
\end{equation}
As $M\gg m$, the corrections to the Wilson coefficients are suppressed by $O(m/M)$. This amplitude now has the Regge intercept $j_0=2$. Since this amplitude is unitary only for $M\geq m$, a similar improvement cannot be performed for the spin $0$ exchange amplitude.

It is sometimes possible to add a contact term to cure the Regge behavior (see for example \cite{Haring:2022sdp}, Appendix A). Here, for the spin $0$, one would need to add $+ \frac{\lambda^2}{M^6 s}$ which corresponds to the massless scalar exchange amplitude $|\beta_\phi|^2$, but taken with the wrong sign and is thus not unitary. We do not know how to `improve' the spin $0$ amplitude such that it satisfies unitarity and has $j_0\leq 2$. 
How is it possible then that this amplitude lies at the boundary of the allowed region in the dual approach? First, while we have not found it, it is possible that an improvement exists such that it does not change the value of the Wilson coefficient by adding a tower of particles of mass $M\gg m$. Second, in the dual formalism,  this amplitude satisfies all the sum rules written and hence is not excluded.

\section{Unitarity of open string finite sums}\label{sec:noFiniteSumAnsatz}
In this section, we extend the argument of Sivers and  Yellin \cite{Sivers:1971ig} to show that any finite ansatz \eqref{eq:sumVenAnsatz} cannot satisfy unitarity. The truncated ansatz reads 
\begin{equation}\label{eq:Topen_imax}
     T_{\Nmax}(s,t) = \sum_{i=0}^{\Nmax}\sum_{k=0}^{i}b_{ik}\frac{\Gamma(i-s)\Gamma(i-t)}{\Gamma(i+k-s-t)}\,,
\end{equation}
where by assumption $b_{ik}$ are finite real coefficients.

The residue of a single term in the ansatz is 
\begin{equation}
    \frac{\G(i-s)\G(i-t)}{\G(i+k-s-t)}\underset{s\to n}{\sim} -\frac{R_n^{(i,k)}(t)}{s-n}
\end{equation}
where
\begin{equation}\label{eq:ResForSinglenik}
    R_n^{(i,k)}(t) = \frac{(-1)^{i+n} \Gamma (i-t)}{(n-i)! \Gamma (i+k-n-t)}
\end{equation}
is a polynomial of degree $n-k$ in $t$.

As a first step, we show that a single term with $i\neq 0$ does not satisfy unitarity. To this end, we will show that in the partial wave expansion 
\begin{equation}\label{eq:Rnik_PWexpansion}
     R_n^{(i,k)}(t) = \sum_{J=0}^{n-k} c^{(i,k)}_{n,J} P_J\left(1+ \frac{2t}{n}\right)\,
\end{equation}
the $J=n-k$ and $J=n-k-1$ terms have opposite signs and hence unitarity cannot be satisfied for all $n,J$.  Using, \eqref{eq:ResForSinglenik}, we can expand in $x=1+\frac{2t}{n}$ and focus on the leading power
\begin{equation}\label{eq:Rnik_leadingx}
    (n-i)!R_n^{(i,k)}(t)\Big|_{t=\frac{n}{2}(x-1)} = (-1)^{i+k} \left(\frac{n}{2}\right)^{n-k}\left[x^{n-k} + \frac{n-k}{n}(1-k-2i) x^{n-k-1} + \dots \right]\,.
\end{equation}
Importantly, $(1-k-2i)<0$ for all $i\geq k>0$ and thus the coefficients in front of $x^{n-k}$ and $x^{n-k-1}$ have opposite signs. Moreover, we know that the Legendre polynomials $P_J(x)$ are expansion in odd/even powers of $x$ for odd/even $J$. This implies that $x^{n-k}$ and $x^{n-k-1}$ contributes to different spin. Moreover, using that the coefficient of $x^J$ in $P_J(x)$ is always positive, comparing \eqref{eq:Rnik_PWexpansion} and \eqref{eq:Rnik_leadingx} we can thus conclude that $ c^{(i,k)}_{n,n-k}$ and $ c^{(i,k)}_{n,n-k-1}$ have opposite signs. We can also write them explicitly
\begin{equation}\label{eq:cikn_explicit}
    \left\{ \begin{array}{cc}
        c^{(i,k)}_{n,n-k} &= (-1)^{i+k} \left(\frac{n}{4}\right)^{n-k}\frac{\sqrt{\pi } \Gamma (n-k+1)}{(n-i)! \Gamma \left(n-k+\frac{1}{2}\right)}   \\
        c^{(i,k)}_{n,n-k-1} &=  (-1)^{i+k+1}\left(\frac{n}{4}\right)^{n-k} \frac{2 \sqrt{\pi } (2 i+k-1)   \Gamma (n-k+1)}{n (n-i)! \Gamma \left(n-k-\frac{1}{2}\right)}
    \end{array}\right.\,,
\end{equation}
which makes it clear that they have opposite signs. This concludes the proof showing that a single term cannot satisfy unitarity as it requires $c^{(i,k)}_{n,J}\geq 0$.

Let us go back to a finite sum \eqref{eq:Topen_imax}, \eqref{eq:ResForSinglenik} implies that at large $n$
\begin{equation}
    n!\, R_n^{(i,k)}(t) \sim n^{i}\,t^{n-k}\,,
\end{equation}
and thus the terms with $\max(i)$ dominates at large enough $n$. Therefore, it is enough to consider an ansatz at fixed $i$ and show that it does not satisfy unitarity. We call the fixed $i$ amplitude
\begin{equation}
    T^{(i)} = \sum_{k=0}^i b_{ik}\frac{\G(i-s)\G(i-t)}{\G(i+k-s-t)}
\end{equation}
Considering the residue at $n$ of this amplitude
\begin{equation}\label{eq:residuesGeneral}
     R^{(i)}_n(t)=\sum_{J=0} c^{(i)}_{n,J} P_J\left(1+ \frac{2t}{n}\right)\,,
\end{equation} 
where unitarity for the fixed $i$ sum requires $c^{(i)}_{n,J}\geq 0$.
It is straightforward to see from \eqref{eq:ResForSinglenik} that
\begin{equation}
     c^{(i)}_{n,J} = \sum_{k=0}^{n-J} c_{n,J}^{(i,k)} b_{ik}
\end{equation}
and only $b_{i,0}$ contribute to the residue at $J=n$,  $(b_{i,0},b_{i,1})$ contribute to the residue at $J=n-1$ and so on. 

Consider first the residue at $J=n$, from \eqref{eq:cikn_explicit} it is clear that all $c_{n,n}^{(i,0)}$ have the same sign. This fixed the sign of $b_{i0}$ and we can also normalize it to $b_{i0}=\pm 1$. The other option is $b_{i0}=0$ and we will come back to this later.

We turn now to the residue at $J=n-1$
\begin{equation}
    c^{(i)}_{n,n-1} = c_{n,n-1}^{(i,0)} b_{i0} +  c_{n,n-1}^{(i,1)} b_{i1}
\end{equation}
Using \eqref{eq:cikn_explicit}, we have that at large $n$
\begin{equation}\label{eq:resRatioGrowthToshow}
    \left|\frac{c_{n,n-1}^{(i,0)}}{c_{n,n-1}^{(i,1)}}\right|\sim n
\end{equation}
and thus $c^{(i)}_{n,n-1}\geq 0$ implies 
\begin{equation}\label{eq:bi10largengrowth}
    \left|\frac{b_{i1}}{b_{i0}}\right|\gtrsim n
\end{equation}
For $b_{i0}=\pm 1$ this leads to a contradiction with having a regular finite ansatz. Indeed, \eqref{eq:bi10largengrowth}  must be true for all $n$, it implies $b_{i1}\to \infty$ which is not compatible with having a well-defined finite ansatz. 

We showed that $b_{i0}$ cannot be finite or the residue $c_{n,n-1}^{(i)}$ cannot be positive. The other option is $b_{i0}=0$. In such a case the residue at $J=n-1$ fixed the sign of $b_{i1}$, and we can choose normalization $b_{i1}=\pm 1$. Now looking at $J=n-2$, only $b_{i1}$ and $b_{i2}$ contributes. Repeating the argument above, it is easy to see that $b_{i2}\gtrsim\, n\, b_{i1}$ and hence the only option is $b_{i1}=0$. By iteration, we obtain that $b_{i\,k<i}=0$. 

What remains is the term with $k=i$. However, we showed above that a single term cannot satisfy unitarity. 
This concludes the proof that a finite sum ansatz cannot satisfy unitarity.


\section{Completeness argument for the open string case}
\label{app:completeness}

Here we would like to comment on the completeness of the ansatz \eqref{eq:sumVenAnsatz} following Khuri \cite{khuri1969derivation}. 
The fact that all particles live on equidistant linear Regge trajectories translates to the following statement
\be
T(s,t) \sim {1 \over \Gamma(1+t) \sin \pi t} \Big( a_0(t) (-s)^t + a_1(t) (-s)^{t-1} + ... \Big) + ... ,
\ee
where the last $...$ stands for the RSR violating contributions ${1 \over s^n}$ that vanish in the $s \to \infty$ limit. The first observation is that
$a_k(t)$ are entire functions. Moreover, if we consider the residue at $t=n$ it should become polynomial,
therefore
\be
a_k(n) = 0, ~~~ k>n.
\ee

We now consider the ansatz
\be
  \tilde T(s,t) &= \sum_{i=0}^{\infty}\sum_{k=0}^{i}b_{ik}\frac{\Gamma(i-s)\Gamma(i-t)}{\Gamma(i+k-s-t)} .
\ee
The basic idea is that by choosing $b_{ik}$ we can reproduce a given set of entire functions $a_k(t)$. For 
example, for the leading one we get the following equation
\be
\label{eq:leadingaexp}
a_0(t)  = \sum_{i=0}^\infty b_{i 0}  {\Gamma(i-t) \over \Gamma(-t)} ,
\ee
where in writing \eqref{eq:leadingaexp} we expanded each term under the series. Eq. \eqref{eq:leadingaexp} expresses the entire function
$a_0(t)$ in terms of Newton polynomials with interpolating points chosen to be nonnegative integer $t=n$.  

A sufficient condition for convergence for such an expansion was derived by Buck \cite{buck1948interpolation}. Let us introduce the growth indicator of an entire function $f(t)$
\be
h(\theta, f) = \lim_{r \to \infty} \sup {1 \over r} \log | f(r e^{i \theta}) | .
\ee
Then Buck has proven that the expansion \eqref{eq:leadingaexp} converges if 
\be
\label{eq:buckcondition}
h(\theta, f) < \cos \theta \log (2 \cos \theta) + \theta \sin \theta ,~~~ | \theta | < \pi /2 \ .
\ee
Similar conditions hold for subleading trajectories.

We expect that \eqref{eq:buckcondition} follows from consistency in the semiclassical limit $s,t \gg 1$ as discussed in \cite{Caron-Huot:2016icg}. Note that in all known examples $h(\theta,f) \leq 0$ for $|\theta| < \pi/2$ and therefore the bound \eqref{eq:buckcondition} is trivially satisfied. Assuming this is the case, we consider next the difference
\be
\delta T(s,t) = T(s,t) - \tilde T(s,t).
\ee
It is an entire function that vanishes at infinity. Therefore $\delta T(s,t) = 0$. We do not have an analogous argument for the closed string ansatz.

\section{Bound on the asymptotic form of the amplitude}
\label{app:asybound}

We consider a stringy amplitude at large $s,t \to \infty$. We focus on the discontinuity that takes the form
\be \label{eq:discT_polynomial}
T_s (s,t) = \sum_i \delta(s- m_i^2) \sum_J^{j(s)} c_{i,J} P_J\left(1+{2 t \over m_i^2}\right) \ . 
\ee
The RHS is a polynomial that is characterized by a set of zeros. If we now perform an average over many poles, for example, by considering $T(s(1+i \eps), t(1+i \eps))$, it was argued in \cite{Caron-Huot:2016icg} that it is these zeros, called the excess zeros, that control the amplitude asymptotically. 

It is convenient to introduce a distribution of zeros $\rho(z,\bar z)$ and write for the asymptotic form of the amplitude
\be
\log T(s,t) = c_0 t^k \int d^2 z \rho(z, \bar z) \log \left(1 - {s \over t z} \right) ,
\ee
where we assumed the asymptotic form of the Regge trajectory takes the form $j_{\text{asy}}(s) = c_0 s^k$. Our task is then
to find the distribution of zeros that arises from the sum over Legendre polynomials with positive coefficients,
such that
\be
\label{eq:normaliz}
\int d^2 z \rho(z, \bar z) = 1 , ~~~ \rho(z, \bar z) \geq 0 ,
\ee
which satisfies crossing that takes the following form
\be
\int d^2 z \rho(z, \bar z) \left( \beta^k \log \left(1-{1 \over \beta z}\right) + \log \left(1 - {\beta \over z}\right) \right) = 0 \ ,~~~ \beta >0, 
\ee
where we introduced $\beta = t/s$.

Moreover, the distribution of zeros should come from a positive sum of Legendre polynomials and should
correctly reproduce the Regge limit behavior. Introducing the `electric field' produced by the excess zeros
\be
f(\beta) \equiv \int d^2 z {\rho(z , \bar z) \over \beta - z} ,
\ee
one can show that
\be
0 \leq \sqrt{\beta(1+\beta)} f(\beta) &< 1 , \\
\partial_\beta \left( \sqrt{\beta(1+\beta)} f(\beta) \right) &\geq 0 \ . 
\ee
In addition to that, consistency with the Regge limit implies that
\be
f(\beta) = - k \log \beta \beta^{k-1} + (k+1) M_1 \beta^k+ ... ,
\ee
where $M_1 = - \int d^2 z \rho(z, \bar z) z$ is the dipole moment of the distribution. The asymptotic above is only 
consistent with the formulas above for $k>1/2$. 

\subsection{Support of the distribution of excess zeros}

To make further progress \cite{Caron-Huot:2016icg} had to make an assumption on the support of the distribution of zeros $\rho(z, \bar z)$ that arises from the 
sum over Legendre polynomials with positive coefficients. The first, rather weak, assumption that zeros are localized for ${\rm Re} z \leq 0$
leads to an additional constraint
\be
k \leq 1 .
\ee 
Making a stronger assumption that the zeros are located inside an ellipse that touches the real axis at ${\rm Re} z = 0,-1$, \cite{Caron-Huot:2016icg} then argued
that $k=1$ and that the amplitude is given by the asymptotic limit of the Veneziano amplitude.

\subsection{Extending the support of the distribution}

It is clear from the results of this paper that the assumption above about the effective support of zeros is too restrictive. Let us consider the generalized Veneziano amplitude $T_{c_0, c_1 , \lambda}$ with $c_1 \neq 0$. There is a nontrivial range of parameters for which it satisfies unitarity and crossing and takes the following
form in the asymptotic region $s,t \to \infty$
\be
\log T_{c_0, c_1 , \lambda} &= (s+t) \log (s+t) - s \log s - t \log t \nn \\
&+ c_1 \Big( t \log {1 \over 2} \Big( 1 - \tilde \lambda {s-t \over s+t} +\sqrt{1- \tilde \lambda} \sqrt{1 -  \tilde \lambda {(s-t)^2 \over (s+t)^2} } \Big) + \{ s \leftrightarrow t \} \Big) , \nn
\ee
where the second line vanishes for $\tilde \lambda \equiv 4 \lambda (1-\lambda) = 0$. 

In this case the amplitude can be written as follows
\be
\log T_{c_0, c_1 , \lambda} = t \int_{{\sqrt{\tilde \lambda}+1  \over \sqrt{ \tilde \lambda}-1}}^0 d x \rho(x) \log \left(1 - {s \over t x} \right) .
\ee
In particular, we have
\be
\lim_{\tilde \lambda \to 1} {\sqrt{\tilde \lambda}  + 1 \over \sqrt{ \tilde \lambda}-1} \to \infty . 
\ee
Therefore, we see that positive sums over Legendre polynomials consistent with the Regge limit can generate distributions of zeros that `spills' arbitrarily far beyond $-1\leq x \leq 0$ considered in \cite{Caron-Huot:2016icg}. 

\subsection{Maximal value of the amplitude}

To make some progress it is interesting to consider a dual formulation of \cite{Caron-Huot:2016icg}. We consider the case of a linear Regge trajectory $j(t) =\alpha_{\text{asy}}' t$, $k=1$, and we ask the following question: what is the maximal value that the amplitude can attain at the crossing-symmetric point $\log T(s,s) \leq \alpha_{{\rm max}}s$?

We assume that all zeros are localized along the negative real axis parameterized by $z \leq 0$ and we set $s=1$ so that everything only depends on $\beta = {t \over s}$. Let us quickly demonstrate that such a bound exists. To do it we introduce a set of `null constraints' by expanding the crossing equation around $\beta=1$
\be
\beta \log \left(1-{1 \over \beta z}\right) + \log \left(1 - {\beta \over z}\right) \Big|_{\beta = 1 - \eps} = \sum_{i=1}^\infty n_i(z) \eps^i ,
\ee
such that
\be
\int_{- \infty}^0 d z \rho(z) n_i(z) = 0 . 
\ee
One can check that not all of the null constraints are linearly independent. We find that a convenient choice is to consider $(n_1, n_2, n_{3}, n_5, n_7, ...)$. As an example
\be
n_1(z) &= \log\left(1-{1 \over z}\right) - {2 \over 1-z} \ , \\
n_2(z) &= {1 \over 6} {1+3 z \over (z-1)^3} \ . 
\ee
To derive a bound on the amplitude
\be
\log T(1, 1) \leq \alpha ,
\ee
we look for a functional, or, in other words, an $\alpha$ and a set of $d_i$'s, such that
\be
\label{eq:functional}
1 - {1 \over \alpha} \log \left(1-{1 \over z}\right) + \sum_{i=1}^\infty d_i n_i(z) \geq 0 , ~~~~ z \leq 0 .
\ee
Indeed, imagine that we have found a functional with this property. We can then integrate the equation above against the density of zeros to get
\be
\label{eq:funcasympt}
\int_{- \infty}^0 d z \ \rho(z) \Bigg( 1 - {1 \over \alpha}  \log \left(1-{1 \over z}\right) + \sum_{i=1}^\infty d_i n_i(z) \Bigg) = 1 - {1 \over \alpha} \log T(1,1) \geq 0 ,
\ee
where we used the fact that $\rho(z) \geq 0$, the normalization condition \eqref{eq:normaliz}, and, of course, \eqref{eq:functional}. 

It is not immediately obvious that functionals with the property \eqref{eq:functional} exist, so let us demonstrate it explicitly. We take $\alpha=2$ and $d_1={1 \over 2}$
to get
\be
1 -  {1 \over 2}  \log \left(1-{1 \over z}\right)  + {1 \over 2} \left(  \log\left(1-{1 \over z}\right) - {2 \over 1-z}  \right) =  {- z \over 1 - z} \geq 0 , ~~~ z \leq 0 ,
\ee
which immediately tells us that 
\be
\log T(s,s) \leq 2 s . 
\ee

\subsection{Extremal functional and extremality of the Veneziano amplitude}

\begin{figure}[h!]
    \centering
    \includegraphics[scale=0.7]{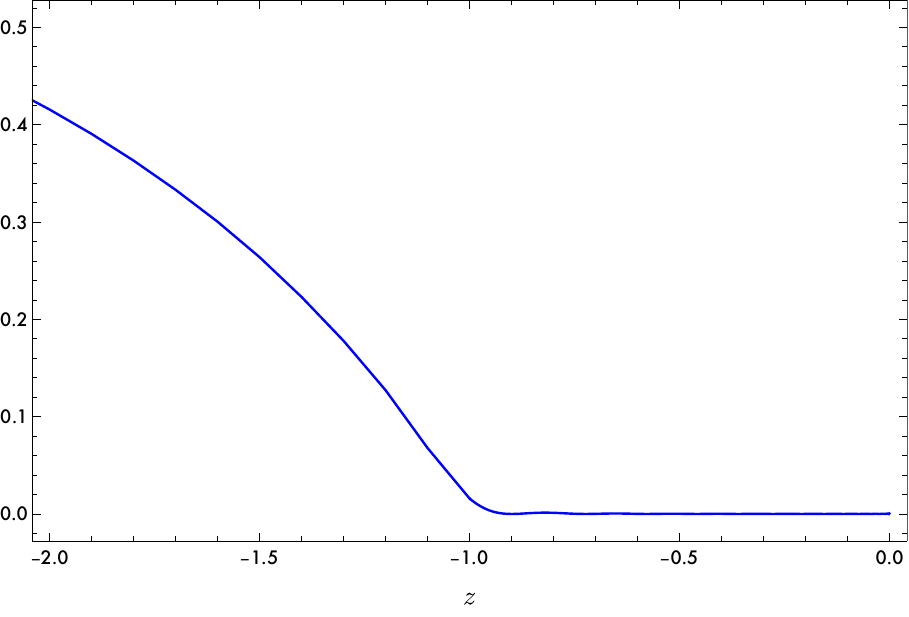}
    \caption{We plot the optimal functional \eqref{eq:funcasympt} obtained using $n_1, ... , n_{83}$ null constraints. It produces $\alpha=0.721138$ which corresponds to the upper bound $\log T(s,s) \leq 1.38671 s$.}
    \label{fig:plotfunc}
\end{figure}

A simple bound above was derived using a single null constraint. We can set a numerical scheme that employs more and more null constraints. 
As a result, we get an extremal functional that tends to zero for $-1 \leq z \leq 0$ and is positive otherwise. We plot the result for the functional obtained using null constraints up to $n_{83}$ in \figref{fig:plotfunc}, which produces the bound $\log T(s,s) \leq 1.38671 s$, whereas the Veneziano amplitude at this point takes the value $2 \log 2 \ s \approx 1.38629 s$.

Therefore, we see that the extremal amplitude that saturates the bound should have the support of excess zeros $\rho(z)$ only for $-1 \leq z \leq 0$. These are precisely the types of amplitudes considered in \cite{Caron-Huot:2016icg}. It was argued there that such amplitudes are unique for any $k$, however only for $k=1$ it can come from the positive sum of Legendre polynomials. 

Our conclusion here instead is that the asymptotic limit of the Veneziano amplitude maximizes the bound on the amplitude in the region $s,t \to \infty$, and in this sense, the Veneziano amplitude is an extremal (but not unique) solution to the axioms considered in \cite{Caron-Huot:2016icg}, so that in the limit $s,t \gg 1$ we have
\be
\label{eq:asyboundAPP}
\log T(s,t) \leq \alpha_{\text{asy}}' \Big( (s+t) \log (s+t) - s \log s - t \log t \Big)
\ee
The question of uniqueness and which extra conditions are needed to obtain it, e.g., the extra asymptotic crossing condition considered in \cite{Sever:2017ylk}, requires further investigation.

\subsection{Distribution of excess zeros}

Here we present some results on the distribution of the excess zeros of the open string amplitudes $T_{c_0,c_1,\lambda}(s,t)$. For our purposes,
the excess zeros $z_i(t)$ are defined as follows. We consider the residue of the amplitude
\be
\label{eq:excess}
-{\rm Res}_{t = n} T_{c_0,c_1,\lambda}(s,t) = \sum_{J=0}^{j(n)} c_{n,J} P_J\left(1+{2 s \over n}\right) \propto \prod_{i=1}^{j(n)}\left(1-{s \over z_i(n)}\right).
\ee
By taking the logarithm of this formula, we can rewrite it as follows
\be
\log \prod_{i=1}^{j(t)}(s-z_i(t)) = \int d^2z \rho(t, z, \bar z) \left( \log (z - s) - \log z \right) . 
\ee
We then take the large $s,t \gg 1$ limit of this formula. We define the asymptotic distribution
\be
\label{eq:rescalezeros}
\rho(t,z, \bar z) = {j_{\text{asy}}(t) \over t^2} \rho_{\text{asy}}\left({z \over t}, {\bar z \over t}\right) + ... \ ,
\ee
where $...$ includes contributions that do not contribute to the limit, e.g., some of the zeros could escape to infinity. Plugging this into the formula above and rescaling the integration variables, we get the following representation for the asymptotic amplitude
\be
\log T(s,t) \simeq j_{\text{asy}}(t) \int d^2 z \rho_{\text{asy}}(z, \bar z) \log \left( 1 - {1 \over \beta z} \right) , ~~~ \beta = {t \over s}. 
\ee
By taking the asymptotic Regge limit $s \gg t$ we get that
\be
\int d^2 z \rho_{\text{asy}}(z, \bar z) = 1.
\ee
We are interested in $j_{\text{asy}}(t)  \sim t$ therefore the crossing equation $\log T(s,t) = \log T(t,s)$ becomes
\be
\beta \int d^2 z \rho_{\text{asy}}(z, \bar z) \log \left( 1 - {1 \over \beta z} \right) = \int d^2 z \rho_{\text{asy}}(z, \bar z) \log \left( 1 - {\beta \over z} \right) .
\ee
We can therefore write the following representation of the amplitude 
\be
\log T(1,\beta)=\int d^2 z \rho_{\text{asy}}(z, \bar z) \log \left( 1 - {\beta \over z} \right).
\ee
By taking the derivative with respect to $\beta$, we get the relationship between the asymptotic distribution of zeros $\rho_{\text{asy}}(z, \bar z)$ and the discontinuity of $\partial_\beta \log T(1,\beta)$ 
\be
\label{eq:contourint}
\partial_\beta \log T(1,\beta)=\int d^2 z {\rho_{\text{asy}}(z, \bar z) \over \beta - z} \ . 
\ee

\begin{figure}
    \centering
    \includegraphics[scale=1.1]{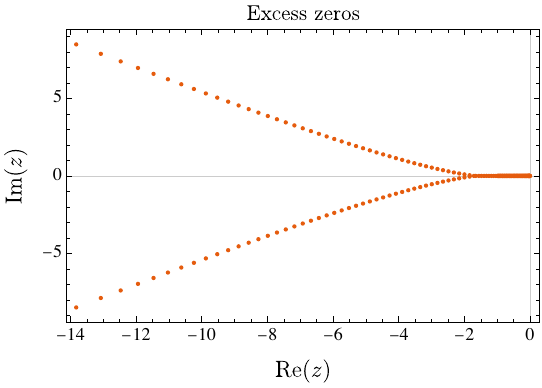}
    \caption{We plot the distribution of excess zeros, see \eqref{eq:excess}, for $n=200$, $c_0 = -3/10$, $c_1 = 7/20$, $\lambda = 1/2$. We rescaled them by $1 \over n$ as in \eqref{eq:rescalezeros}. As we increase the energy or $n$, they reach out further and further into the complex plane.}
    \label{fig:excesszeros}
\end{figure}

Let us now consider the distribution of zeros in a concrete example. We take $n=200$, $c_0 = -3/10$, $c_1 = 7/20$, $\lambda = 1/2$. One can numerically check that at this point the logarithm of the amplitude is  well captured by the asymptotic formula. The distribution of zeros rescaled by $200$ is shown in Figure \ref{fig:excesszeros}.

\begin{figure}[h]
    \centering
    \includegraphics[scale=0.5]{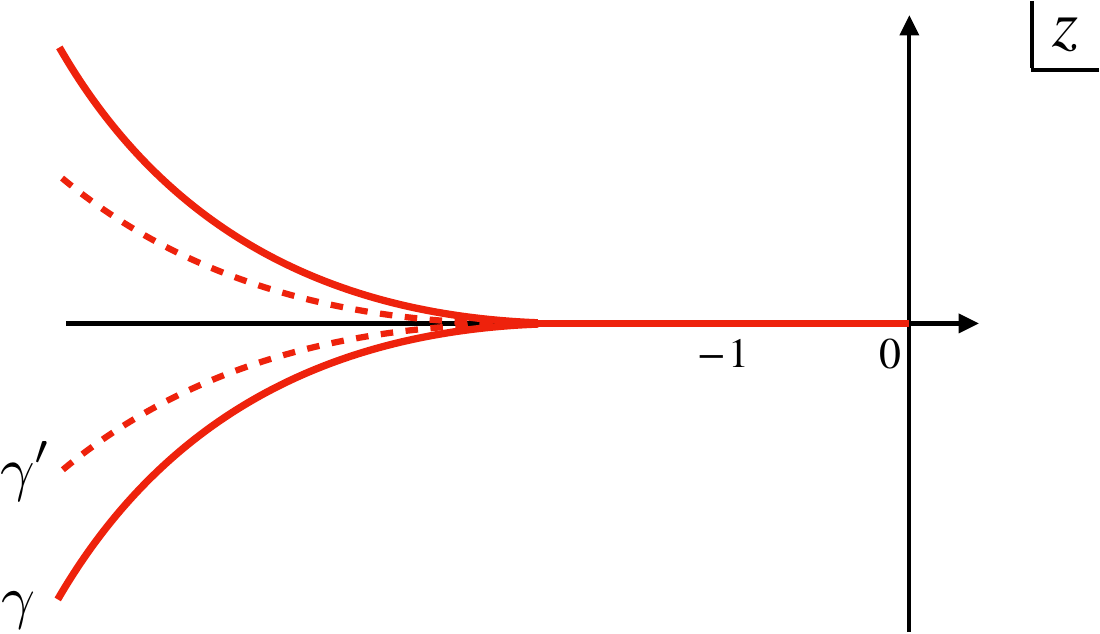}
    \caption{Nonuniqueness of the asymptotic distribution of zeros $\rho_{\text{asy}}(z)$. Given a nonnegative analytic distribution of zeros in \eqref{eq:contourint} along the contour $\gamma$ we can use the Cauchy theorem to deform the integral to the location $\gamma'$ (which in particular we can choose to be along the negative axis).}
    \label{fig:excesszeroscontour}
\end{figure}

We see that it has an interesting shape that branches into the complex plane. Moreover, by increasing $n$ we see that the rescaled zeros go further and further in the complex plane. We are thus led to the following picture of the asymptotic distribution
\be
\partial_\beta \log T(1,\beta)={1 \over 2 \pi i}\oint_\gamma dz {\rho_{\text{asy}}(z) \over \beta - z},
\ee
where the contour $\gamma$ is shown in Figure \ref{fig:excesszeroscontour}. 

\begin{figure}[h]
    \centering
    \includegraphics[scale=1.2]{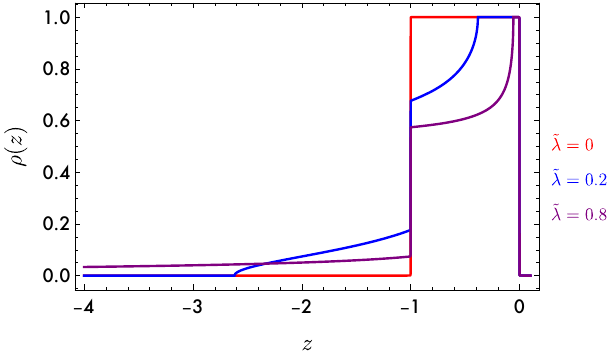}
    \caption{Here we plot the asymptotic density of zeros for the amplitude with $c_1=1/2$ and different $\tilde \lambda$. The homogeneous distribution between $[-1,0]$ that corresponds to $\tilde \lambda =0$ captures the high-energy limit of the Veneziano amplitude. We see that for $\tilde \lambda >0$ the density of zeros has a larger support. Finally, for $\tilde \lambda=1$ some of the excess zeros escape to infinity.}
    \label{fig:distrzeros}
\end{figure}

We can now use our asymptotic result \eqref{eq:fixedangleOS} to find the explicit form of $\rho_{\text{asy}}(z)$ in this case. We get the following result
\be
\partial_\beta \log T(1,\beta)= \log {1+\beta \over \beta} + c_1 \log { 1-\tilde \lambda +(1+\tilde \lambda) \beta + (1 - \tilde \lambda) \sqrt{\left(\beta - {\sqrt{\tilde \lambda}+1 \over \sqrt{\tilde \lambda} - 1}\right)\left(\beta - {\sqrt{\tilde \lambda}-1 \over \sqrt{\tilde \lambda} + 1}\right)}\over 2(1+\beta)} .
\ee
By taking the discontinuity, we can write the following representation for this amplitude 
\be
\label{eq:dbetaT}
\partial_\beta \log T(1,\beta)=\int_{-1}^0 dz {1 \over \beta - z}+ c_1  \int_{{\sqrt{\tilde \lambda}+ 1\over \sqrt{\tilde \lambda} - 1}}^{{\sqrt{\tilde \lambda}- 1 \over \sqrt{\tilde \lambda} + 1}} dz {\rho_{\text{asy}}(\tilde \lambda, z) \over \beta - z} ,
\ee
where $\rho_{\text{asy}}(z)$ can be readily computed by taking the discontinuity of \eqref{eq:dbetaT}. It has the following properties
\be
\label{eq:asymptoticnegdistr}
\int_{{\sqrt{\tilde \lambda}+ 1\over \sqrt{\tilde \lambda} - 1}}^{{\sqrt{\tilde \lambda}- 1 \over \sqrt{\tilde \lambda} + 1}} dz \rho_{\text{asy}}(z) &=0, ~~~~ 1>\tilde \lambda \geq 0, \\
\int_{{\sqrt{\tilde \lambda}+ 1\over \sqrt{\tilde \lambda} - 1}}^{{\sqrt{\tilde \lambda}- 1 \over \sqrt{\tilde \lambda} + 1}} dz \rho_{\text{asy}}(z) &=-1, ~~~~ \tilde \lambda = 1 \ . 
\ee
It is also bounded from below $\rho_{\text{asy}}(z) \geq -1$ for $-1 \leq z \leq 0$, and it is nonnegative $\rho_{\text{asy}}(z) \geq 0$ for $z<-1$. Therefore we see that as we turn on $\tilde \lambda$ zeros `spill' outside the $-1 \leq z \leq 0$ region. Moreover, as we set $\tilde \lambda = 1$ they escape to infinity. We plot examples of distributions of zeros in Figure \ref{fig:distrzeros}.

The reader might wonder how the distribution that we explicitly got in Figure \ref{fig:excesszeros} and the asymptotic distribution \eqref{eq:asymptoticnegdistr} are consistent with each other. In fact, the two representations can be deformed into one another using the Cauchy theorem, see Figure \ref{fig:excesszeroscontour}. Therefore we see that there is no unique way to read off the distribution of zeros starting from the known form of the amplitude for $\beta > 0$. In \eqref{eq:asymptoticnegdistr} we chose to deform the contour all the way to lie across the negative axis.

Imagine now we start with a positive $\rho(z) \geq 0$ analytic density of zeros along some contour $\gamma$ in the complex plan and we deform it to the negative axis. It is not clear a priori that after the deformation the new effective density of zeros has to be positive. We, however, observed this to be the case in the example above.

\section{Stringy Cerulus-Martin bound}
\label{app:CMstringy}

The Cerulus-Martin bound on the high-energy scattering at fixed angles \cite{Cerulus:1964cjb,Tourkine:2023xtu,Buoninfante:2023dyd} effectively expresses Mandelstam analyticity in the following form
\be
{\rm Regge} \leq ({\rm Fixed \ real \ angle}) \times ({\rm Fixed \  complex \ angle}) .
\ee
It is usually presented as a lower bound on scattering as follows
\be
({\rm Fixed \ real \ angle}) \geq {{\rm Regge} \over ({\rm Fixed \  complex \ angle})} \ . 
\ee
In nonperturbative QFT, for example in QCD, we do not have a bound on scattering at complex angles, 
therefore it is not a rigorous lower bound in this case. 

We would like next derive a lower bound on the scattering at physical fixed scattering angle for stringy amplitudes. Our input will be the following: an upper bound on the amplitude discussed in the section above, and the polynomial nature of the Regge limit.

We consider fixed-angle scattering so that
\be
t = - {s \over 2} (1-z) ,
\ee
and we would like to derive a lower bound of the following type
\be
\max_{|z| \leq z_0} |T(s,z)| \geq T_0(s,z) . 
\ee
We first start with the following simple observation 
\be
\max_{|z| \leq z_0} |T(s,z)| \geq \max_{|z| \leq z_0} |{\rm Im}T(s,z)| = \max_{|z| \leq z_0}|T_s(s,z)| .
\ee
Next, we notice that the discontinuity of the amplitude is simply a polynomial (as described above in \eqref{eq:discT_polynomial})
\be
\label{eq:impartapparg}
T_s(s,z) \sim \sum_{J=0}^{j(s)} c_{i,J} P_J(z) , ~~~ c_{i,J} \geq 0 \ , 
\ee
and thus, it is an analytic function in the $z$-plane. To derive a lower bound, we consider the following mapping
\be
w(z) = {z + \sqrt{z^2-z_0^2} \over z_0} .
\ee
Under this mapping, the real-line segment $- z_0 \leq z \leq z_0$ is mapped into a unit circle in the $w$-plane. 

We now consider three circles in the $w$-plane: $|w|=1$, $|w|=r_2$ and $|w|=r_3$, such that $r_3>r_2>1$. The discontinuity of the amplitude is an analytic function in the annulus $1 \leq |w|\leq r_3$.
We also introduce the following notation
\be
M_r = \max_{|w|=r} |T_s(s,z)| .
\ee
We then have the three-circle theorem that states the following. For a function analytic inside the annulus and bounded on its boundary, we have
\be
M_{r_2} \leq M_1^{1-{\log r_2 \over \log r_3}} M_{r_3}^{{\log r_2 \over \log r_3}} \ . 
\ee
We choose $r_2$ at fixed $t$ such that the circle includes the Regge limit of the amplitude, which is $\sim 1$, where the equivalence relation means
`modulo powers'. We then have
\be
1 \leq M_1^{1-{\log w(1) \over \log r_3}} M_{r_3}^{{\log w(1) \over \log r_3}} , 
\ee
where we set $r_2 = w(1)$ which is its leading large $s$ behavior. 

We next rewrite this bound as follows
\be
M_1 \geq (M_{r_3})^{-{\log w(1) \over \log r_3}/(1-{\log w(1) \over \log r_3})} ,
\ee
where the LHS is related to fixed-angle scattering for physical angles, whereas the RHS is related to scattering at complex angles.

We next notice the following simple fact
\be
\Big| \sum_J^{j(s)} c_{i,J} P_J(z) \Big| \leq  \sum_J^{j(s)} c_{i,J} \Big| P_J(z) \Big| \leq  \sum_J^{j(s)} c_{i,J} P_J(\sqrt{|z|^2 + 1}) , ~~~ |z| \geq 1 ,
\ee
where we used that $c_{i,J}  \geq 0$. We then get 
\be
M_{r_3} \leq T_s (s, \sqrt{z_*^2 + 1}) ,
\ee
where
\be
z_* \equiv \max_{|w|=r_3} | z(w) |  = z_0 {1+r_3^2 \over 2 r_3} . 
\ee
Next, we use the bound discussed in the previous section \eqref{eq:asyboundAPP} to get
\be
T_s (s, z)  &\lesssim e^{\alpha_{\text{asy}'}s f_{\text{Ven}}(z)}, \nn \\
f_{\text{Ven}}(z) &\equiv {z+1 \over 2} \log {z+1 \over 2} - {z-1 \over 2} \log {z-1 \over 2} ,
\ee
where we are working modulo power-like corrections.

Combining all the inequalities above, we get that for any $r_3$
\be
\max_{|z| \leq z_0} |T(s,z)| \geq (e^{ \alpha_{\text{asy}'} s f_{\text{Ven}}(\sqrt{z_*^2 + 1})})^{-{\log w(1) \over \log r_3}/(1-{\log w(1) \over \log r_3})} . 
\ee
To optimize the bound, we would like to maximize the RHS. We find that the maximum is attained at $r_3=\infty$ which finally gives
\be
\max_{|z| \leq z_0} |T(s,z)| \gtrsim e^{- \alpha_{\text{asy}'} s \log {1 + \sqrt{1-z_0^2} \over z_0}} \ . 
\ee
This constitutes a stringy generalization of the Cerulus-Martin lower bound on scattering at fixed angles. In the main text, we discussed the relationship between the asymptotic Regge trajectory $j_{\text{asy}} \simeq \alpha_{\text{asy}} t$, and the leading Regge trajectory $j(t) \simeq \alpha' t$, and argued that $\alpha'_{\text{asy}} \leq \alpha'$. We can, therefore, write the bound in terms of the leading Regge trajectory
\be
\label{eq:boundappCM}
\max_{|z| \leq z_0} |T(s,z)| \gtrsim e^{- \alpha' s \log {1 + \sqrt{1-z_0^2} \over z_0}} \ ,
\ee
which is the bound we quoted in the main text.

Let us comment on the following technical subtlety in the argument above. Strictly speaking, the imaginary part we considered above in \eqref{eq:impartapparg} is $\propto \delta(s-m^2)$, and therefore our bound above directly applies to the residues of the amplitude only. The standard way to cure this problem (as well as to use the bound \eqref{eq:boundappCM}) is to take the high-energy limit away from the real axis $s \to s(1+i \epsilon)$. We expect that in this limit, the imaginary part of the amplitude is still effectively given by \eqref{eq:impartapparg}, see \cite{Caron-Huot:2016icg} for the discussion of this point, and therefore our arguments apply.

\newpage
\bibliographystyle{JHEP}
\bibliography{refs}

\providecommand{\href}[2]{#2}\begingroup\raggedright\begin{thebibliography}{10}

\bibitem{Eden:1971fm}
R.~J. Eden, \emph{{Theorems on high energy collisions of elementary
  particles}}, \href{http://dx.doi.org/10.1103/RevModPhys.43.15}{\emph{Rev.
  Mod. Phys.} {\bf 43} (1971) 15--35}.

\bibitem{Jaffe:1967nb}
A.~M. Jaffe, \emph{{High-energy behavior in quantum field theory. I. Strictly
  localizable fields}},
  \href{http://dx.doi.org/10.1103/PhysRev.158.1454}{\emph{Phys. Rev.} {\bf 158}
  (1967) 1454--1461}.

\bibitem{Polchinski:2001tt}
J.~Polchinski and M.~J. Strassler, \emph{{Hard scattering and gauge / string
  duality}}, \href{http://dx.doi.org/10.1103/PhysRevLett.88.031601}{\emph{Phys.
  Rev. Lett.} {\bf 88} (2002) 031601},
  [\href{https://arxiv.org/abs/hep-th/0109174}{{\tt hep-th/0109174}}].

\bibitem{Gross:1987kza}
D.~J. Gross and P.~F. Mende, \emph{{The High-Energy Behavior of String
  Scattering Amplitudes}},
  \href{http://dx.doi.org/10.1016/0370-2693(87)90355-8}{\emph{Phys. Lett. B}
  {\bf 197} (1987) 129--134}.

\bibitem{Gross:1989ge}
D.~J. Gross and J.~L. Manes, \emph{{The High-energy Behavior of Open String
  Scattering}},
  \href{http://dx.doi.org/10.1016/0550-3213(89)90435-5}{\emph{Nucl. Phys. B}
  {\bf 326} (1989) 73--107}.

\bibitem{Caron-Huot:2016icg}
S.~Caron-Huot, Z.~Komargodski, A.~Sever and A.~Zhiboedov, \emph{{Strings from
  Massive Higher Spins: The Asymptotic Uniqueness of the Veneziano Amplitude}},
  \href{http://dx.doi.org/10.1007/JHEP10(2017)026}{\emph{JHEP} {\bf 10} (2017)
  026}, [\href{https://arxiv.org/abs/1607.04253}{{\tt 1607.04253}}].

\bibitem{Amati:1987uf}
D.~Amati, M.~Ciafaloni and G.~Veneziano, \emph{{Classical and Quantum Gravity
  Effects from Planckian Energy Superstring Collisions}},
  \href{http://dx.doi.org/10.1142/S0217751X88000710}{\emph{Int. J. Mod. Phys.
  A} {\bf 3} (1988) 1615--1661}.

\bibitem{Arkani-Hamed:2007ryv}
N.~Arkani-Hamed, S.~Dubovsky, A.~Nicolis, E.~Trincherini and G.~Villadoro,
  \emph{{A Measure of de Sitter entropy and eternal inflation}},
  \href{http://dx.doi.org/10.1088/1126-6708/2007/05/055}{\emph{JHEP} {\bf 05}
  (2007) 055}, [\href{https://arxiv.org/abs/0704.1814}{{\tt 0704.1814}}].

\bibitem{Giddings:2009gj}
S.~B. Giddings and R.~A. Porto, \emph{{The Gravitational S-matrix}},
  \href{http://dx.doi.org/10.1103/PhysRevD.81.025002}{\emph{Phys. Rev. D} {\bf
  81} (2010) 025002}, [\href{https://arxiv.org/abs/0908.0004}{{\tt
  0908.0004}}].

\bibitem{Bah:2022uyz}
I.~Bah, Y.~Chen and J.~Maldacena, \emph{{Estimating global charge violating
  amplitudes from wormholes}},
  \href{http://dx.doi.org/10.1007/JHEP04(2023)061}{\emph{JHEP} {\bf 04} (2023)
  061}, [\href{https://arxiv.org/abs/2212.08668}{{\tt 2212.08668}}].

\bibitem{Chen:2022nym}
H.~Chen, A.~L. Fitzpatrick and D.~Karateev, \emph{{Nonperturbative bounds on
  scattering of massive scalar particles in d \ensuremath{\geq} 2}},
  \href{http://dx.doi.org/10.1007/JHEP12(2022)092}{\emph{JHEP} {\bf 12} (2022)
  092}, [\href{https://arxiv.org/abs/2207.12448}{{\tt 2207.12448}}].

\bibitem{EliasMiro:2022xaa}
J.~Elias~Miro, A.~Guerrieri and M.~A. Gumus, \emph{{Bridging positivity and
  S-matrix bootstrap bounds}},
  \href{http://dx.doi.org/10.1007/JHEP05(2023)001}{\emph{JHEP} {\bf 05} (2023)
  001}, [\href{https://arxiv.org/abs/2210.01502}{{\tt 2210.01502}}].

\bibitem{Cheung:2023adk}
C.~Cheung and G.~N. Remmen, \emph{{Stringy Dynamics from an Amplitudes
  Bootstrap}},  \href{https://arxiv.org/abs/2302.12263}{{\tt 2302.12263}}.

\bibitem{Veneziano:1968yb}
G.~Veneziano, \emph{{Construction of a crossing - symmetric, Regge behaved
  amplitude for linearly rising trajectories}},
  \href{http://dx.doi.org/10.1007/BF02824451}{\emph{Nuovo Cim. A} {\bf 57}
  (1968) 190--197}.

\bibitem{Coon:1969yw}
D.~D. Coon, \emph{{Uniqueness of the veneziano representation}},
  \href{http://dx.doi.org/10.1016/0370-2693(69)90106-3}{\emph{Phys. Lett. B}
  {\bf 29} (1969) 669--672}.

\bibitem{Baker:1970vxk}
M.~Baker and D.~D. Coon, \emph{{Dual resonance theory with nonlinear
  trajectories}}, \href{http://dx.doi.org/10.1103/PhysRevD.2.2349}{\emph{Phys.
  Rev. D} {\bf 2} (1970) 2349--2358}.

\bibitem{Coon:1972qz}
D.~D. Coon, U.~P. Sukhatme and J.~Tran Thanh~Van, \emph{{Duality and
  proton-proton scattering at all angles}},
  \href{http://dx.doi.org/10.1016/0370-2693(73)90205-0}{\emph{Phys. Lett. B}
  {\bf 45} (1973) 287--291}.

\bibitem{Adams:2006sv}
A.~Adams, N.~Arkani-Hamed, S.~Dubovsky, A.~Nicolis and R.~Rattazzi,
  \emph{{Causality, analyticity and an IR obstruction to UV completion}},
  \href{http://dx.doi.org/10.1088/1126-6708/2006/10/014}{\emph{JHEP} {\bf 10}
  (2006) 014}, [\href{https://arxiv.org/abs/hep-th/0602178}{{\tt
  hep-th/0602178}}].

\bibitem{Lucini:2001ej}
B.~Lucini and M.~Teper, \emph{{SU(N) gauge theories in four-dimensions:
  Exploring the approach to N = infinity}},
  \href{http://dx.doi.org/10.1088/1126-6708/2001/06/050}{\emph{JHEP} {\bf 06}
  (2001) 050}, [\href{https://arxiv.org/abs/hep-lat/0103027}{{\tt
  hep-lat/0103027}}].

\bibitem{Veneziano:2017cks}
G.~Veneziano, S.~Yankielowicz and E.~Onofri, \emph{{A model for pion-pion
  scattering in large-N QCD}},
  \href{http://dx.doi.org/10.1007/JHEP04(2017)151}{\emph{JHEP} {\bf 04} (2017)
  151}, [\href{https://arxiv.org/abs/1701.06315}{{\tt 1701.06315}}].

\bibitem{Cheung:2023uwn}
C.~Cheung and G.~N. Remmen, \emph{{Bespoke dual resonance}},
  \href{http://dx.doi.org/10.1103/PhysRevD.108.086009}{\emph{Phys. Rev. D} {\bf
  108} (2023) 086009}, [\href{https://arxiv.org/abs/2308.03833}{{\tt
  2308.03833}}].

\bibitem{Arkani-Hamed:2020blm}
N.~Arkani-Hamed, T.-C. Huang and Y.-t. Huang, \emph{{The EFT-Hedron}},
  \href{http://dx.doi.org/10.1007/JHEP05(2021)259}{\emph{JHEP} {\bf 05} (2021)
  259}, [\href{https://arxiv.org/abs/2012.15849}{{\tt 2012.15849}}].

\bibitem{Bern:2021ppb}
Z.~Bern, D.~Kosmopoulos and A.~Zhiboedov, \emph{{Gravitational effective field
  theory islands, low-spin dominance, and the four-graviton amplitude}},
  \href{http://dx.doi.org/10.1088/1751-8121/ac0e51}{\emph{J. Phys. A} {\bf 54}
  (2021) 344002}, [\href{https://arxiv.org/abs/2103.12728}{{\tt 2103.12728}}].

\bibitem{Caron-Huot:2022ugt}
S.~Caron-Huot, Y.-Z. Li, J.~Parra-Martinez and D.~Simmons-Duffin,
  \emph{{Causality constraints on corrections to Einstein gravity}},
  \href{http://dx.doi.org/10.1007/JHEP05(2023)122}{\emph{JHEP} {\bf 05} (2023)
  122}, [\href{https://arxiv.org/abs/2201.06602}{{\tt 2201.06602}}].

\bibitem{Chiang:2022jep}
L.-Y. Chiang, Y.-t. Huang, W.~Li, L.~Rodina and H.-C. Weng,
  \emph{{(Non)-projective bounds on gravitational EFT}},
  \href{https://arxiv.org/abs/2201.07177}{{\tt 2201.07177}}.

\bibitem{Albert:2023jtd}
J.~Albert and L.~Rastelli, \emph{{Bootstrapping Pions at Large $N$. Part II:
  Background Gauge Fields and the Chiral Anomaly}},
  \href{https://arxiv.org/abs/2307.01246}{{\tt 2307.01246}}.

\bibitem{Albert:2022oes}
J.~Albert and L.~Rastelli, \emph{{Bootstrapping pions at large N}},
  \href{http://dx.doi.org/10.1007/JHEP08(2022)151}{\emph{JHEP} {\bf 08} (2022)
  151}, [\href{https://arxiv.org/abs/2203.11950}{{\tt 2203.11950}}].

\bibitem{Fernandez:2022kzi}
C.~Fernandez, A.~Pomarol, F.~Riva and F.~Sciotti, \emph{{Cornering
  large-N$_{c}$ QCD with positivity bounds}},
  \href{http://dx.doi.org/10.1007/JHEP06(2023)094}{\emph{JHEP} {\bf 06} (2023)
  094}, [\href{https://arxiv.org/abs/2211.12488}{{\tt 2211.12488}}].

\bibitem{Li:2023qzs}
Y.-Z. Li, \emph{{Effective field theory bootstrap, large-N $\chi$PT and
  holographic QCD}},  \href{https://arxiv.org/abs/2310.09698}{{\tt
  2310.09698}}.

\bibitem{Huang:2020nqy}
Y.-t. Huang, J.-Y. Liu, L.~Rodina and Y.~Wang, \emph{{Carving out the Space of
  Open-String S-matrix}},
  \href{http://dx.doi.org/10.1007/JHEP04(2021)195}{\emph{JHEP} {\bf 04} (2021)
  195}, [\href{https://arxiv.org/abs/2008.02293}{{\tt 2008.02293}}].

\bibitem{Chiang:2023quf}
L.-Y. Chiang, Y.-t. Huang and H.-C. Weng, \emph{{Bootstrapping string theory
  EFT}},  \href{https://arxiv.org/abs/2310.10710}{{\tt 2310.10710}}.

\bibitem{Berman:2023jys}
J.~Berman, H.~Elvang and A.~Herderschee, \emph{{Flattening of the EFT-Hedron:
  Supersymmetric Positivity Bounds and the Search for String Theory}},
  \href{https://arxiv.org/abs/2310.10729}{{\tt 2310.10729}}.

\bibitem{Figueroa:2022onw}
F.~Figueroa and P.~Tourkine, \emph{{Unitarity and Low Energy Expansion of the
  Coon Amplitude}},
  \href{http://dx.doi.org/10.1103/PhysRevLett.129.121602}{\emph{Phys. Rev.
  Lett.} {\bf 129} (2022) 121602},
  [\href{https://arxiv.org/abs/2201.12331}{{\tt 2201.12331}}].

\bibitem{Geiser:2022icl}
N.~Geiser and L.~W. Lindwasser, \emph{{Properties of infinite product
  amplitudes: Veneziano, Virasoro, and Coon}},
  \href{http://dx.doi.org/10.1007/JHEP12(2022)112}{\emph{JHEP} {\bf 12} (2022)
  112}, [\href{https://arxiv.org/abs/2207.08855}{{\tt 2207.08855}}].

\bibitem{Chakravarty:2022vrp}
J.~Chakravarty, P.~Maity and A.~Mishra, \emph{{On the positivity of Coon
  amplitude in D = 4}},
  \href{http://dx.doi.org/10.1007/JHEP10(2022)043}{\emph{JHEP} {\bf 10} (2022)
  043}, [\href{https://arxiv.org/abs/2208.02735}{{\tt 2208.02735}}].

\bibitem{Bhardwaj:2022lbz}
R.~Bhardwaj, S.~De, M.~Spradlin and A.~Volovich, \emph{{On unitarity of the
  Coon amplitude}},
  \href{http://dx.doi.org/10.1007/JHEP08(2023)082}{\emph{JHEP} {\bf 08} (2023)
  082}, [\href{https://arxiv.org/abs/2212.00764}{{\tt 2212.00764}}].

\bibitem{Jepsen:2023sia}
C.~B. Jepsen, \emph{{Cutting the Coon amplitude}},
  \href{http://dx.doi.org/10.1007/JHEP06(2023)114}{\emph{JHEP} {\bf 06} (2023)
  114}, [\href{https://arxiv.org/abs/2303.02149}{{\tt 2303.02149}}].

\bibitem{Geiser:2022exp}
N.~Geiser and L.~W. Lindwasser, \emph{{Generalized Veneziano and Virasoro
  amplitudes}}, \href{http://dx.doi.org/10.1007/JHEP04(2023)031}{\emph{JHEP}
  {\bf 04} (2023) 031}, [\href{https://arxiv.org/abs/2210.14920}{{\tt
  2210.14920}}].

\bibitem{Cheung:2022mkw}
C.~Cheung and G.~N. Remmen, \emph{{Veneziano variations: how unique are string
  amplitudes?}}, \href{http://dx.doi.org/10.1007/JHEP01(2023)122}{\emph{JHEP}
  {\bf 01} (2023) 122}, [\href{https://arxiv.org/abs/2210.12163}{{\tt
  2210.12163}}].

\bibitem{Geiser:2023qqq}
N.~Geiser, \emph{{The Baker-Coon-Romans $N$-point amplitude and an exact field
  theory limit of the Coon amplitude}},
  \href{https://arxiv.org/abs/2311.04130}{{\tt 2311.04130}}.

\bibitem{McPeak:2023wmq}
B.~McPeak, M.~Venuti and A.~Vichi, \emph{{Adding subtractions: comparing the
  impact of different Regge behaviors}},
  \href{https://arxiv.org/abs/2310.06888}{{\tt 2310.06888}}.

\bibitem{Mizera:2021fap}
S.~Mizera, \emph{{Crossing symmetry in the planar limit}},
  \href{http://dx.doi.org/10.1103/PhysRevD.104.045003}{\emph{Phys. Rev. D} {\bf
  104} (2021) 045003}, [\href{https://arxiv.org/abs/2104.12776}{{\tt
  2104.12776}}].

\bibitem{Hebbar:2020ukp}
A.~Hebbar, D.~Karateev and J.~Penedones, \emph{{Spinning S-matrix bootstrap in
  4d}}, \href{http://dx.doi.org/10.1007/JHEP01(2022)060}{\emph{JHEP} {\bf 01}
  (2022) 060}, [\href{https://arxiv.org/abs/2011.11708}{{\tt 2011.11708}}].

\bibitem{Igi:1962zz}
K.~Igi, \emph{{pi-N Scattering Length and Singularities in the Complex J
  Plane}}, \href{http://dx.doi.org/10.1103/PhysRevLett.9.76}{\emph{Phys. Rev.
  Lett.} {\bf 9} (1962) 76--79}.

\bibitem{Logunov:1967dy}
A.~A. Logunov, L.~D. Soloviev and A.~N. Tavkhelidze, \emph{{Dispersion sum
  rules and high-energy scattering}},
  \href{http://dx.doi.org/10.1016/0370-2693(67)90487-X}{\emph{Phys. Lett. B}
  {\bf 24} (1967) 181--182}.

\bibitem{Igi:1967zza}
K.~Igi and S.~Matsuda, \emph{{New Sum Rules and Singularities in the Complex J
  Plane}}, \href{http://dx.doi.org/10.1103/PhysRevLett.18.625}{\emph{Phys. Rev.
  Lett.} {\bf 18} (1967) 625--627}.

\bibitem{Gatto:1967zza}
R.~Gatto, \emph{{New Sum Rules for Superconvergence}},
  \href{http://dx.doi.org/10.1103/PhysRevLett.18.803}{\emph{Phys. Rev. Lett.}
  {\bf 18} (1967) 803--806}.

\bibitem{Dolen:1967zz}
R.~Dolen, D.~Horn and C.~Schmid, \emph{{Prediction of Regge Parameters of rho
  Poles from Low-Energy pi N Data}},
  \href{http://dx.doi.org/10.1103/PhysRevLett.19.402}{\emph{Phys. Rev. Lett.}
  {\bf 19} (1967) 402--407}.

\bibitem{Dolen:1967jr}
R.~Dolen, D.~Horn and C.~Schmid, \emph{{Finite energy sum rules and their
  application to pi N charge exchange}},
  \href{http://dx.doi.org/10.1103/PhysRev.166.1768}{\emph{Phys. Rev.} {\bf 166}
  (1968) 1768--1781}.

\bibitem{Ademollo:1967zz}
M.~Ademollo, H.~R. Rubinstein, G.~Veneziano and M.~A. Virasoro,
  \emph{{Bootstraplike Conditions from Superconvergence}},
  \href{http://dx.doi.org/10.1103/PhysRevLett.19.1402}{\emph{Phys. Rev. Lett.}
  {\bf 19} (1967) 1402--1405}.

\bibitem{Ademollo:1968cno}
M.~Ademollo, H.~R. Rubinstein, G.~Veneziano and M.~A. Virasoro,
  \emph{{Bootstrap of meson trajectories from superconvergence}},
  \href{http://dx.doi.org/10.1103/PhysRev.176.1904}{\emph{Phys. Rev.} {\bf 176}
  (1968) 1904--1925}.

\bibitem{Mukhametzhanov:2018zja}
B.~Mukhametzhanov and A.~Zhiboedov, \emph{{Analytic Euclidean Bootstrap}},
  \href{http://dx.doi.org/10.1007/JHEP10(2019)270}{\emph{JHEP} {\bf 10} (2019)
  270}, [\href{https://arxiv.org/abs/1808.03212}{{\tt 1808.03212}}].

\bibitem{Strominger:2017zoo}
A.~Strominger, \emph{{Lectures on the Infrared Structure of Gravity and Gauge
  Theory}}.
\newblock 3, 2017.

\bibitem{Caron-Huot:2021enk}
S.~Caron-Huot, D.~Mazac, L.~Rastelli and D.~Simmons-Duffin, \emph{{AdS bulk
  locality from sharp CFT bounds}},
  \href{http://dx.doi.org/10.1007/JHEP11(2021)164}{\emph{JHEP} {\bf 11} (2021)
  164}, [\href{https://arxiv.org/abs/2106.10274}{{\tt 2106.10274}}].

\bibitem{Chang:2023szz}
C.-H. Chang, Y.~Landau and D.~Simmons-Duffin, \emph{{Spinning dispersive CFT
  sum rules and bulk scattering}},
  \href{https://arxiv.org/abs/2311.04271}{{\tt 2311.04271}}.

\bibitem{Altarelli:1969ck}
G.~Altarelli and H.~R. Rubinstein, \emph{{Closed forms for the scattering
  amplitudes and bootstrap based on sum rules}},
  \href{http://dx.doi.org/10.1103/PhysRev.178.2165}{\emph{Phys. Rev.} {\bf 178}
  (1969) 2165--2166}.

\bibitem{Virasoro:1969me}
M.~A. Virasoro, \emph{{Alternative constructions of crossing-symmetric
  amplitudes with regge behavior}},
  \href{http://dx.doi.org/10.1103/PhysRev.177.2309}{\emph{Phys. Rev.} {\bf 177}
  (1969) 2309--2311}.

\bibitem{Shapiro:1970gy}
J.~A. Shapiro, \emph{{Electrostatic analog for the virasoro model}},
  \href{http://dx.doi.org/10.1016/0370-2693(70)90255-8}{\emph{Phys. Lett. B}
  {\bf 33} (1970) 361--362}.

\bibitem{Polchinski:1998rq}
J.~Polchinski, \emph{{String theory. Vol. 1: An introduction to the bosonic
  string}}.
\newblock Cambridge Monographs on Mathematical Physics. Cambridge University
  Press, 12, 2007,
  \href{http://dx.doi.org/10.1017/CBO9780511816079}{10.1017/CBO9780511816079}.

\bibitem{Martin:1970hmp}
A.~D. Martin and T.~D. Spearman, \emph{{Elementary Particle Theory}}.
\newblock North-Holland Publishing Co., Amsterdam, 1970.

\bibitem{Kawai:1985xq}
H.~Kawai, D.~C. Lewellen and S.~H.~H. Tye, \emph{{A Relation Between Tree
  Amplitudes of Closed and Open Strings}},
  \href{http://dx.doi.org/10.1016/0550-3213(86)90362-7}{\emph{Nucl. Phys. B}
  {\bf 269} (1986) 1--23}.

\bibitem{Simmons-Duffin:2015qma}
D.~Simmons-Duffin, \emph{{A Semidefinite Program Solver for the Conformal
  Bootstrap}}, \href{http://dx.doi.org/10.1007/JHEP06(2015)174}{\emph{JHEP}
  {\bf 06} (2015) 174}, [\href{https://arxiv.org/abs/1502.02033}{{\tt
  1502.02033}}].

\bibitem{Landry:2019qug}
W.~Landry and D.~Simmons-Duffin, \emph{{Scaling the semidefinite program solver
  SDPB}},  \href{https://arxiv.org/abs/1909.09745}{{\tt 1909.09745}}.

\bibitem{gurobi}
{Gurobi Optimization, LLC}, \emph{{Gurobi Optimizer Reference Manual}},  2023.

\bibitem{Camanho:2014apa}
X.~O. Camanho, J.~D. Edelstein, J.~Maldacena and A.~Zhiboedov, \emph{{Causality
  Constraints on Corrections to the Graviton Three-Point Coupling}},
  \href{http://dx.doi.org/10.1007/JHEP02(2016)020}{\emph{JHEP} {\bf 02} (2016)
  020}, [\href{https://arxiv.org/abs/1407.5597}{{\tt 1407.5597}}].

\bibitem{khuri1969derivation}
N.~Khuri, \emph{Derivation of a veneziano series from the regge
  representation}, {\emph{Physical Review} {\bf 185} (1969) 1876}.

\bibitem{Matsuda:1969zz}
S.~Matsuda, \emph{{MODEL FOR INFINITE VENEZIANO SERIES}}, .

\bibitem{Mandelstam:1968czc}
S.~Mandelstam, \emph{{Veneziano formula with trajectories spaced by two
  units}}, \href{http://dx.doi.org/10.1103/PhysRevLett.21.1724}{\emph{Phys.
  Rev. Lett.} {\bf 21} (1968) 1724--1728}.

\bibitem{Goddard:1972iy}
P.~Goddard and C.~B. Thorn, \emph{{Compatibility of the Dual Pomeron with
  Unitarity and the Absence of Ghosts in the Dual Resonance Model}},
  \href{http://dx.doi.org/10.1016/0370-2693(72)90420-0}{\emph{Phys. Lett. B}
  {\bf 40} (1972) 235--238}.

\bibitem{Arkani-Hamed:2022gsa}
N.~Arkani-Hamed, L.~Eberhardt, Y.-t. Huang and S.~Mizera, \emph{{On unitarity
  of tree-level string amplitudes}},
  \href{http://dx.doi.org/10.1007/JHEP02(2022)197}{\emph{JHEP} {\bf 02} (2022)
  197}, [\href{https://arxiv.org/abs/2201.11575}{{\tt 2201.11575}}].

\bibitem{Sever:2017ylk}
A.~Sever and A.~Zhiboedov, \emph{{On Fine Structure of Strings: The Universal
  Correction to the Veneziano Amplitude}},
  \href{http://dx.doi.org/10.1007/JHEP06(2018)054}{\emph{JHEP} {\bf 06} (2018)
  054}, [\href{https://arxiv.org/abs/1707.05270}{{\tt 1707.05270}}].

\bibitem{Costa:2017twz}
M.~S. Costa, T.~Hansen and J.~a. Penedones, \emph{{Bounds for OPE coefficients
  on the Regge trajectory}},
  \href{http://dx.doi.org/10.1007/JHEP10(2017)197}{\emph{JHEP} {\bf 10} (2017)
  197}, [\href{https://arxiv.org/abs/1707.07689}{{\tt 1707.07689}}].

\bibitem{Paulos:2017fhb}
M.~F. Paulos, J.~Penedones, J.~Toledo, B.~C. van Rees and P.~Vieira, \emph{{The
  S-matrix bootstrap. Part III: higher dimensional amplitudes}},
  \href{http://dx.doi.org/10.1007/JHEP12(2019)040}{\emph{JHEP} {\bf 12} (2019)
  040}, [\href{https://arxiv.org/abs/1708.06765}{{\tt 1708.06765}}].

\bibitem{Cerulus:1964cjb}
F.~A. Cerulus and A.~Martin, \emph{{A lower bound for large-angle elastic
  scattering at high energies}},
  \href{http://dx.doi.org/10.1016/0031-9163(64)90807-8}{\emph{Phys. Lett.} {\bf
  8} (1964) 80--82}.

\bibitem{Caron-Huot:2022jli}
S.~Caron-Huot, Y.-Z. Li, J.~Parra-Martinez and D.~Simmons-Duffin,
  \emph{{Graviton partial waves and causality in higher dimensions}},
  \href{http://dx.doi.org/10.1103/PhysRevD.108.026007}{\emph{Phys. Rev. D} {\bf
  108} (2023) 026007}, [\href{https://arxiv.org/abs/2205.01495}{{\tt
  2205.01495}}].

\bibitem{Bachu:2022gof}
B.~Bachu and A.~Hillman, \emph{{Stringy Completions of the Standard Model from
  the Bottom Up}},  \href{https://arxiv.org/abs/2212.03871}{{\tt 2212.03871}}.

\bibitem{Arkani-Hamed:2020gyp}
N.~Arkani-Hamed, M.~Pate, A.-M. Raclariu and A.~Strominger, \emph{{Celestial
  amplitudes from UV to IR}},
  \href{http://dx.doi.org/10.1007/JHEP08(2021)062}{\emph{JHEP} {\bf 08} (2021)
  062}, [\href{https://arxiv.org/abs/2012.04208}{{\tt 2012.04208}}].

\bibitem{Fubini:1969qb}
S.~Fubini and G.~Veneziano, \emph{{Level structure of dual-resonance models}},
  \href{http://dx.doi.org/10.1007/BF02758835}{\emph{Nuovo Cim. A} {\bf 64}
  (1969) 811--840}.

\bibitem{Gross:1969db}
D.~J. Gross, \emph{{Factorization and the generalized veneziano model with
  satellites}},
  \href{http://dx.doi.org/10.1016/0550-3213(69)90248-X}{\emph{Nucl. Phys. B}
  {\bf 13} (1969) 467--476}.

\bibitem{Bianchi:2020cfc}
M.~Bianchi, D.~Consoli and P.~Di~Vecchia, \emph{{On the N-pion extension of the
  Lovelace-Shapiro model}},
  \href{http://dx.doi.org/10.1007/JHEP03(2021)119}{\emph{JHEP} {\bf 03} (2021)
  119}, [\href{https://arxiv.org/abs/2002.05419}{{\tt 2002.05419}}].

\bibitem{Caron-Huot:2020cmc}
S.~Caron-Huot and V.~Van~Duong, \emph{{Extremal Effective Field Theories}},
  \href{http://dx.doi.org/10.1007/JHEP05(2021)280}{\emph{JHEP} {\bf 05} (2021)
  280}, [\href{https://arxiv.org/abs/2011.02957}{{\tt 2011.02957}}].

\bibitem{Haring:2022cyf}
K.~H\"aring and A.~Zhiboedov, \emph{{Gravitational Regge bounds}},
  \href{https://arxiv.org/abs/2202.08280}{{\tt 2202.08280}}.

\bibitem{Correia:2020xtr}
M.~Correia, A.~Sever and A.~Zhiboedov, \emph{{An analytical toolkit for the
  S-matrix bootstrap}},
  \href{http://dx.doi.org/10.1007/JHEP03(2021)013}{\emph{JHEP} {\bf 03} (2021)
  013}, [\href{https://arxiv.org/abs/2006.08221}{{\tt 2006.08221}}].

\bibitem{Gribov:2003nw}
V.~N. Gribov, \emph{{The theory of complex angular momenta: Gribov lectures on
  theoretical physics}}.
\newblock Cambridge Monographs on Mathematical Physics. Cambridge University
  Press, 6, 2007,
  \href{http://dx.doi.org/10.1017/CBO9780511534959}{10.1017/CBO9780511534959}.

\bibitem{wignerE}
M.~Andrews and J.~Gunson, \emph{{Complex Angular Momenta and Many‐Particle
  States. I. Properties of Local Representations of the Rotation Group}},
  \href{http://dx.doi.org/10.1063/1.1704074}{\emph{Journal of Mathematical
  Physics} {\bf 5} (12, 2004) 1391--1400}.

\bibitem{Haring:2022sdp}
K.~H\"aring, A.~Hebbar, D.~Karateev, M.~Meineri and J.~a. Penedones,
  \emph{{Bounds on photon scattering}},
  \href{https://arxiv.org/abs/2211.05795}{{\tt 2211.05795}}.

\bibitem{Sivers:1971ig}
D.~Sivers and J.~Yellin, \emph{{Review of recent work on narrow resonance
  models}}, \href{http://dx.doi.org/10.1103/RevModPhys.43.125}{\emph{Rev. Mod.
  Phys.} {\bf 43} (1971) 125--188}.

\bibitem{buck1948interpolation}
R.~C. Buck, \emph{Interpolation series}, {\emph{Transactions of the American
  Mathematical Society} {\bf 64} (1948) 283--298}.

\bibitem{Tourkine:2023xtu}
P.~Tourkine and A.~Zhiboedov, \emph{{Scattering amplitudes from dispersive
  iterations of unitarity}},
  \href{http://dx.doi.org/10.1007/JHEP11(2023)005}{\emph{JHEP} {\bf 11} (2023)
  005}, [\href{https://arxiv.org/abs/2303.08839}{{\tt 2303.08839}}].

\bibitem{Buoninfante:2023dyd}
L.~Buoninfante, J.~Tokuda and M.~Yamaguchi, \emph{{New lower bounds on
  scattering amplitudes: non-locality constraints}},
  \href{https://arxiv.org/abs/2305.16422}{{\tt 2305.16422}}.

\end{thebibliography}\endgroup

\end{document}